\documentclass[journal,10pt]{IEEEtran}
\ifCLASSINFOpdf
\else
\fi

\usepackage[cmex10]{amsmath}
\usepackage{amssymb,amsthm,amsfonts}
\usepackage{stackrel,graphicx,subfig}
\usepackage{color}
\usepackage{amssymb,mathrsfs}
\usepackage{soul,dsfont,enumerate}
\usepackage[export]{adjustbox}
\usepackage{booktabs}
\usepackage{multirow}
\usepackage{algorithm}
\usepackage{algpseudocode}
\usepackage{pifont,dsfont}
\usepackage[outdir=./Figures/]{epstopdf}

\newcommand{\domain}{\Omega}
\newcommand{\Curve}{\mathcal C}

\newcommand{\ud}{\, \mathrm{d}}
\newcommand{\R}{ \mathbb{R}}
\newcommand{\Z}{ \mathbb{Z}}

\newcommand{\dd}{ {\rm d} }

\newcommand{\ind}{\mathds{1}}

\theoremstyle{plain}\newtheorem{theo}{Theorem}
\theoremstyle{plain}\newtheorem{defi}{Definition}
\theoremstyle{plain}
\theoremstyle{plain}\newtheorem{remark}{Remark}
\theoremstyle{plain}

\begin{document}

\title{Sampling and Reconstruction of Shapes with Algebraic Boundaries}
\author{Mitra~Fatemi,
		Arash~Amini,
        and~Martin~Vetterli
\thanks{M. Fatemi and M. Vetterli are with the Department
of Computer and Communication Sciences, \'Ecole Polytechnique F\'ed\'eral de Lausanne (e-mail: \{mitra.fatemi, martin.vetterli\}@epfl.ch).}
\thanks{A. Amini is with the Department of Electrical Engineering, Sharif University of Technology (e-mail: aamini@sharif.edu).}}

\maketitle

\begin{abstract}
We present a sampling theory for a class of binary images with finite rate of innovation (FRI). Every image in our model is the restriction of $\ind_{\{p\leq0\}}$ to the image plane, where $\ind$ denotes the indicator function and $p$ is some real bivariate polynomial. This particularly means that the boundaries in the image form a subset of an algebraic curve with the implicit polynomial $p$. We show that the image parameters --i.e., the polynomial coefficients-- satisfy a set of linear annihilation equations with the coefficients being the image moments. The inherent sensitivity of the moments to  noise makes the reconstruction process numerically unstable and narrows the choice of the sampling kernels to polynomial reproducing kernels. As a remedy to these problems, we replace conventional moments with more stable \emph{generalized moments} that are adjusted to the given sampling kernel. The benefits are threefold: (1) it relaxes the requirements on the sampling kernels, (2) produces annihilation equations that are robust at numerical precision, and (3) extends the results to images with unbounded boundaries. We further reduce the sensitivity of the reconstruction process to noise by taking into account the sign of the polynomial at certain points, and sequentially enforcing  measurement consistency. We consider various numerical experiments to demonstrate the performance of our algorithm in reconstructing binary images, including low to moderate noise levels and a range of realistic sampling kernels.
\end{abstract}


\IEEEpeerreviewmaketitle

\section{Introduction}\label{sec:intro}
In today's digital world, sampling is a key block of any signal acquisition device: the device senses and stores analog signals at certain points and uses the samples later for the representation of the analog signal (possibly after some post-processing). 
The main concern here is whether and how the collected samples provide a fair representation of the original signal.
Hence, as a first step in the design of acquisition devices, we should develop suitable sampling and reconstruction techniques for the target class of signals. 

The classical Shannon sampling theory and its variations present sampling strategies for bandlimited signals and more generally the class of signals living in a shift-invariant space \cite{Shannon48,Jerri77,Unser2000,Vetterli-Book2014}. Still many crucial signals stay out of reach of this class. Among them are signals which can be described with a finite number of parameters, hence called signals with finite rate of innovation (FRI). In \cite{Vetterli2002}, a study of one-dimensional (1D) FRI signals was presented. This work then evolved to include more general FRI signals such as piecewise polynomials \cite{Vetterli2002,Maravic2005}, streams of Diracs \cite{Vetterli2002,Maravic2005,Blu2008} and piecewise sinusoids \cite{Berent2010}. 

Extension of sampling schemes to images is an essential but challenging problem. Because of the sharp intensity transitions along edges, images are non-bandlimited. Also, the diverse geometry of the edges in typical images excludes them from the known shift-invariant spaces. Some preliminary efforts to generalize the FRI framework to images led to the sampling schemes with adequate sampling kernels for step-edge images and polygons \cite{Maravic2004,Shukla2007,Chen2012}. In a recent work, an FRI-based sampling scheme is presented for images with more versatile edge geometries \cite{Pan2014}. The curves in this model are zero level sets of a mask function that is a linear combination of a finite number of two-dimensional (2D) exponentials. The curve parameters are shown to satisfy an annihilation system of equations which could be solved directly, or more robustly as a minimization problem.

The curve model introduced in \cite{Pan2014} is novel and further investigation is needed to reveal its descriptive power --i.e., the range of shape geometries and the number of free parameters required for generating a given shape in the range. On the other hand, a rich parametric model for 2D curves already exists in the literature: algebraic curves \cite{Walker50,Fulton89,Arbarello2011}. An algebraic curve is the zero level set of a bivariate polynomial of a finite degree. Algebraic curves can be decomposed into a finite number of smooth arcs. Nevertheless, they are dense, in the Hausdorff metric, among all smooth curves. Hence, every curve can be approximated by a sequence of algebraic curves arbitrarily closely \cite{Gustafsson83}. This characteristic makes them an excellent candidate in modeling the general image boundaries.

We call a subset of the 2D plane with an algebraic boundary curve an \emph{algebraic domain} and the restriction of it to the image plane an \emph{algebraic shape}. According to a classical result \cite{Karlin66}, an algebraic domain of degree $n$ can be uniquely determined from its set of 2D moments of order less than or equal to $n$. But as stated in  \cite{Milanfar2000}, ``there has been so far no constructive way of passing from the given moments to the unique algebraic domain, or equivalently to the defining polynomial". In \cite{Milanfar2000} and \cite{Gustafsson2000}, the authors present an algorithm for the reconstruction of a subset of bounded algebraic domains --called quadrature domains-- from their moments. However, moments are inherently very sensitive to noise and consequently, the suggested algorithm (as noted by the authors) suffers from severe numerical instabilities.

Moments have been used as the standard descriptors of 2D shapes in \cite{Rad91,Desai94,Tsirikolias93}. Also, there are some works on the exact calculation of moments of the shapes with parametric boundary curves in terms of the curve parameters, through nonlinear equations. Examples are \cite{Singer93} for polygonal shapes and \cite{Jacob2001} for shapes with wavelet and spline curves.

\subsection{Contribution}
In this paper, we propose sampling and reconstruction techniques for algebraic shapes. We first derive a set of linear annihilation equations for the shape parameters with the coefficients being factors of 2D moments of the image. We prove that any solution of these equations will lead to a polynomial that vanishes on the boundaries of the original shape. By employing sampling kernels that reproduce polynomials like the well-known B-splines \cite{Unser99}, we are able to calculate the shape moments from the samples. 

Moments are inherently very sensitive to the noise and the reason is that noise in the image or the samples is boosted by polynomial factors before it contaminates the moments. To overcome this difficulty, we replace moments with some generalized moments that are still reproducible from the samples but do not amplify the noise. This is achieved by multiplying the monomials in the conventional moments with a function that is adjusted to the sampling kernel and decays at the image borders. The advantages are threefold: we get more stable moments that can be reproduced by a wider range of sampling kernels. Furthermore, we can extend our model to algebraic shapes with unbounded boundaries.

In any sampling problem, consistency of the reconstruction with noiseless samples is a crucial constraint \cite{Thao94,Unser94,Unser98}. It is also proved to be a strong tool for recovering binary images in the absence of a parametric model \cite{Fatemi2015}. In this paper, we further improve the stability of our reconstruction by enforcing measurement (or sample) consistency to the recovered algebraic shape. This results in a reconstruction algorithm that is robust to moderate noise levels in the samples.

\subsection{Organization of the paper}
The paper is organized as follows. In Section \ref{sec:definition}, we first define the image model and study algebraic curves in details. Then, we explicitly define the sampling problem. We derive the annihilation equations for the shape parameters in Section \ref{sec:recons} and present a perfect reconstruction algorithm for the noiseless scenario. In Section \ref{sec:stable_recovery}, we develop a stable reconstruction algorithm. For this purpose, we introduce the notion of generalized moments and present an algorithm for generating the adequate generalized moments corresponding to the given sampling kernel. Also, we prove that any solution of the annihilation equations formed from (generalized) moments generates the original shape boundaries. We present some experimental results with different curves in the noiseless and noisy scenarios in Section \ref{sec:experiments} and conclude in Section \ref{sec:conclusion}.

\section{Sampling of algebraic shapes}\label{sec:definition}
\subsection{Image model}
Consider a bivariate polynomial of degree $n$ with real coefficients $a_{i,j}$ 
\begin{align}\label{eq:polynomial}
p(x,y)=\sum_{{0\leq i,j,~i+j\leq n}}a_{i,j}x^iy^j.
\end{align}
The set of points $\{(x,y)\in\R^2:\,p(x,y)\leq 0\}$ defines an algebraic domain. The boundary of this domain, defined by the zero level set of $p$, is an algebraic curve of degree $n$,
\begin{align*}
\Curve=\{(x,y)\in\R^2:\,p(x,y)=0\}.
\end{align*}
Let $\Omega$ denote a closed domain in $\R^2$ modeling the image plane. Without loss of generality, we take $\Omega=[-L,L]^2$ for some $L\in\mathbb{Z}^+$. We define an algebraic shape in $\domain$ as the binary image
\begin{align}\label{eq:AS}
I(x,y)=\ind_{\{(p(x,y)\leq 0\}},~(x,y)\in\domain,
\end{align}
where $\ind$ denotes the indicator function. This means that the edges of $I$ are contained in the algebraic curve $\Curve$.

An algebraic shape of degree $n$ is specified with $\binom{n+2}{2}$ parameters (the coefficients in \eqref{eq:polynomial}). In developing the annihilation equations of Section \ref{sec:recons}, we assume that the algebraic shapes have closed boundaries. This restricts the polynomial degree to the even integers. We later remove this assumption by introducing generalized moments in Section
\ref{sec:stable_recovery}.

\begin{figure}
\centering
\includegraphics[width=0.85\linewidth]{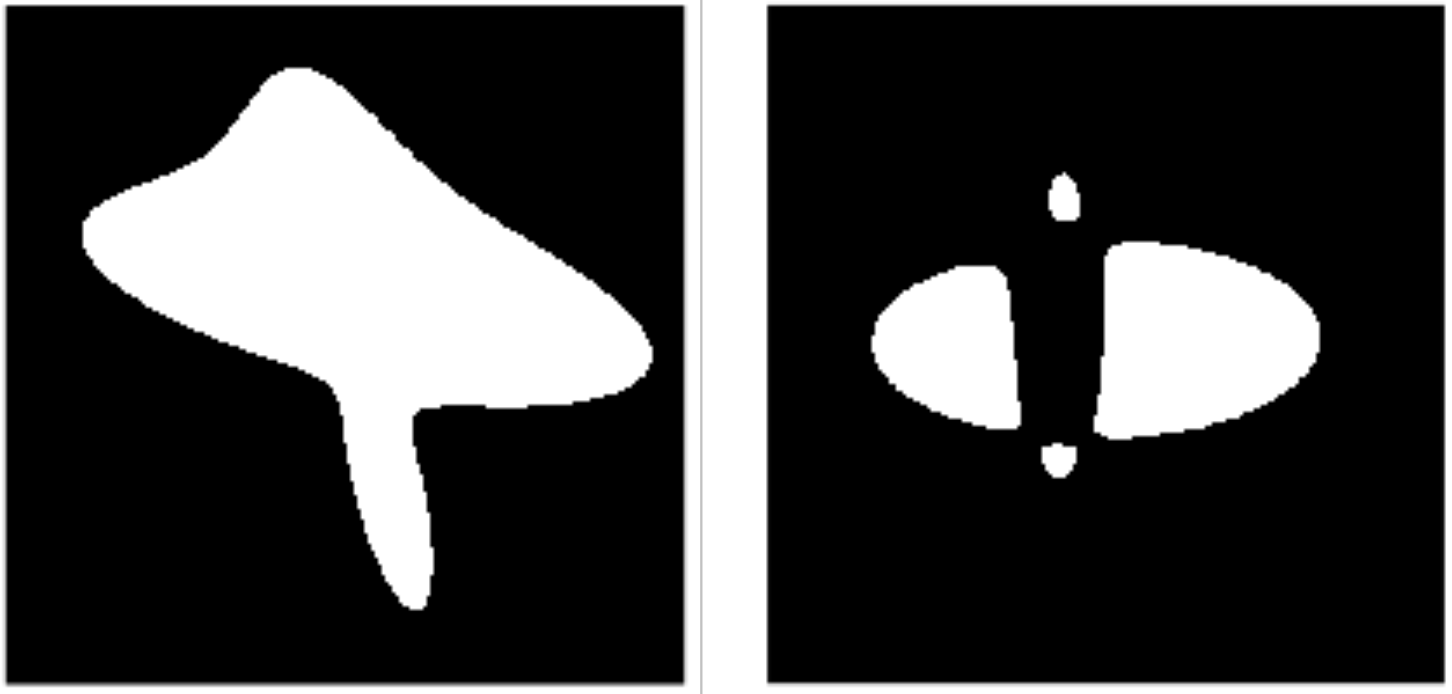}
\caption{Algebraic domains of degree 4.}\label{fig:ACexamples}
\end{figure}

Typical examples of algebraic domains of degree 2 are circles and ellipses. Figure \ref{fig:ACexamples} displays two algebraic domains of degree 4. We see in this figure that an algebraic domain of degree 4 can have four disconnected components. The following remark asserts that this is an upper bound.

\begin{remark}[\cite{Keren94}]
An algebraic domain of degree $n$ cannot have more than $n$ disconnected closed components.
\end{remark}

This remark is a consequence of Bezout's theorem \cite{Walker50}. We will also make use of this theorem in Section \ref{sec:stable_recovery} to prove our result. 

\begin{theo}[Bezout]
Two algebraic curves of degree $n$ and $m$ that do not share a common component intersect in at most $mn$ points.
\end{theo}

\begin{figure}
\centering
\includegraphics[width=0.85\linewidth]{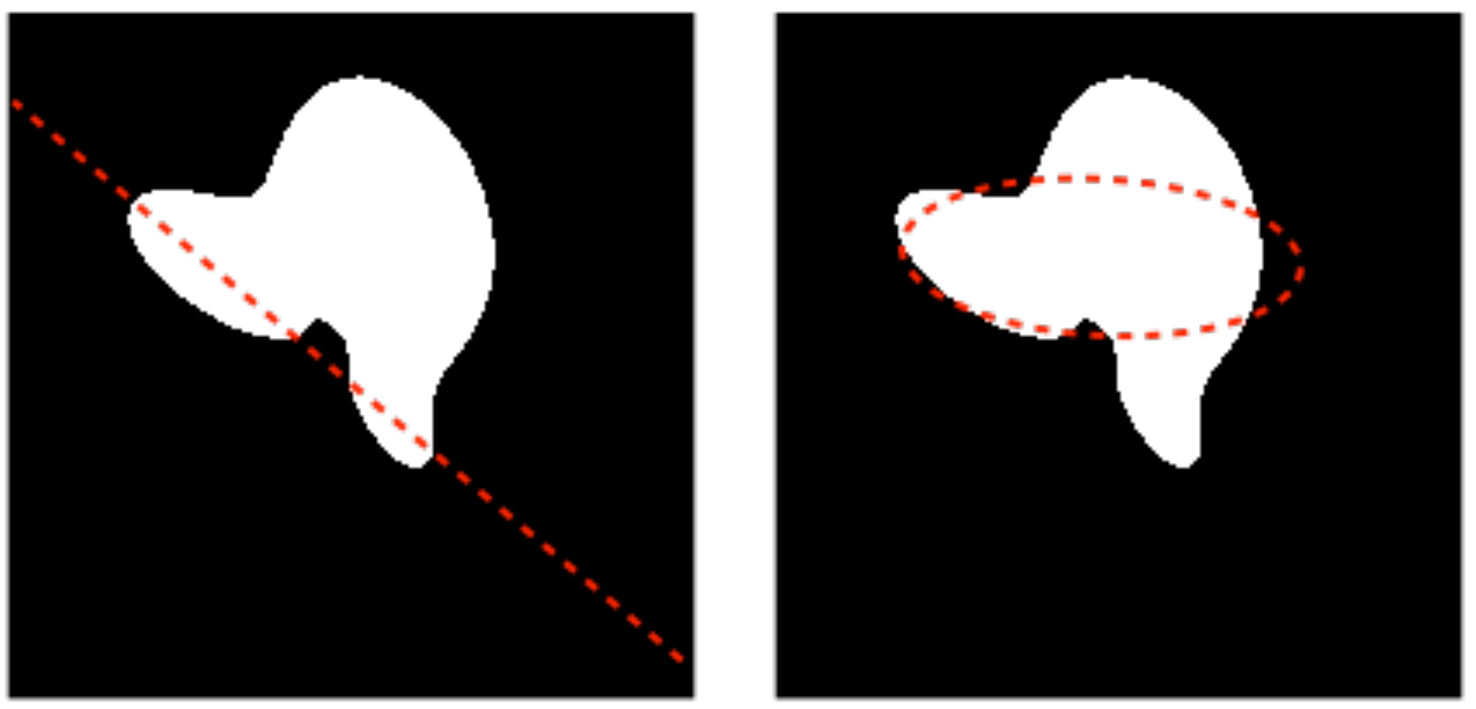}
\caption{An algebraic shape of degree at least 4.}\label{fig:Bezout}
\end{figure}

Bezout's theorem also provides us a handy tool to roughly estimate the degree of an algebraic shape. Consider a shape image $I$ with boundary $\Curve$. $\Curve$ should have a degree of at least $n$ if it intersects a line (a first-degree polynomial) at $n$ points or if it intersects an ellipse (a polynomial of degree 2) at $2n$ points. This is illustrated with an example in Figure \ref{fig:Bezout}.

Algebraic curves have been studied and applied to data fitting in computer vision (\emph{e.g.} \cite{Keren94,Taubin94}). This rather long history of application has revealed that polynomials of modest degree (\emph{e.g.} degree 4 with 15 parameters) have enough descriptive power to generate a diverse range of curve geometries. Hence, in the rest of this paper, we mostly consider $n\leq4$. Nevertheless, all results remain valid for higher degree polynomials. 

\subsection{Sampling}
\begin{figure}
\centering
\includegraphics[width=0.7\linewidth]{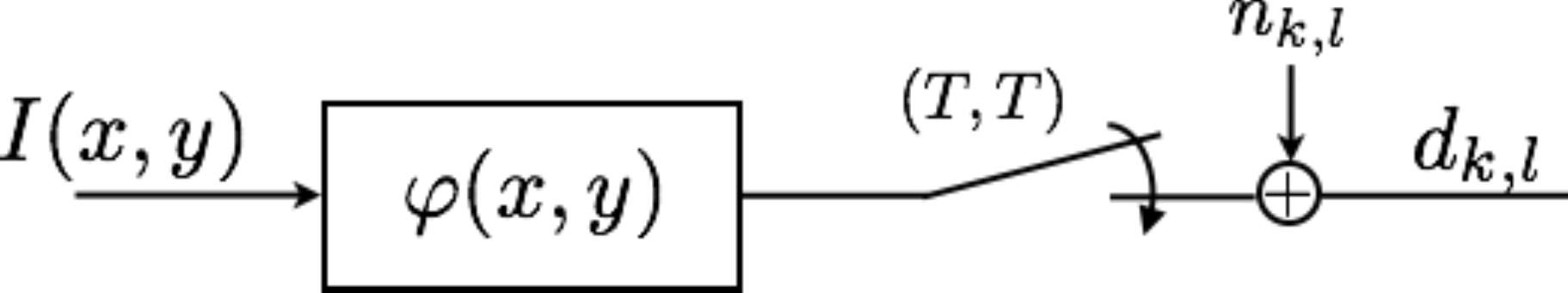}
\caption{In a typical sampling scenario, the image goes through convolution with a 2D kernel and spacial sampling to generate the measurements.}\label{fig:sampling}
\end{figure}

In a typical sampling setup (Figure \ref{fig:sampling}), the image is first convolved with a 2D kernel and then sampled at a uniform grid to generate the samples
\begin{align*}
d_{k,l}=\frac{1}{T^2}\iint_\Omega I(x,y)\varphi\Big(\frac{x}{T}-k,\frac{y}{T}-l\Big)\ud x\ud y.
\end{align*}
In a noisy setup, the noise vector will be added to the measurements after spacial sampling. The sampling kernel $\varphi(x,y)$ is determined by the physics of the sampling device but in most cases it can be considered as a separable kernel $\varphi(x)\varphi(y)$. In the first part of this paper, we consider separable kernels that can reproduce polynomials up to some degree. $\varphi(x)$ is a polynomial reproducing kernel of degree $\mathcal{N}$ if there exist coefficients $c_k^{(i)}$ such that \cite{Dragotti2007}
\begin{align*}
\sum_{k\in\mathbb{Z}}c_k^{(i)}\varphi(x-k)=x^i,~~i=0,...,\mathcal{N}.
\end{align*}
B-splines are well-known examples of polynomial reproducing kernels \cite{Unser99}. A zero order B-spline $\beta^{(0)}(x)$ is defined as
\begin{align*}
\beta^{(0)}(x) =\left\{\begin{array}{ll}
1,&-0.5<x<0.5\\
0.5,&|x|=0.5\\
0,&\text{otherwise.}\end{array}\right.
\end{align*}
A B-spline of order $m$ is obtained by convolving $m+1$ kernel $\beta^{(0)}(x)$
\begin{align*}
\beta^{(m)}(x)=\underbrace{\beta^{(0)}*\beta^{(0)}*...*\beta^{(0)}}_{m+1~\text{times}}.
\end{align*}
The B-spline kernel $\beta^{(m)}$ can reproduce monomials up to degree $m$ and the corresponding coefficients are obtained as 
\begin{align*}
c_k^{(i)}=\langle x^i,\tilde{\beta}^{(m)}(x-k)\rangle,
\end{align*} 
where $\tilde{\beta}^{(m)}(.)$ is the dual of $\beta^{(m)}(.)$ \cite{Unser99}.

\begin{figure}
\centering
\subfloat[]{\includegraphics[width=0.35\linewidth]{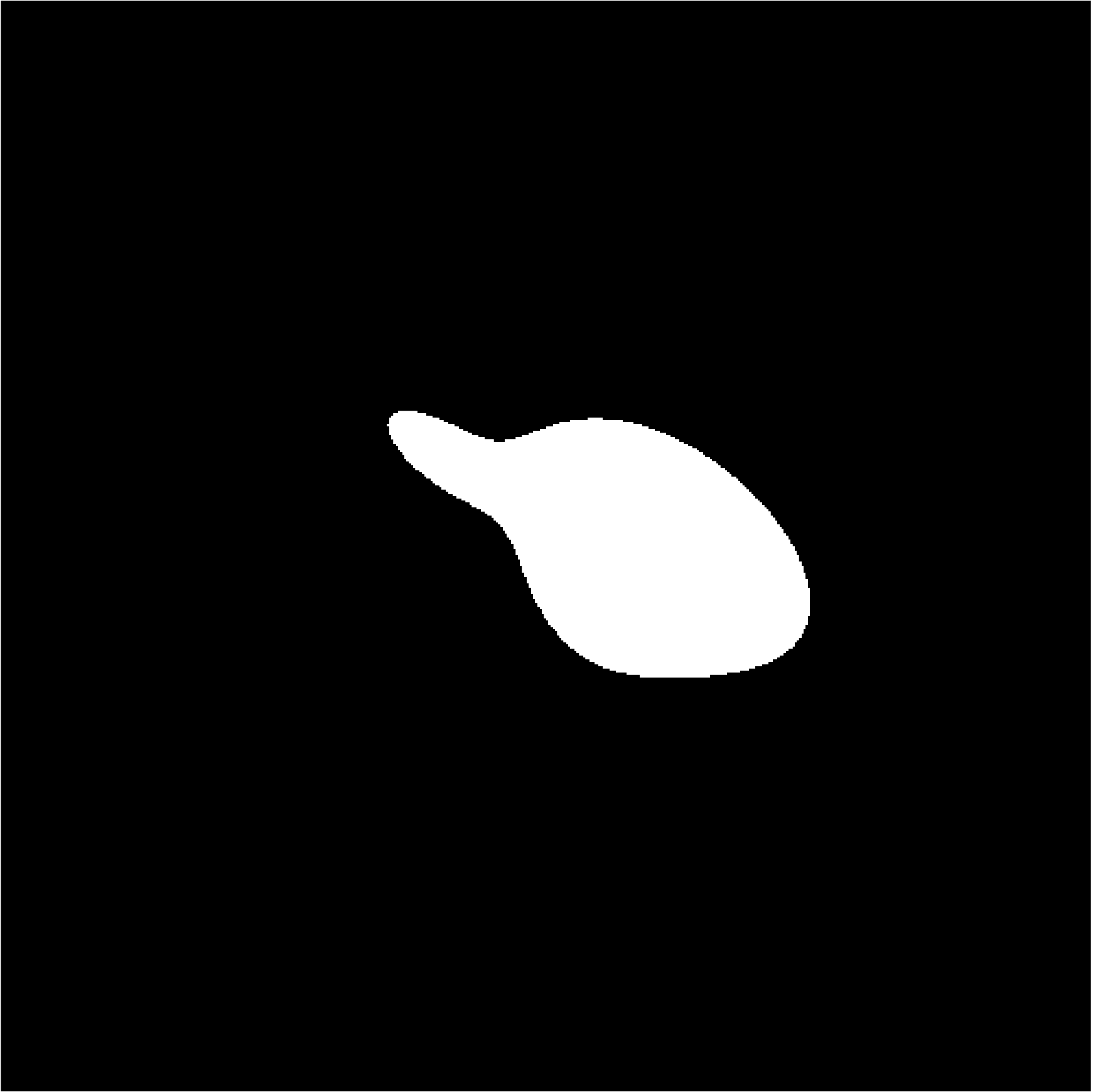}}
\hspace{10pt}
\subfloat[]{\includegraphics[width=0.35\linewidth]{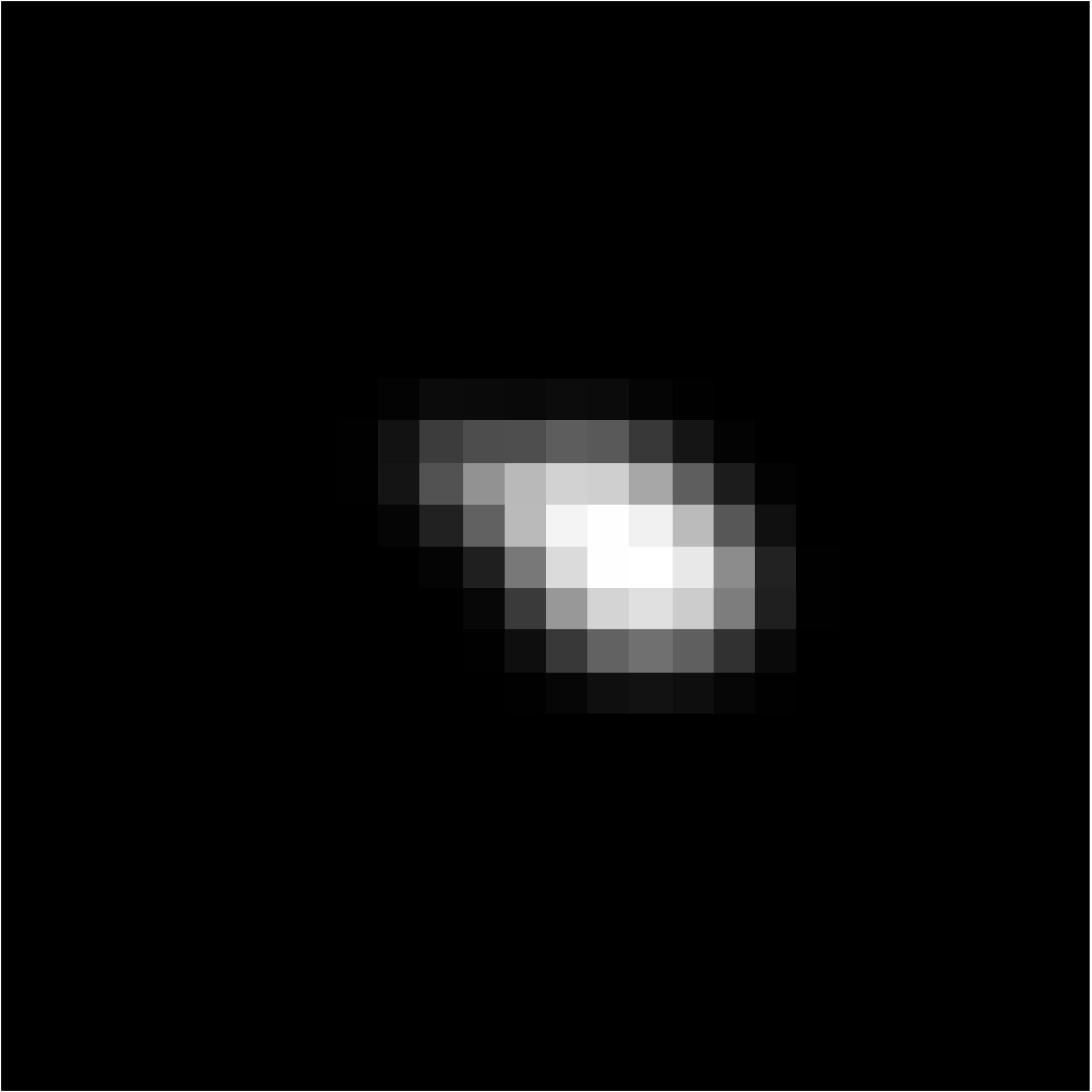}}
\caption{(a) An algebraic shape. (b) Samples generated with the tensor product of B-spline kernels of order 6.}\label{fig:samples}
\end{figure}

In any image sampling scenario, the question is whether and how we can reconstruct the original image $I(x,y)$ from a finite number of samples $d_{k,l}$ (Figure \ref{fig:samples}). In the next sections, we present a technique for the reconstruction of the boundary curve $\Curve$ and hence the algebraic shape $I(x,y)$ from adequate noiseless or noisy samples $d_{k,l}$.

\section{Reconstruction from Moments}\label{sec:recons}
For an exact reconstruction of an algebraic shape image, we should estimate its boundary --the algebraic curve $\Curve$-- from the samples. In the sequel, we first derive some annihilating equations for the curve parameters based on the shape moments. Then, we use the existing FRI techniques \cite{Dragotti2007} to calculate shape moments from the samples. The overall procedure is summarized in Algorithm \ref{alg:AS_recons_noiseless}.

\subsection{Annihilation equations}\label{sec:annihilation_eq}
Consider a closed algebraic curve $\Curve$ inside the domain $\Omega$ and the corresponding shape image $I$. We can rewrite $I$ in equation \eqref{eq:AS} as 
\begin{align*}
I(x,y)=\left\{\begin{array}{ll}
1,&(x,y)\in \overline{Int(\Curve)}\\
0,&\text{otherwise,}
\end{array}\right.
\end{align*} 
where $\overline{Int(\Curve)}$ denotes the closure of the interior of $\Curve$. This equation explains that the partial derivatives $\frac{\partial I(x,y)}{\partial x}$ and $\frac{\partial I(x,y)}{\partial y}$ vanish everywhere in $\Omega$ except possibly on $\Curve$, where they behave like the Dirac $\delta$ function. So, similar to the equation $x\delta(x) = 0$, we conclude that
\begin{align}
p(x,y)\,\frac{\partial I(x,y)}{\partial x}\equiv0,\label{eq:curve_anihilating1}\\
p(x,y)\,\frac{\partial I(x,y)}{\partial y}\equiv0,\label{eq:curve_anihilating2}
\end{align}
inside $\Omega$. 

We can multiply the above equations with $x^ry^s$ for any $r,s\in\Z^{\geq0}$ and integrate over the domain to obtain the equations
\begin{align}
\iint_\domain x^ry^sp(x,y)\frac{\partial I(x,y)}{\partial x}\ud x\ud y=0,\label{eq:basic1}\\
\iint_\domain x^ry^sp(x,y)\frac{\partial I(x,y)}{\partial y}\ud x\ud y=0.\label{eq:basic2}
\end{align}
By substituting $p(x,y)$ from equation \eqref{eq:polynomial} in \eqref{eq:basic1} and using integration by parts, we get
\small
\begin{align}\label{eq:basic1-reform}
\sum_{\stackrel{0\leq i,j}{i+j\leq n}}(i+r)\,a_{i,j}\iint_\domain x^{(i+r-1)}y^{(j+s)}~I(x,y)\ud x\ud y=0.
\end{align} 
\normalsize
In the derivation of \eqref{eq:basic1-reform}, we also used the fact that $\Curve$ is a closed curve inside $\Omega$ and hence, $I$ is zero at the domain borders. 

\begin{algorithm}[t!]
\caption{Algebraic shape reconstruction from noiseless samples}
\label{alg:AS_recons_noiseless}
\floatname{algorithm}{Procedure}
\renewcommand{\algorithmicrequire}{\textbf{Input:}}
\renewcommand{\algorithmicensure}{\textbf{Output:}}
\algorithmicrequire ~noiseless samples $d_{k,l}$, degree $n$ of the algebraic shape, polynomial reproducing coefficients $c_k^{(i)}$ of the sampling kernel.\\
\algorithmicensure ~boundary curve $\Curve$.
\begin{algorithmic}[1]
\item Calculate shape moments $M_{i,j}$ from samples for any $0\leq i,j\leq 3n/2$, according to equation \eqref{eq:moments_from_samples}.
\item Form the annihilation equations \eqref{eq:moment1} and \eqref{eq:moment2} for any $0\leq r,s\leq n/2$ and put them into a linear system of the form $\mathbf{Ma=0}$ .
\item Solve $\mathbf{Ma=0}$ for the polynomial coefficients $\mathbf{a}$ with the constraint $\mathbf a[0]= a_{0,0}= 1$.
\item Form the polynomial $p(x,y)$ from the coefficients in $\mathbf{a}$ according to \eqref{eq:polynomial}.
\item Set $\Curve$ equal to the zero level set of $p(x,y)$ inside $\Omega$.
\end{algorithmic}
\end{algorithm}

The integrals in equation \eqref{eq:basic1-reform} represent 2D moments of the image $I$
\begin{align*}
M_{i,j}=\iint_\domain x^iy^j\,I(x,y)\ud x\ud y.
\end{align*}
Hence, we can rewrite equation \eqref{eq:basic1-reform} as 
\begin{align}\label{eq:moment1}
\sum_{{0\leq i,j,~i+j\leq n}}(i+r)\,M_{i+r-1,j+s}\,a_{i,j}=0.
\end{align} 
We can similarly modify equation \eqref{eq:basic2} to derive the additional equation
\begin{align}\label{eq:moment2}
\sum_{{0\leq i,j,~i+j\leq n}}(j+s)\,M_{i+r,j+s-1}\,a_{i,j}=0.
\end{align}

For any pair of $(r,s)$, formula \eqref{eq:moment1} and \eqref{eq:moment2} give us two linear annihilation equations for the $\binom{n+1}{2}$ coefficients $a_{i,j}$, in terms of the image moments. We get enough equations to build a linear system of the form
\begin{align}\label{eq:LSE}
\mathbf{Ma=0}
\end{align}
and derive the curve parameters, if we consider all pairs  $(r,s),~0\leq r,s\leq n/2$. This implies that we require all image moments of degree up to $3n/2$, i.e., $M_{i,j},~0\leq i,j\leq 3n/2$.

To avoid the trivial solution $\mathbf{a=0}$, we set the term corresponding to $x^0y^0$ to 1. We recall that a scaling of the polynomial coefficients does not change its level sets. In Theorem \ref{theo:mainTheorem}, we prove that the zero level set of the polynomial $q(x,y)$ formed by any solution of \eqref{eq:LSE} contains $\Curve$. This specifically means that although the system of equations in \eqref{eq:LSE} might have a null space with dimension larger than 1, any vector $\mathbf a$ in  this null space generates a polynomial that vanishes on the boundary of $I$. Hence, we can recover the boundary curve $\Curve$ and the algebraic shape $I$ from any solution of \eqref{eq:LSE}.

\begin{figure}[t]
\centering
\includegraphics[width=0.9\linewidth]{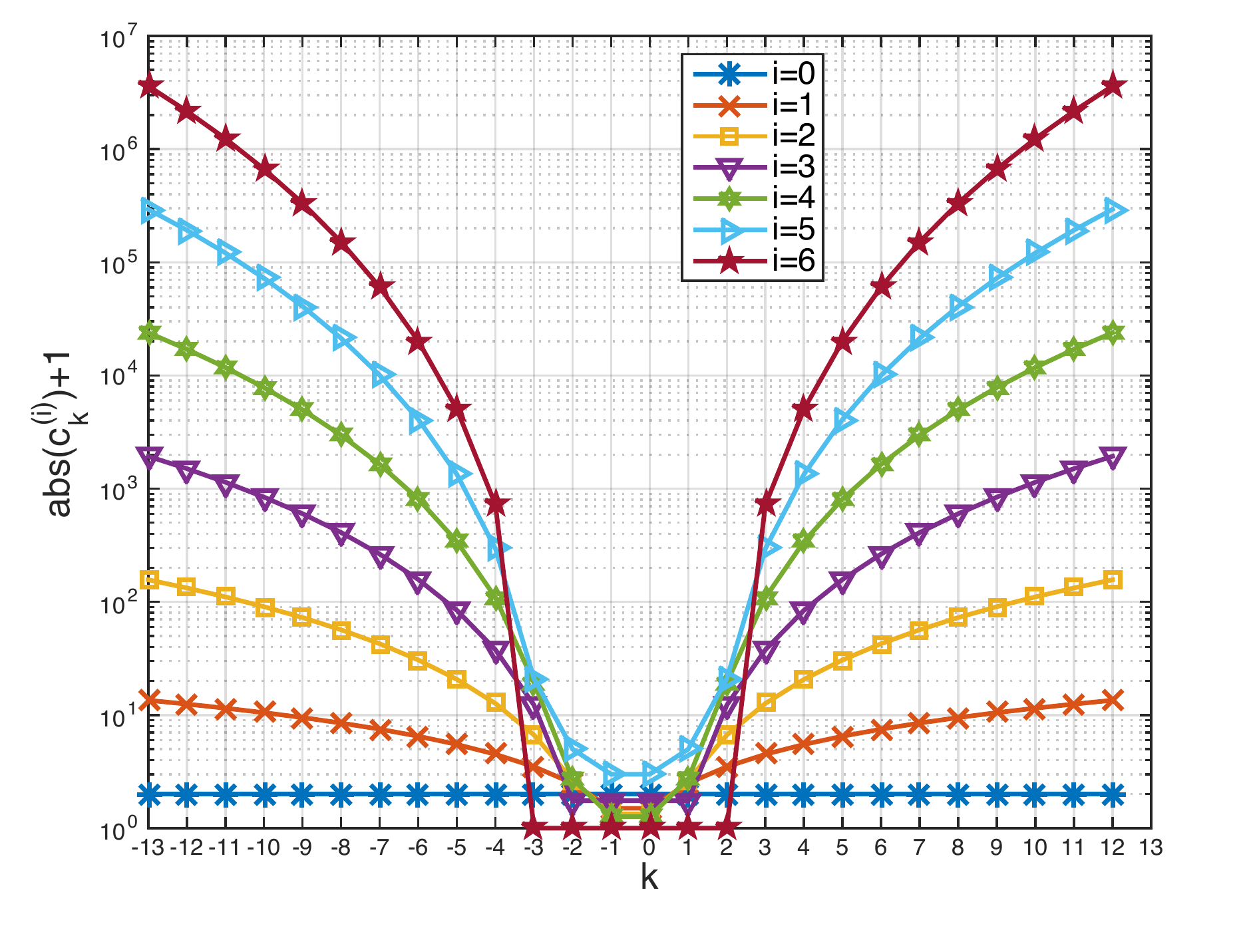}
\caption{The exponential growth rate of polynomial reproducing coefficients of the B-spline kernel $\beta^{(6)}(x)$.}\label{fig:poly_rep_coef}
\end{figure}

Finally, it remains to retrieve moments from the samples. Suppose that the kernel $\varphi(x)$ can reproduce polynomials up to degree $3n/2$, with the corresponding coefficients $c_k^{(i)},~i=0,...,3n/2$. The 2D moments of the image can be calculated as
\begin{align}\label{eq:moments_from_samples}
M_{i,j}&=\iint_\Omega x^iy^jI(x,y)\ud x\ud y\nonumber\\
&=\iint_\Omega\sum_{k\in\mathbb{Z}}c_k^{(i)}\varphi(x-k)\sum_{l\in\mathbb{Z}}c_l^{(j)}\varphi(y-l)~I(x,y)\ud x\ud y\nonumber\\
&=\sum_{k\in\mathbb{Z}}\sum_{l\in\mathbb{Z}}c_k^{(i)}c_l^{(j)}\iint_\Omega\varphi(x-k)\varphi(y-l)~I(x,y)\ud x\ud y\nonumber\\
&=\sum_{k\in\mathbb{Z}}\sum_{l\in\mathbb{Z}}c_k^{(i)}c_l^{(j)} d_{k,l}=\sum_{k\in\mathcal{K}}\sum_{l\in\mathcal{L}}c_k^{(i)}c_l^{(j)} d_{k,l},
\end{align}
where $\mathcal{K}$ and $\mathcal{L}$ indicate the set of indices $k$ and $l$ such that $\varphi(x-k)\varphi(y-l)$ is nonzero over $\Omega$.

\subsection{Stability}
Algorithm \ref{alg:AS_recons_noiseless} restores the exact algebraic curve when it has access to the noiseless samples. But it breaks down in the presence of noise. The reason is that the polynomial reproducing coefficients $c_k^{(i)}$ have the same growth rate as the polynomials, i.e., they grow like $|k|^i$. (To illustrate this, we show the polynomial reproducing coefficients $c_k^{(i)}$ of a 1D 6th order B-spline kernel for $i=0,..,6$ in Figure \ref{fig:poly_rep_coef}.) This specially implies that in equation \eqref{eq:moments_from_samples}, the weight of samples that are away from the image center are considerably larger than the weight of the central samples. But for images in our model, samples at the image borders mostly contain noise. This transfers an amplified noise to the moments and results in severely degraded moments SNR. The noise boosting effect becomes more critical as the order of moments grow. This makes Algorithm \ref{alg:AS_recons_noiseless} unstable even at a sample SNR as high as 100 dB.

We recall that in the related works of \cite{Baboulaz2009} and \cite{Chen2012}, only the first order moment are required as they focus on first degree polynomials (step edges). Hence, the aforementioned noise boosting effect is not an issue.

In the next section, we introduce some generalized moments that have slower growth rates and discard the noise at the image borders. Above all, they are still reproducible from the samples generated with a wider range of sampling kernels.

\section{Stable Recovery}\label{sec:stable_recovery}
The sampling scheme of Section \ref{sec:recons} has some limitations: $(i)$ the reconstruction algorithm succeeds only in the absence of noise; $(ii)$ the acceptable sampling kernels $\varphi(\cdot)$ are limited to the ones that exactly reproduce polynomials; and $(iii)$ the algebraic shapes should have closed boundary curves. In this section, we modify Algorithm \ref{alg:AS_recons_noiseless} in three steps to resolve these limitations:

First and foremost, we introduce a fast decaying (or even compact-support) function $g(x,y)$ in the integrands of equations \eqref{eq:basic1} and \eqref{eq:basic2} to reduce the growth rate of polynomials, especially at the borders of $\Omega$. This translates into the annihilation equations as replacing moments with some generalized moments. We prove in Theorem \ref{theo:mainTheorem} that under noiseless samples, the resulting annihilation equations restore the exact boundary curve of any algebraic shape. Our proof is general and includes the case $g(x,y)=1$ which leads to conventional moments. Next, we describe the requirements for $g(x,y)$ to ensure stable generalized moments and we propose an optimization procedure for finding the best candidate $g$ that pairs with a given sampling kernel. Interestingly, the inclusion of $g$ allows for extension of the image model to algebraic shapes with open boundaries.

For our second step, we note that the image moments do not take full advantage of the available samples. For instance, the samples allow for prediction of the sign of the implicit polynomial on a subset of the sampling grid points and this prediction is fairly robust against noise. To further improve the reconstruction, we enforce sign consistency of the polynomial with the prediction of the available samples.

In our last step, we encourage full measurement consistency (not just sign) through bounded changes in the coefficients of the implicit polynomial.

\subsection{Annihilation equations with generalized moments}
We developed the annihilation equations of Section \ref{sec:annihilation_eq} by multiplying equations \eqref{eq:curve_anihilating1} and \eqref{eq:curve_anihilating2} with $x^ry^s$. This caused the image moments to appear in the equations. To control the growth rate of the polynomials and hence the moments, we replace $x^ry^s$ with $g(x,y)x^ry^s$ for an appropriate function $g$. 
 
\begin{defi}\label{defi:generalized_moments}
For any bivariate function $g(.,.)$ and integers $i,j\geq 0$, we call
\begin{align}\label{eq:generalized_moments}
M_{i,j}^{g(x,y)} = \int_{-\infty}^{+\infty}\int_{-\infty}^{+\infty} x^iy^jg(x,y)I(x,y)\ud x\ud y.
\end{align}
a 2D generalized moment of $I$, associated with $g$.
\end{defi}

Having separable sampling kernels, we also take $g(x,y)$ to be separable of the form $g(x,y) = g(x)g(y)$. Though, the results can be similarly extended to the non-separable kernels. In the following, we derive the new annihilation equations and discuss the requirements on $g$ afterwards. 

We multiply equations \eqref{eq:curve_anihilating1} and \eqref{eq:curve_anihilating2} with $g(x)g(y)x^ry^s$ and repeat similar steps as in section \ref{sec:annihilation_eq} to obtain 
\begin{align}
\sum_{\stackrel{0\leq i,j}{i+j\leq n}}a_{i,j}\iint_\domain \frac{\partial x^{r+i}g(x)}{\partial x}y^{s+j}g(y)I(x,y)\ud x\ud y=0,\label{eq:annihilation_new1}\\
\sum_{\stackrel{0\leq i,j}{i+j\leq n}}a_{i,j}\iint_\domain x^{r+i}g(x)\frac{\partial y^{s+j}g(y)}{\partial y}I(x,y)\ud x\ud y=0.\label{eq:annihilation_new2}
\end{align}
In Section \ref{sec:recons}, we had to assume that $I$ is zero at the borders of the image plane in order to use integration by parts. Here, we assume that $g(\cdot)$ is either zero outside $(-L,L)$ or decays so fast that the integral outside of this interval becomes negligible. This allows $I$ to take non-zero values at the borders of $\Omega$; consequently, $I$ can represent an unbounded shape. 

We can further simplify equations \eqref{eq:annihilation_new1} and \eqref{eq:annihilation_new2} and substitute the integrals with generalized moments to get the new annihilation equations 
\begin{align}
\sum_{\stackrel{0\leq i,j}{i+j\leq n}} \Big((i+r)M^{g(x)g(y)}_{i+r-1,j+s}+M^{g'(x)g(y)}_{i+r,j+s}\Big)a_{i,j}=0,\label{eq:annihilation_generalized1}\\
\sum_{\stackrel{0\leq i,j}{i+j\leq n}} \Big((j+s)M^{g(x)g(y)}_{i+r,j+s-1}+M^{g(x)g'(y)}_{i+r,j+s}\Big)a_{i,j}=0.\label{eq:annihilation_generalized2}
\end{align}
The above equations are valid for any $0\leq r,s$. Note that $g=1$ restores the annihilation equations \eqref{eq:moment1} and \eqref{eq:moment2} when $I$ represents a closed shape. In Theorem \ref{theo:mainTheorem} we state a unified result for recovery of algebraic shapes either from conventional annihilation equations or the generalizations in \eqref{eq:annihilation_generalized1} and \eqref{eq:annihilation_generalized2}. The proof of this theorem is provided in the appendix.

\begin{figure*}[!t]
\centering
\subfloat[]{\includegraphics[width=0.3\linewidth]{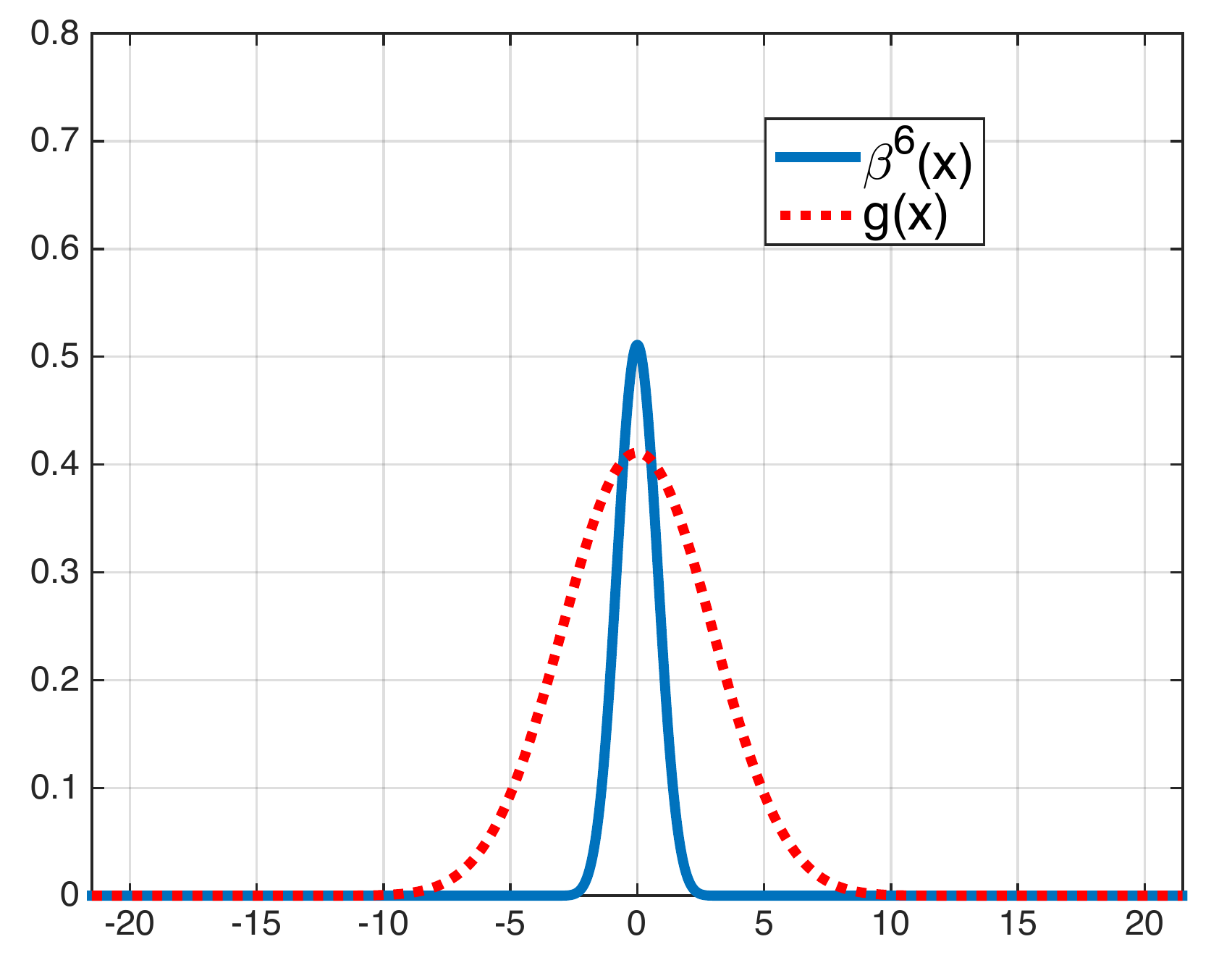}}
\hspace{5pt}
\subfloat[]{\includegraphics[width=0.3\linewidth]{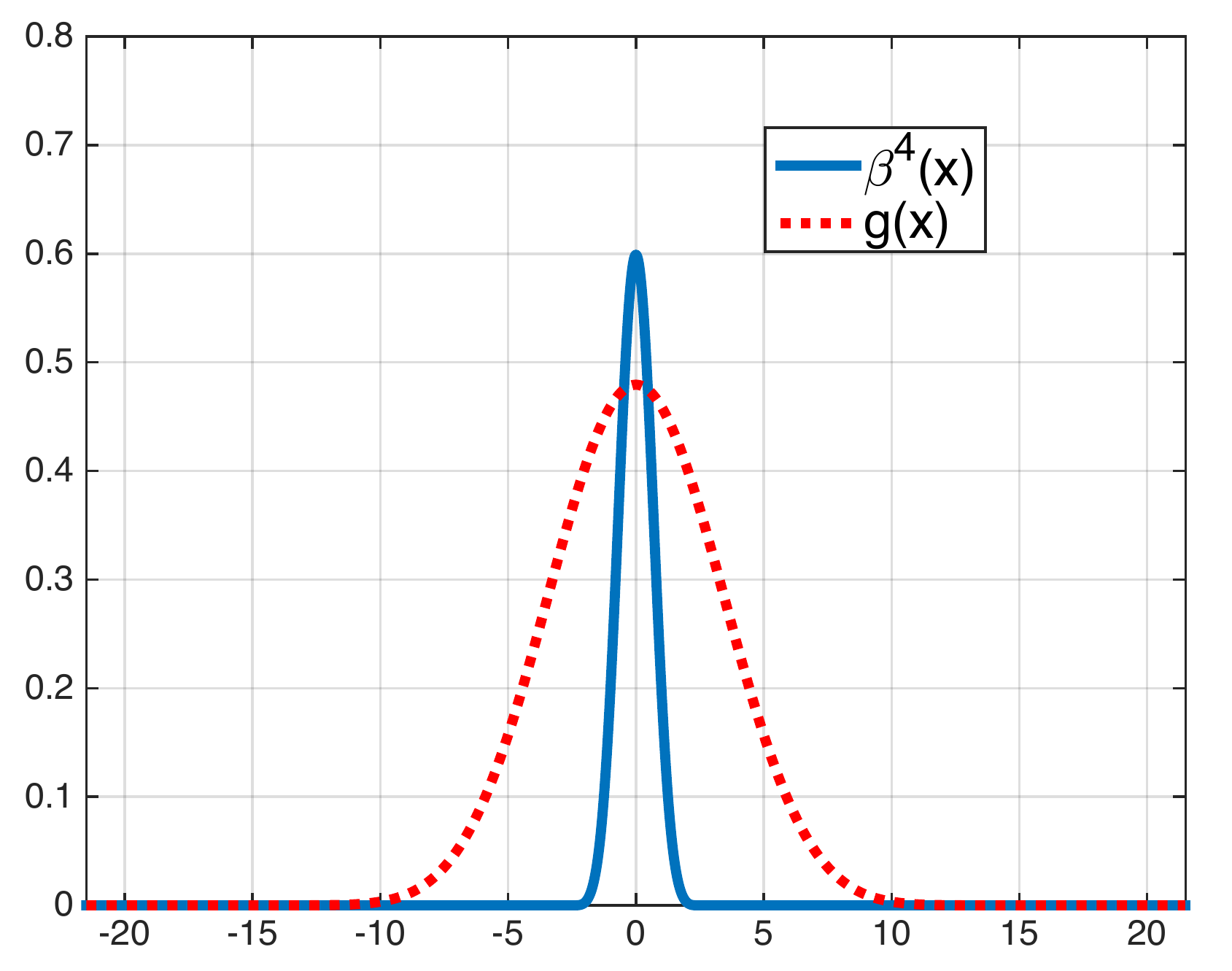}}
\hspace{5pt}
\subfloat[]{\includegraphics[width=0.3\linewidth]{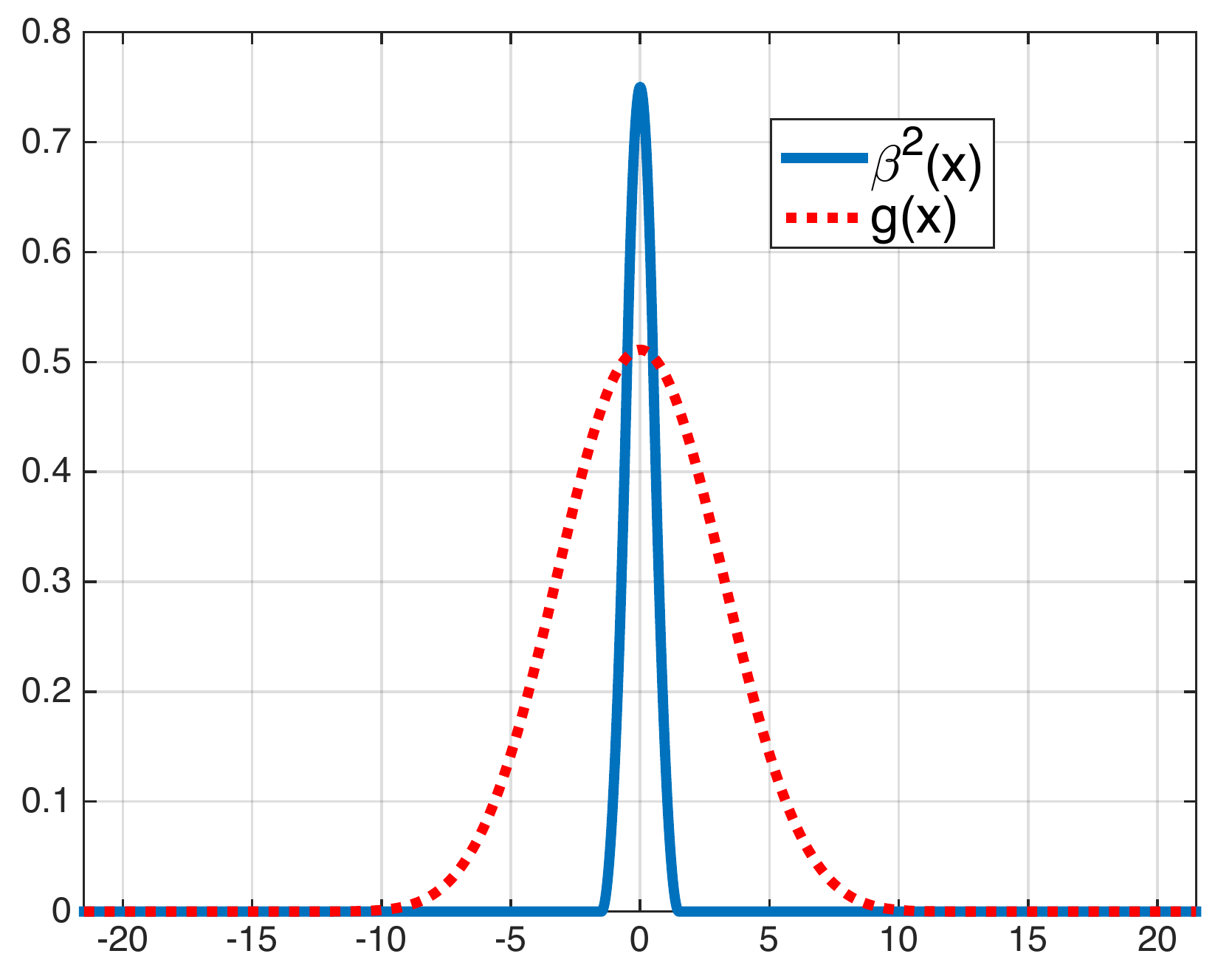}}
\caption{B-spline kernels and their associated $g$'s for reproducing stable generalized moments of order lass than or equal to 6. The indices ($\mathcal{I}$) of the contributing kernels in equations \eqref{eq:g_repro} and \eqref{eq:g_prime_repro} and the minimum number of required samples ($m$) are (a) $\mathcal I=\{-13,-12,\dots,13\}$ and $m=27$, (b) $\mathcal I=\{-14,-13,\dots,14\}$ and $m=29$, (c) $\mathcal I=\{-20,-19,\dots,20\}$ and $m=41$, respectively.}
\label{fig:g_functions}
\end{figure*}

\begin{theo}\label{theo:mainTheorem}
Let $I$ denote an algebraic shape of degree $n$ defined on $\Omega$ without singular edges\footnote{We call an edge singular if the image color does not change on either of its sides; for instance the image associated with $\ind_{\{(x-y)^2\leq 0\}}$ has a singular edge at points with equal coordinates.}. Also, let $M^{g(x)g(y)}_{i,j},M^{g(x)g'(y)}_{i,j}$ and $M^{g'(x)g(y)}_{i,j}$ denote the generalized moments of $I$ (Definition \ref{defi:generalized_moments}) corresponding to a function $g(\cdot)$ for which $I(x,y)g(x)g(y)$ vanishes outside $Int(\Omega)$ and $g(x)g(y)$ remains strictly positive inside $\Omega$. If $\tilde{\mathbf a}=[\tilde a_{i,j}]_{i+j\leq n}\neq\mathbf 0$ satisfies the annihilation equations  \eqref{eq:annihilation_generalized1} and \eqref{eq:annihilation_generalized2} for all $0\leq r,s,~r+s\leq 2n-1$, then, the zero level set of the polynomial 
$$\tilde{p}(x,y)=\sum_{0\leq i,j,~i+j\leq n}\tilde a_{i,j}x^iy^j$$ 
contains the boundaries (edges) of $I$. 
\end{theo}

\begin{remark} Unique recovery of $p(x,y)$ is not generally possible. Obviously, the zero level sets of $p(x,y)$ and $2p(x,y)$ are the same, leading to the same algebraic shapes. However, there are less obvious examples that prevent unique recovery: the zero level sets of both $(x^2 + y^2-1)(x^2-2x + 2)$ and $(x^2+y^2-1)(x^2+y^2+2xy+1)$ coincide with the unit circle, while the two bivariate polynomials have the same degree. The important point in Theorem \ref{theo:mainTheorem} is that the curve $\Curve$ is uniquely determined, but possibly with a different implicit polynomial.
\end{remark}

\begin{remark} 
Theorem \ref{theo:mainTheorem} requires that the coefficients of $\tilde p(x,y)$ satisfy the annihilation equations \eqref{eq:annihilation_generalized1} and \eqref{eq:annihilation_generalized2} for every $0\leq r+s\leq 2n-1$. This generates an over-determined system of the form $\mathbf{Ma=0}$ with about 8 times more rows than columns. In our experiments, we have confirmed successful recovery of algebraic curves from the annihilation equations corresponding to $0\leq r,s\leq n/2$ (yielding an almost balanced system). Our proof technique, however, falls short of this stronger result.
\end{remark}

\subsubsection{Optimal generalized moments}
The primary reason of introducing $g(x)g(y)$ to the equations is to control the growth rate of the monomials $x^ry^s$, especially at the image borders. Ideally, the $g(\cdot)$ function in \eqref{eq:annihilation_new1} and \eqref{eq:annihilation_new2} should be set such that $g$ and $\dot g$ both vanish outside $(-L,L)$. The faster they decay near the borders of $[-L,L]$, the more stable will be the annihilation equations \eqref{eq:annihilation_generalized1} and \eqref{eq:annihilation_generalized2}. However, the bottleneck in setting $g(\cdot)$ is the reproduction of moments from the samples.
That is the functions $x^ig(x)$ and $x^i\dot g(x),~i=0,\dots,\frac{3n}{2}$ should be reproducible by the sampling kernel $\varphi(x)$, i.e., we need coefficients $\{c_k^{(i)}\}$ and $\{\tilde c_k^{(i)}\}$ that satisfy
\begin{align}
\sum_{k\in\mathcal I}c_k^{(i)}\varphi(x-k)&\approx x^ig(x),\label{eq:g_repro}\\
\sum_{k\in\mathcal I}\tilde{c}_k^{(i)}\varphi(x-k)&\approx x^i\dot g(x).\label{eq:g_prime_repro}
\end{align}
Here, $\mathcal I$ represents $k$ values for which $\varphi(x-k)$ has an effective support in $[-L,L]$; this ensures that $g(x)$ and $\dot g(x)$ vanish outside $[-L,L]$.

For recovering an algebraic curve (domain) from samples using the generalized moment technique, we need to linearly combine the samples in correspondence to the coefficients $\{c_k^{(i)}\}$ and $\{\tilde c_k^{(i)}\}$. In other words, we never require the function $g$ explicitly in practice. Consequently, instead of looking for the best $g$ function, we can search for coefficients $\{c_k^{(i)}\}$ and $\{\tilde c_k^{(i)}\}$ such that
\begin{align}
\sum_{k\in\mathcal I}c_k^{(i)}\varphi(x-k) &\approx x\sum_{k\in\mathcal I}c_k^{(i-1)}\varphi(x-k), ~~~ i\geq 1,\nonumber\\
\sum_{k\in\mathcal I}\tilde{c}_k^{(i)}\varphi(x-k) &\approx x\sum_{k\in\mathcal I}\tilde{c}_k^{(i-1)}\varphi(x-k), ~~~ i\geq 1,\nonumber\\
\frac{\dd}{\dd x} \sum_{k\in\mathcal I}c_k^{(i)}\varphi(x-k) &\approx \sum_{k\in\mathcal I}\big( i \, c_k^{(i-1)} + \tilde{c}_k^{(i)}\big)\varphi(x-k) .
\end{align}
To find such coefficients, we introduce the following objective function 
\begin{align*}
&~~~~~~~~~~~~~~~\mathcal G\Big(\{c_k^{(i)}\},\{\tilde{c}_k^{(i)}\}\Big)=\\
&\sum_{i=1}^{3n/2}\|\sum_{k\in\mathcal I}\Big(c_k^{(i)}-xc_k^{(i-1)}\Big)\varphi(x-k)\|^2\\
+&\sum_{i=1}^{3n/2}\|\sum_{k\in\mathcal I}\Big(\tilde c_k^{(i)}-x\tilde c_k^{(i-1)}\Big)\varphi(x-k)\|^2\\
+&\sum_{i=0}^{3n/2}\|\sum_{k\in\mathcal I}c_k^{(i)}\dot\varphi(x-k)-\sum_{k\in\mathcal I}\Big(ic_k^{(i-1)}+\tilde{c}_k^{(i)}\Big)\varphi(x-k)\|^2.
\end{align*}
where $\dot\varphi$ stands for the derivative of $\varphi$. Next, we solve the quadratic program
\begin{eqnarray}\label{eq:optimal_coef}
&\min_{c_k^{(i)},\tilde{c}_k^{(i)}} \mathcal G\Big(\{c_k^{(i)}\},\{\tilde{c}_k^{(i)}\}\Big)&\\
&\text{s.t.}~\left\{\begin{array}{l}
\sum_{k\in\mathcal I} c_k^{(0)} \varphi(x-k)\geq 0, \phantom{\Big|}\\
c_0^{(0)} = 1 \phantom{\Big|}.
\end{array} \right.\nonumber
\end{eqnarray}
The equality constraint in the above minimization is to avoid the trivial zero solution and the inequalities guarantee that $g$ is non-negative. Although solving a quadratic program is computationally manageable, we have frequently observed that \eqref{eq:optimal_coef} is ill-conditioned\footnote{Essentially, the source is the same as the one causing instability in Algorithm \ref{alg:AS_recons_noiseless} except there is no noise here: the error terms corresponding to different $i$'s in the objective function grow polynomially and this makes the problem ill-conditioned.} in the sense that iterative methods are very slow in achieving the global solution, and usually terminate much earlier than desired. This shortcoming could be improved by using a sufficiently good initialization. Furthermore, any set of coefficients which result in a small cost according to the objective function could be used.

We recall that an implicit parameter in this problem is the size of the index set $\mathcal{I}$. This parameter also affects the modeling of $\Omega=[-L,L]^2$ and the minimum required sampling density for this sampling kernel. In fact, by increasing the index set $\mathcal{I}$ the global cost in \eqref{eq:optimal_coef} can only reduce. Thus, the larger the $\mathcal{I}$ the lower the cost. However, larger $\mathcal{I}$ translates into more image samples, and consequently more complexity.  

For the B-spline kernels, we found surprisingly good candidates $g$ that make the objective function almost zero. Figure \ref{fig:g_functions} shows the kernels $\beta^{(6)}(x),\beta^{(4)}(x), \beta^{(2)}(x)$ and their associated $g$'s that reproduce stable generalized moments of order 6 or less. This implies that we can form the annihilation equations and recover algebraic shapes of degree 4 even when the sampling kernel is the tensor product of 2nd order B-splines. The cost is a larger number of required samples.

Our final remark concerns using an asymmetric function $g(x,y)$ in the form $f(x)h(y)$, when satisfying \eqref{eq:g_repro} and \eqref{eq:g_prime_repro} is not possible simultaneously for a single $g(x)$. One can verify that the annihilation equations can still be obtained if 
\begin{align*}
\sum_{k\in\mathcal I}c_k^{(i)}\varphi(x-k)&\approx x^ih(x),\\
\sum_{k\in\mathcal I}\tilde{c}_k^{(i)}\varphi(x-k)&\approx \frac{\dd }{\dd x}\Big(x^if(x)\Big).
\end{align*}
For finding $\{c_k^{(i)}\}$ and $\{\tilde{c}_k^{(i)}\}$, \eqref{eq:optimal_coef} needs to be divided into two quadratic programs that accommodate $\{c_k^{(i)}\}$ and $\{\tilde{c}_k^{(i)}\}$ separately.


\subsubsection{Patch-based recovery}
Equations \eqref{eq:g_repro} and \eqref{eq:g_prime_repro} show that $g(x)$ and consequently $g(x)g(y)$ have compact support. This indicates that the generalized moments are computed from a finite window of the image samples --namely, of size $m\times m$, where $m$ amounts to the number of contributing kernels in $\mathcal I$. Having access to more samples, we can slide a $m\times m$ window over the image samples, calculate 2D moments and form the annihilation equations for each window. This results in a linear system with more equations and improves the noise stability of the reconstruction. 
\begin{figure}
\centering
\includegraphics[scale=0.3]{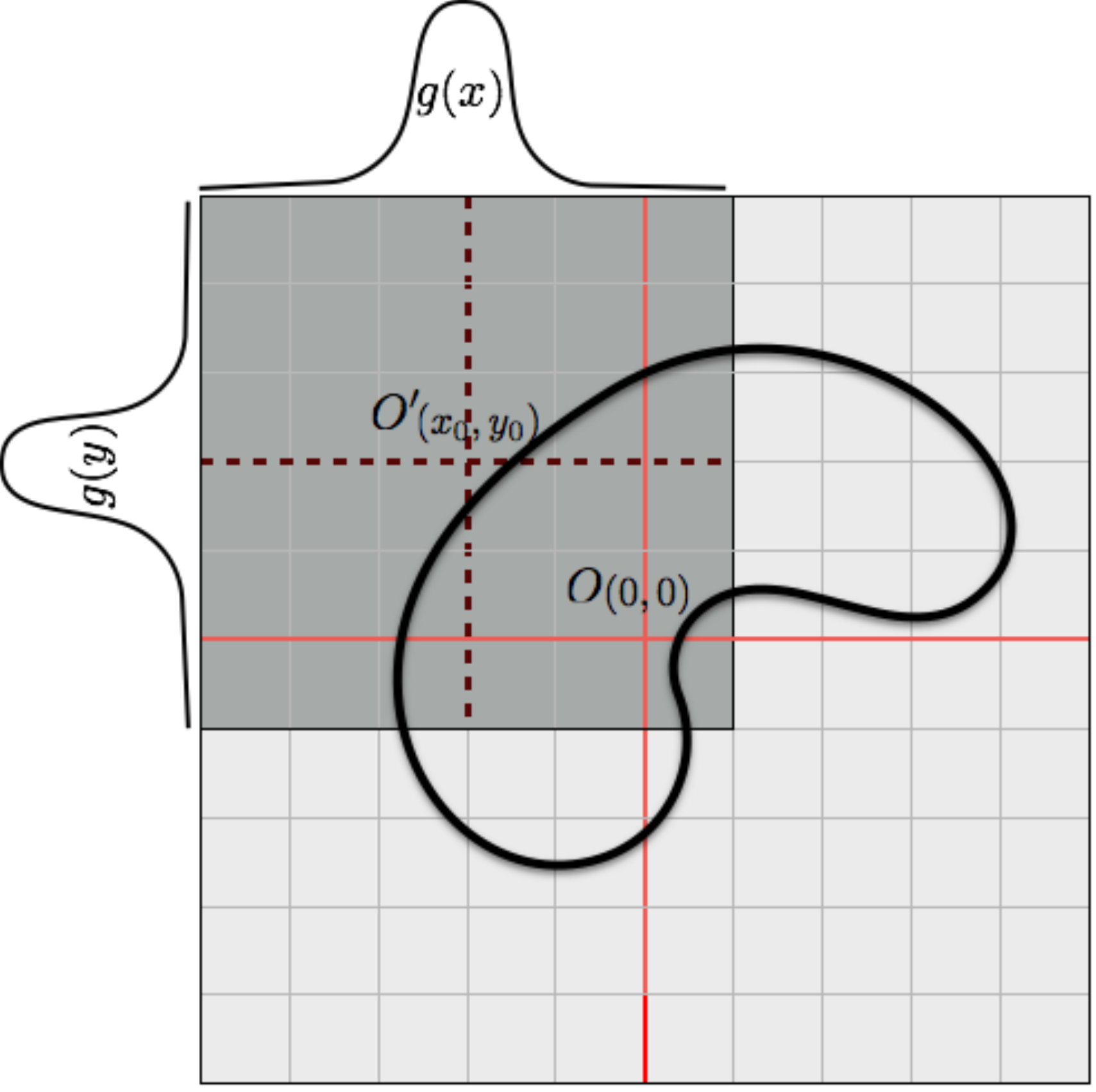}
\caption{A compact support function $g(x,y)$ facilitates the calculation of 2D generalized moments and the annihilation equations for different windows of the image samples. However, the coordinate shifts between different windows should be compensated before concatenating the equations in one system.}\label{fig:patch}
\end{figure}

There is only one issue requiring further attention: in the annihilation equations of each window, the coordinates origin is taken at the window's center (Figure \ref{fig:patch}). This means that the variables of each set of annihilation equations are the coefficients of the polynomial in those coordinates. Hence, we should compensate for the shifts in the coordinates before concatenating the equations of different windows. For this purpose, we choose the reference coordinates as the symmetry axes of the image plane. When the coordinates are shifted by $(x_0,y_0)$, the polynomial $p(x,y)=\sum_{0\leq i,j,i+j\leq n}a_{i,j}x^iy^j$  in the original system shall be mapped to the polynomial
\begin{align*}
\tilde p(x,y) &= p(x+x_0,y+y_0)\\
&=\sum_{0\leq i,j,i+j\leq n}a_{i,j}(x+x_0)^i(y+y_0)^j\\
&=\sum_{0\leq i,j,i+j\leq n}a_{i,j}\sum_{k=0}^i\binom{i}{k}x_0^{(i-k)}\sum_{l=0}^j\binom{j}{l}y_0^{(j-l)}x^ky^l.
\end{align*}
This reveals the mapping between the coefficients of $\tilde p(x,y)$, denoted by $b_{k,l}$, and $a_{i,j}$'s as
\begin{align*}
b_{k,l}=\sum_{\stackrel{k\leq i,l\leq j}{i+j\leq n}}\binom{i}{k}\binom{j}{l}x_0^{(i-k)}y_0^{(j-l)}a_{i,j},
\end{align*}
for any $0\leq k,l,~k+l\leq n$. We can represent the above relations for all polynomial coefficients simultaneously as
\begin{align}
\mathbf{b=B}^{(x_0,y_0)}\mathbf a,\label{eq:coef_adjust}
\end{align}
where $\mathbf{B}^{(x_0,y_0)}$ is an upper triangular square matrix with diagonal entries equal to 1.
This allows us to relate the annihilation equations of a window centered at $(x_0,y_0)$ to the polynomial coefficients $\mathbf a$ in the reference coordinate system through the equation
\begin{align*}
\mathbf M^{(x_0,y_0)}\mathbf b=\mathbf M^{(x_0,y_0)}\mathbf{B}^{(x_0,y_0)}\mathbf a=\mathbf 0.
\end{align*}
In a nutshell, we should multiply the annihilation equations of different windows with the corresponding matrix $\mathbf{B}^{(x_0,y_0)}$ in equation \eqref{eq:coef_adjust} before concatenating them in a bigger system.

\subsection{Constraints on the sign of the polynomial}
So far, we have built a system of equations in terms of the image parameters that is stable at numerical precision. In the presence of noise, the annihilation equations are only approximately singular. In this case, as a common practice, we consider the solution of the least squares minimization problem
\begin{eqnarray}\label{eq:least_squares}
&\min_{\mathbf a} \|\mathbf{Ma}\|^2_2&\\
&s.t.~\mathbf a[0]=a_{0,0}=1.&\nonumber
\end{eqnarray} 
The least squares denoising works well at low noise levels, especially when $\mathbf M$ is a tall matrix. Nevertheless, since algebraic curves are dense among continuous curves, distortion in the image moments (originated from moderate noise levels in the samples) can lead to substantially different solutions. 

Recently, the Cadzow's denoising algorithm \cite{Cadzow88} has been used for denoising of the annihilation equations of 1D \cite{Blu2008} and 2D \cite{Pan2014} FRI signals. The common feature in these works that makes  denoising successful is having  annihilation equations with a Toeplitz structure. Our system of annihilation equations --although almost each element in $\mathbf M$ has a few duplicates-- is not Toeplitz and the Cadzow's denoising algorithm does not help\footnote{In our implementation of Cadzow's algorithm, we observed that it converges to a rank deficient matrix with the expected structure which stays very close to the noisy matrix $\mathbf M$.}.

In our problem, the best reconstruction is an algebraic shape that is as consistent as possible with the image samples (i.e., up to the samples SNR). Theoretically, this can be achieved with a brute-force search over the space of image parameters. But this problem is nonconvex with many parameters and hence, computationally intractable. In the rest of this section, we exploit the local information provided by the samples to improve the reconstruction in the presence of noise.

Sample values represent the area of the intersection of the corresponding kernels with the interior of the shape in a weighted form. For example, $d_{k,l}=1\,(0)$  indicates that $I(x,y)=1\,(0)$ everywhere in the support of $\varphi(x-k,y-l)$\footnote{We assume that $T=1$ and $\varphi(x,y)$ has a unit integral.}. We further incorporate the samples in our reconstruction by interpreting them as the central points of the corresponding kernels lying inside or outside the shape. More precisely, if $d_{k,l}$ is above $1-\epsilon$ for an $\epsilon<0.5$, we assume its center to be inside the shape, i.e., $I(k,l)=1$ or equivalently $p(k,l)\leq 0$. Also, we take $I(k,l)=0$ or $p(k,l)>0$, if $d_{k,l}<\epsilon$. Eventually, we constrain the solution of the least squares problem with the inferred signs:
\begin{eqnarray}\label{eq:quadprog}
&\min_\mathbf{a}\|\mathbf{Ma}\|_2^2,&\\
&s.t.~\begin{aligned}&\mathbf {A}_\text{in}\mathbf{a}\leq\mathbf 0,\\
&\mathbf {A}_\text{out}\mathbf{a}<\mathbf 0,\end{aligned}&\nonumber
\end{eqnarray}
where $\mathbf{A}_\text{in}$ and $\mathbf A_\text{out}$ encode respectively, the normal and sign-negated polynomial evaluation matrices at central locations of the sampling kernels; $\mathbf{A}_\text{in}$ corresponds to locations with large sample values, while  $\mathbf A_\text{out}$ corresponds to locations with small sample values. The minimization problem \eqref{eq:quadprog} can be solved with quadratic programming algorithms.


\subsection{Measurement consistency}
At moderate noise levels (sample SNRs around 25 dBs), the recovered curves from \eqref{eq:quadprog} are close enough to the original boundaries to let us approximate the function mapping the polynomial coefficients to the image samples with a 1st order Taylor expansion around the correct coefficients. We exploit this assumption to improve the measurement consistency of the reconstruction. 

Let $\mathcal{D}$ denote the mapping from the polynomial coefficients into the samples of the algebraic shape. For instance, if $\mathbf{a}^{*}$ stands for the polynomial coefficients associated with an algebraic curve, $\mathbf{d}^{*}=\mathcal{D}(\mathbf{a}^{*})$ represents the vector of noiseless image samples via the sampling kernel.

For a given set of noisy samples $\widetilde{\mathbf{d}^{*}}$, let ${\mathbf a}_{\textrm{cur}}$ be the solution to the sign consistency technique in \eqref{eq:quadprog}, which corresponds to $\mathbf{d}_{\textrm{cur}}=\mathcal{D}(\mathbf{a}_{\textrm{cur}})$. For moderate to low noise levels, we know that ${\mathbf a}_{\textrm{cur}}$ is a good approximation of $\mathbf{a}^{*}$. Thus, we use the linearization of $\mathcal{D}$ around $\mathbf{a}_{\textrm{cur}}$ (1st order Taylor expansion) to write that
\begin{align*}
\mathbf{d}^{*} \approx \mathbf{d}_{\textrm{cur}} + \Big(\frac{\partial }{\partial\mathbf{a}}\mathcal{D}({\mathbf a}_{\textrm{cur}})\Big)(\mathbf{a}^{*}-{\mathbf a}_{\textrm{cur}}),
\end{align*}
where $\frac{\partial }{\partial\mathbf{a}}\mathcal{D}({\mathbf a}_{\textrm{cur}})$ is a matrix that relates the small input variations in $\mathcal{D}$ to its output around the point ${\mathbf a}_{\textrm{cur}}$. In practice, we find $\frac{\partial }{\partial\mathbf{a}}\mathcal{D}({\mathbf a}_{\textrm{cur}})$ by numerically varying ${\mathbf a}_{\textrm{cur}}$ in all directions and observing the corresponding $\mathbf{d}$'s. 
Finally, we improve our current estimate of $\mathbf{a}^{*}$ by
\begin{align}\label{eq:consistency}
{\mathbf a}_{\textrm{new}} = {\mathbf a}_{\textrm{cur}} +\Big(\frac{\partial }{\partial\mathbf{a}}\mathcal{D}({\mathbf a}_{\textrm{cur}})\Big)^{-1}(\widetilde{\mathbf{d}^{*}} - \mathbf{d}_{\textrm{cur}} ).
\end{align}
In our algorithm, we apply few iterations of the above update rule. Each time we evaluate the associated $\mathbf d$ vector and continue the iterations as long as this vector gets closer to $\widetilde{\mathbf{d}^{*}}$.

\section{Experimental Results}\label{sec:experiments}
We evaluate the performance of the proposed algorithm in different scenarios. We select bounded algebraic shapes for most of the experiments. For this purpose, we restrict the polynomial degree to even integers. But a randomly generated even degree polynomial very likely has unbounded level sets. A full characterization as well as a model for the generation of bivariate polynomials of degree 4 with bounded level sets was presented in \cite{Keren94} and \cite{Taubin94}. We adopt this model to generate shapes for our experiments.

\subsection{Noiseless recovery}
\begin{figure}
\subfloat[]{\includegraphics[width=0.3\linewidth]{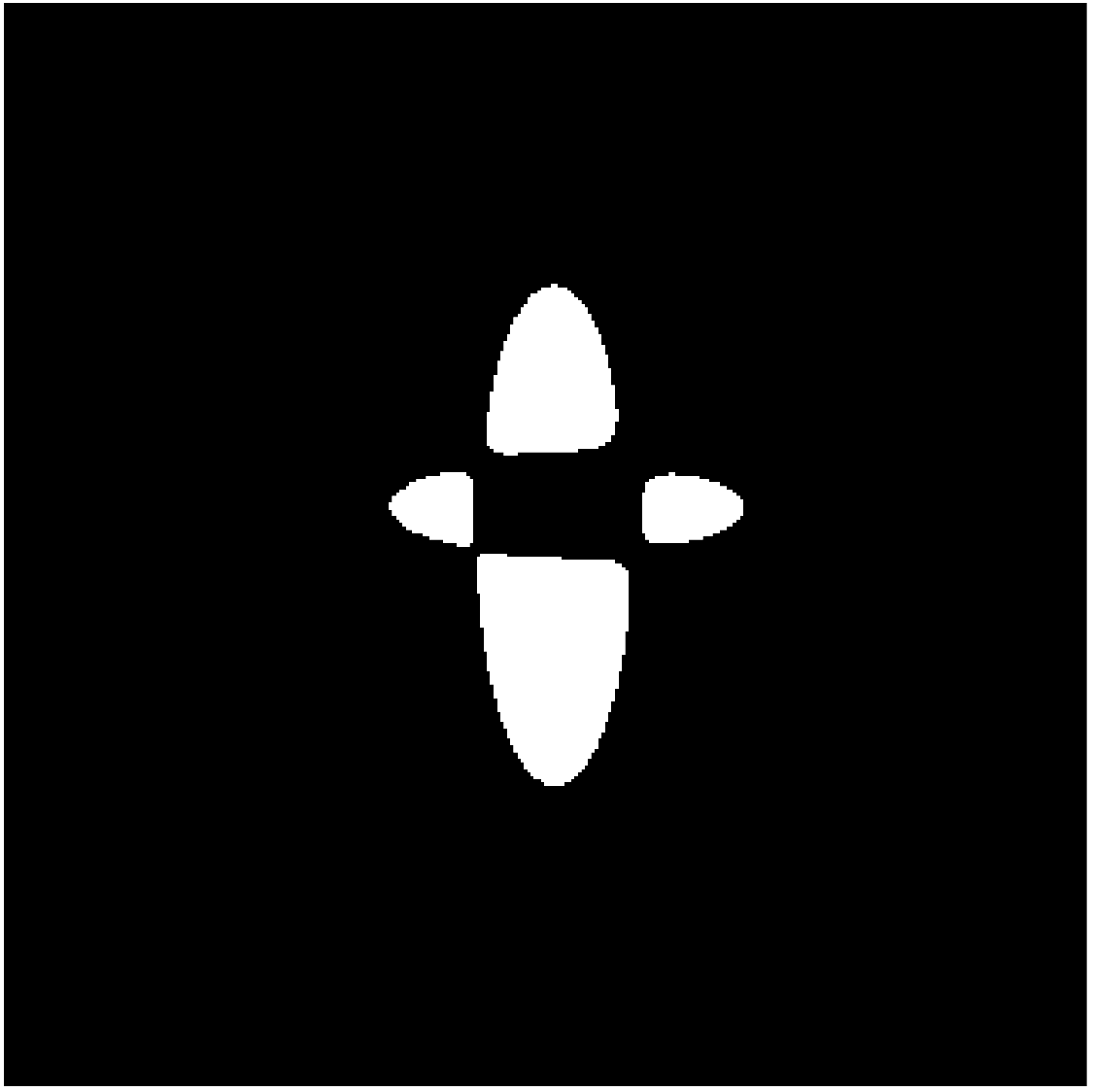}}
\hfill
\subfloat[]{\includegraphics[width=0.3\linewidth]{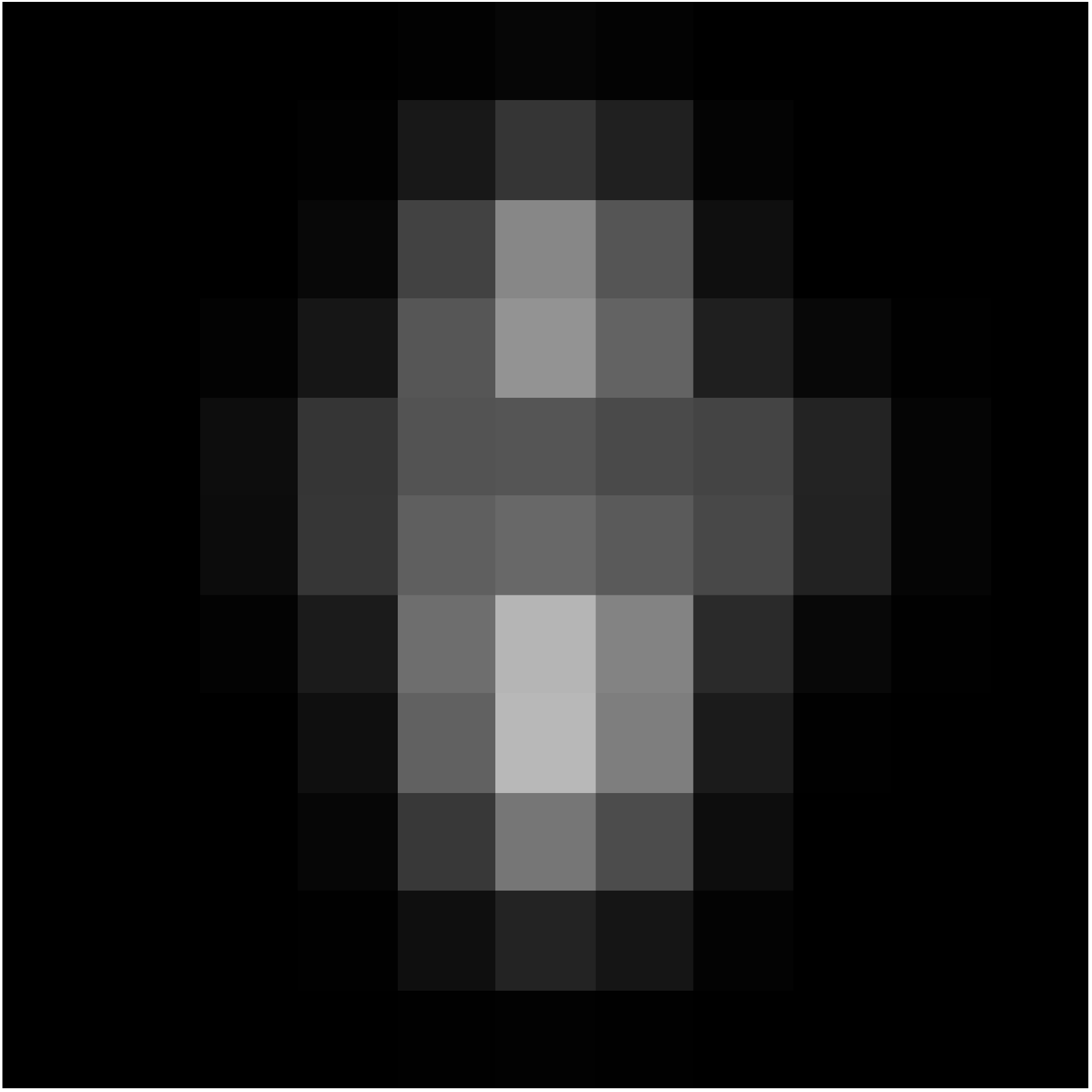}}
\hfill
\subfloat[]{\includegraphics[width=0.3\linewidth]{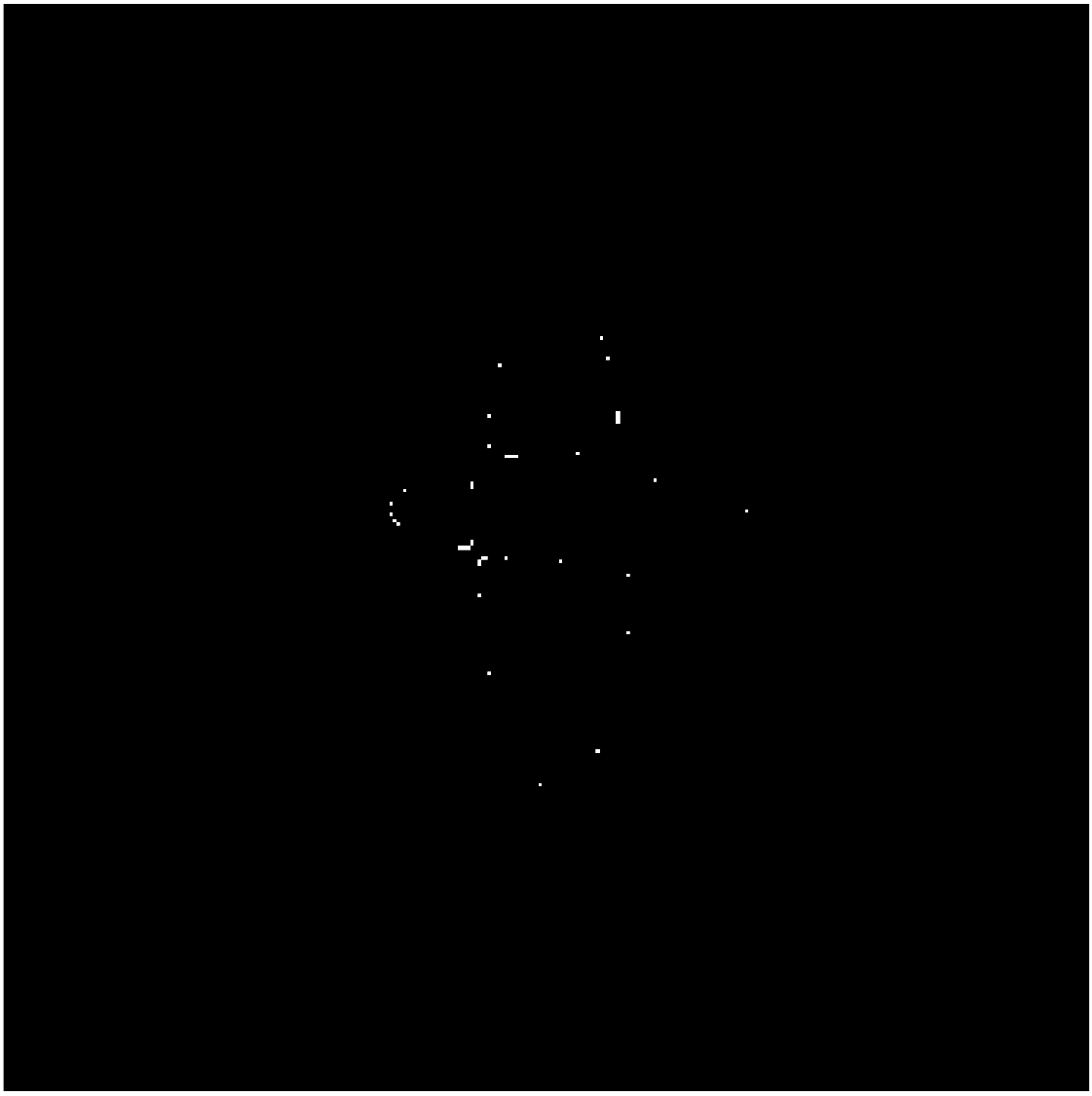}}
\caption{Exact reconstruction of algebraic shapes from noiseless samples. (a) An algebraic shape of degree 4. (b) Noiseless samples (size $11\times 11$), when the sampling kernel is $\varphi(x,y)=\beta^{(6)}(x)\beta^{(6)}(y)$. (c) Absolute difference between the original shape and the least squares solution.}\label{fig:noiseless}
\end{figure}

In the first experiment, we study reconstruction of algebraic shapes from noiseless samples. Recalling the results of the last section, we expect to recover the exact image by solving the least squares problem \eqref{eq:least_squares}. Figure \ref{fig:noiseless} displays perfect reconstruction of an algebraic shape of degree 4, when the sampling kernel is the tensor product of the 6th order B-splines.

\subsection{Recovery in the presence of noise}
\begin{figure}
\centering
\subfloat[]{\includegraphics[width=0.3\linewidth]{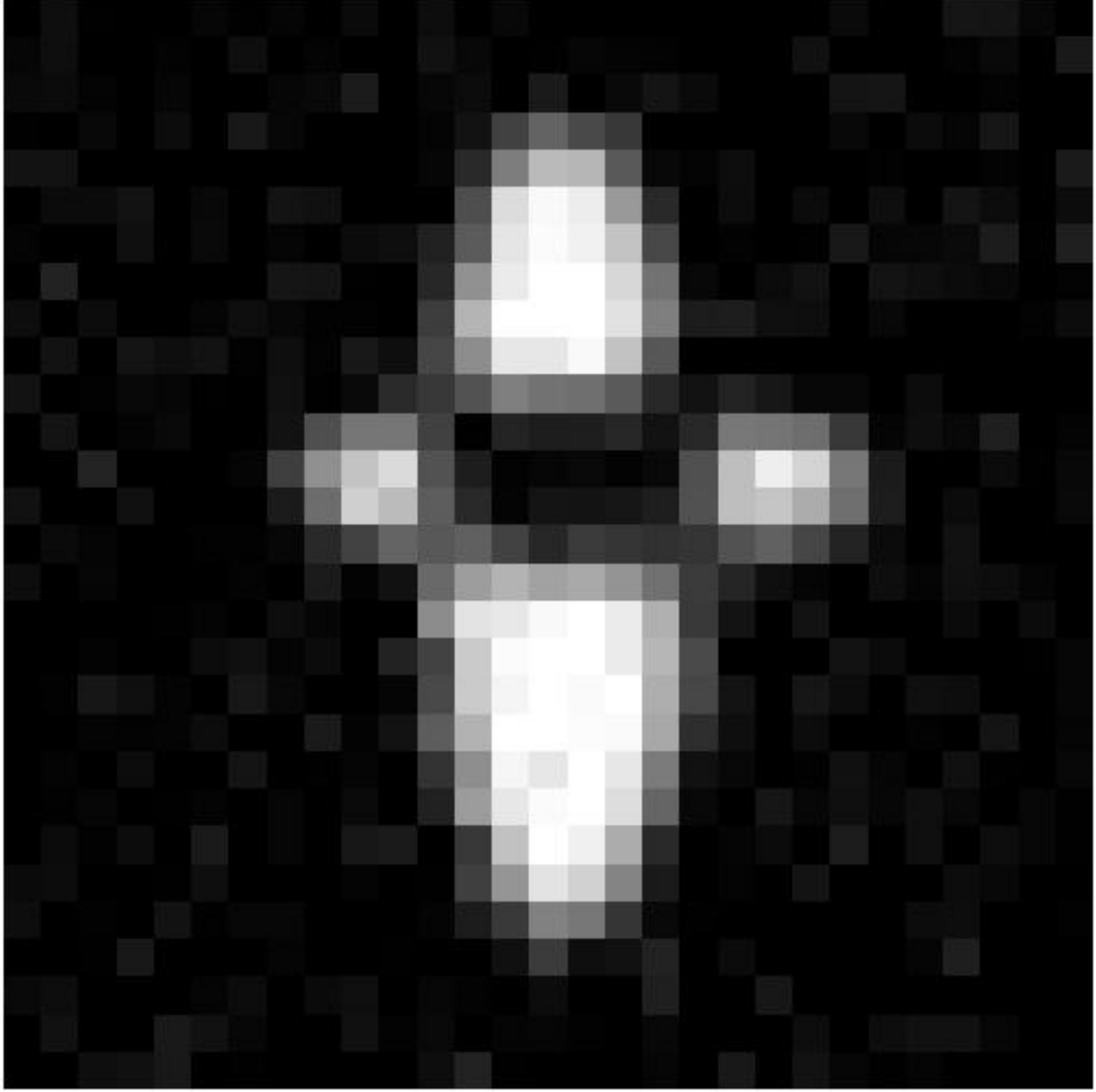}}
\\
\subfloat[]{\includegraphics[width=0.3\linewidth]{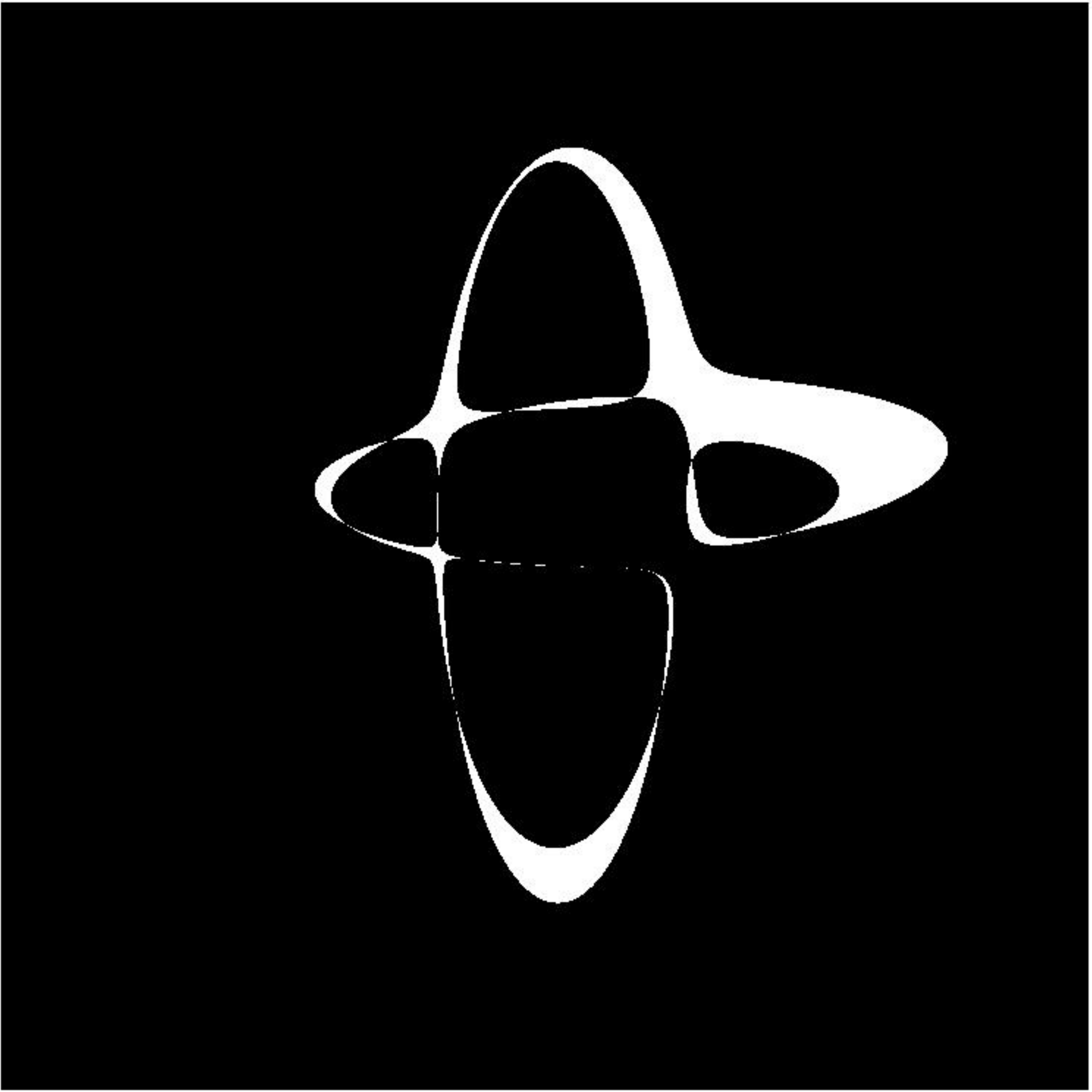}}
\hfill
\subfloat[]{\includegraphics[width=0.3\linewidth]{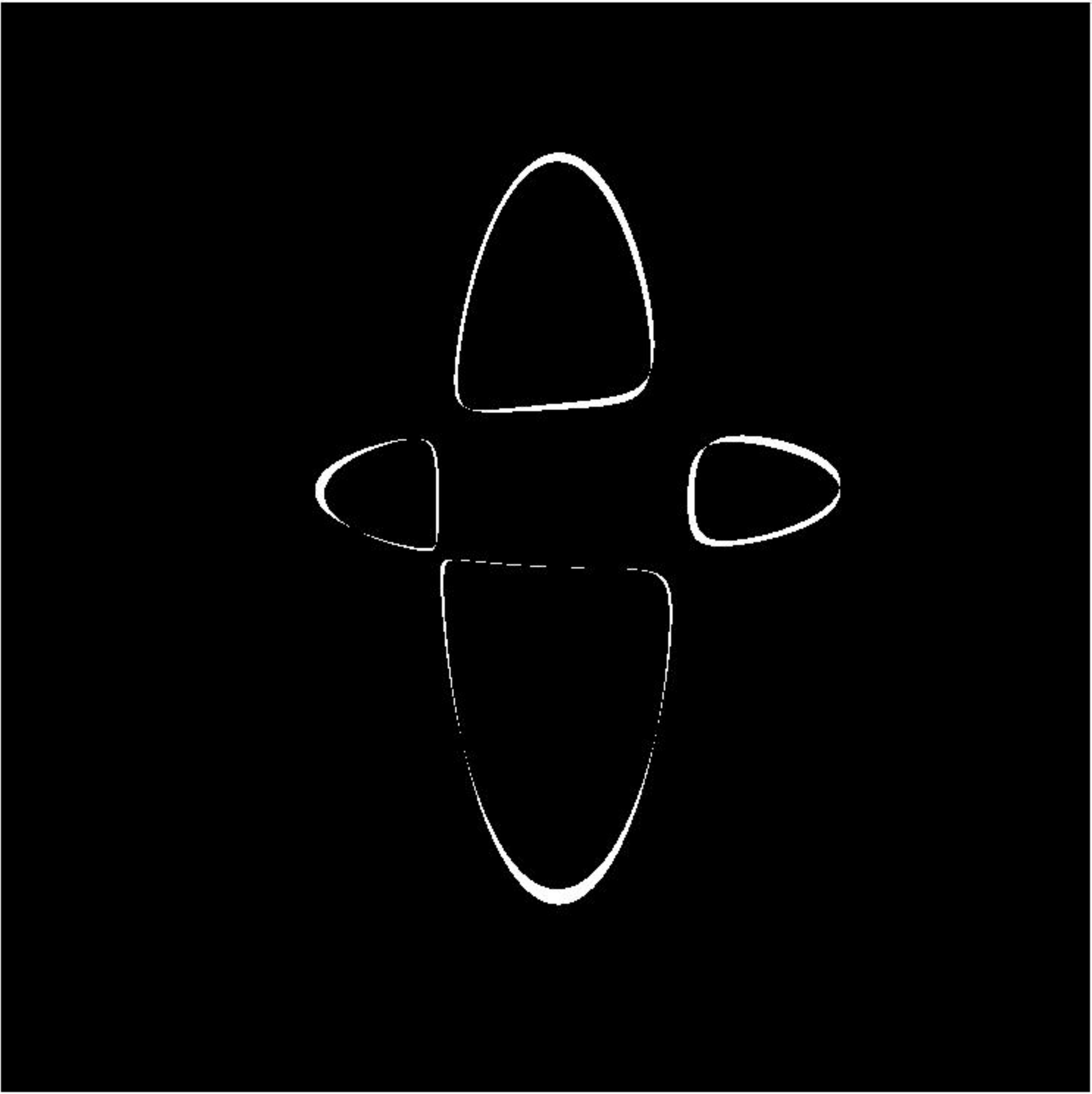}}
\hfill
\subfloat[]{\includegraphics[width=0.3\linewidth]{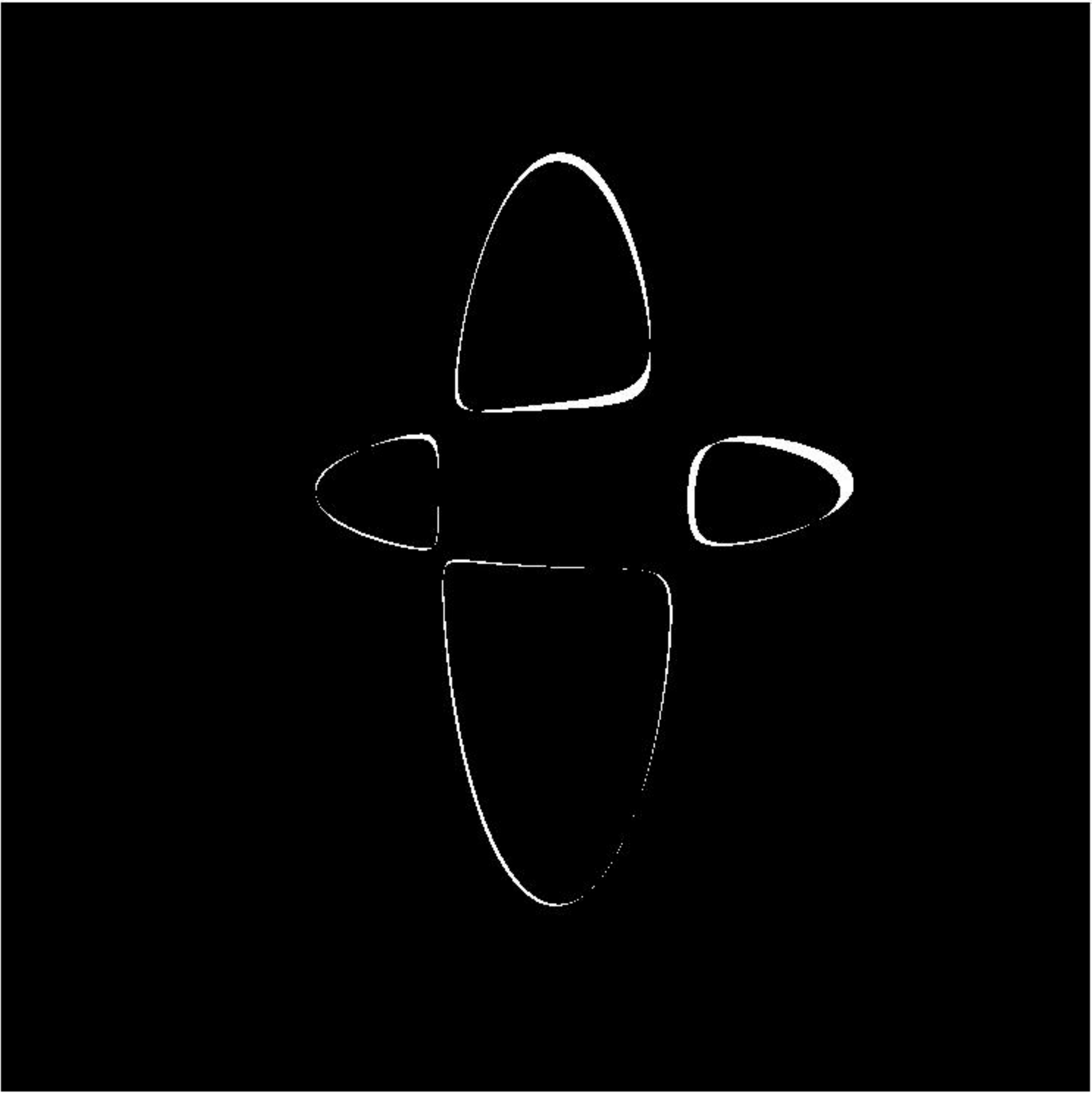}}
\caption{Reconstruction from noisy samples. (a) Noisy samples of the shape in Figure \ref{fig:noiseless}(a) with size $29\times 29$ and SNR = 17 dB. (b) Absolute error of the least squares solution (PSNR = 13.7 dB). (c) Ablsoute error of the quadratic programming (equation \eqref{eq:quadprog}) recontsruction (PSNR = 20.4 dB). (d) Absolute error of the output of the consistency improvement algorithm (PSNR = 21.3 dB). SNR between the samples of the final reconstruction and the noisy samples (a) is 15.4 dB.}\label{fig:noisy1}
\end{figure}

\begin{figure}
\centering
\subfloat[]{\includegraphics[width=0.3\linewidth]{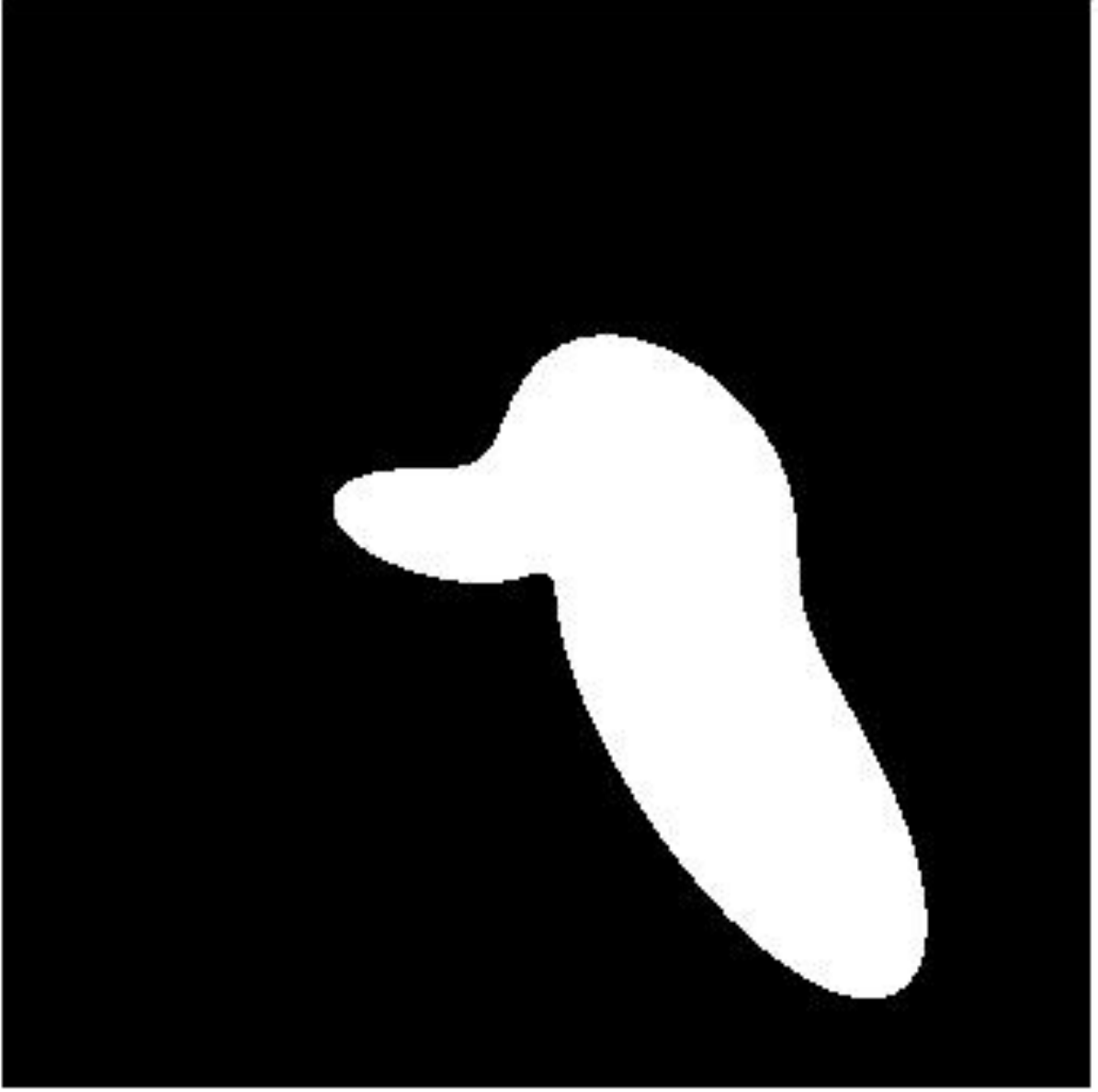}}
\hspace{8pt}
\subfloat[]{\includegraphics[width=0.3\linewidth]{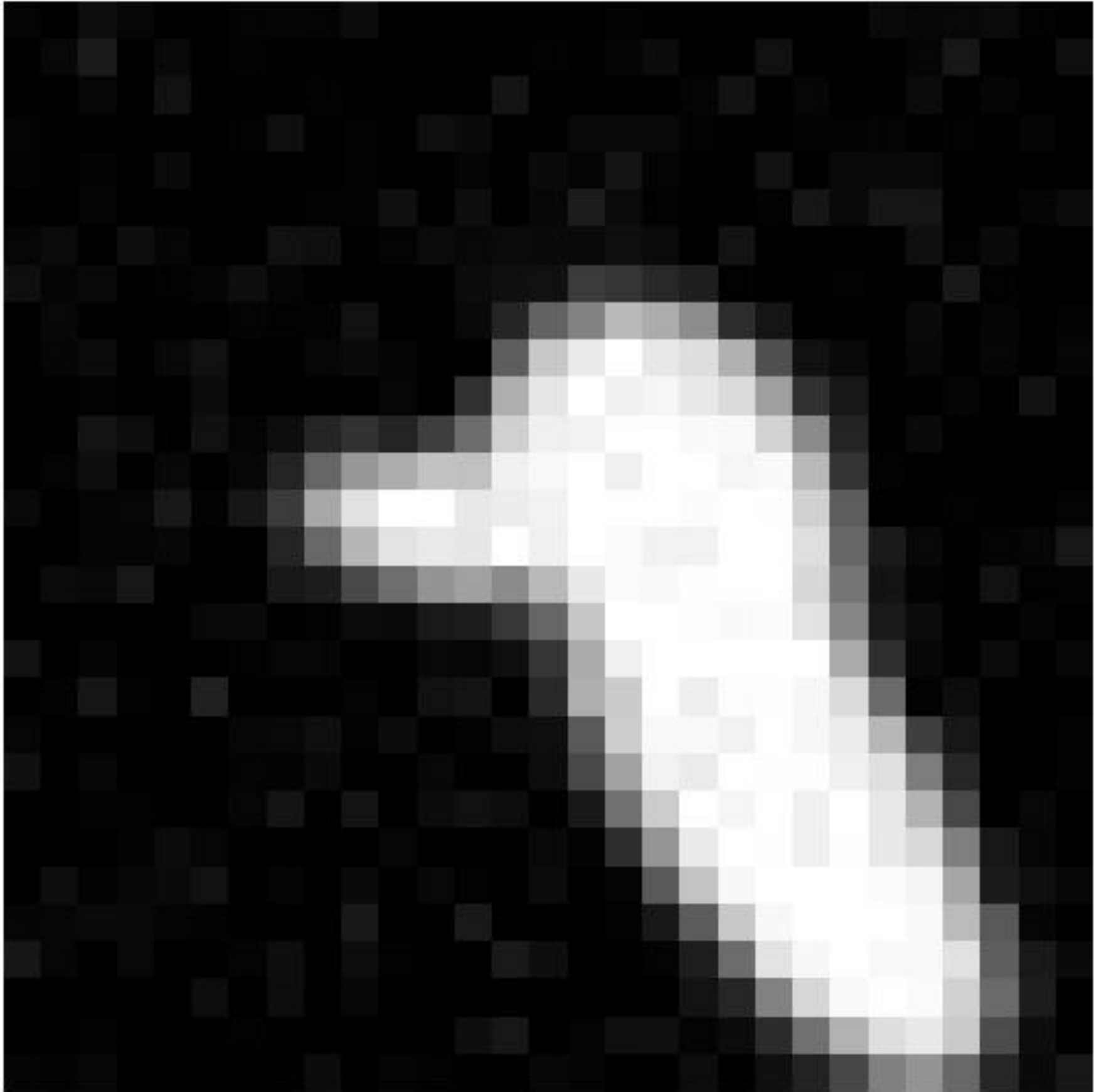}}
\\
\subfloat[]{\includegraphics[width=0.3\linewidth]{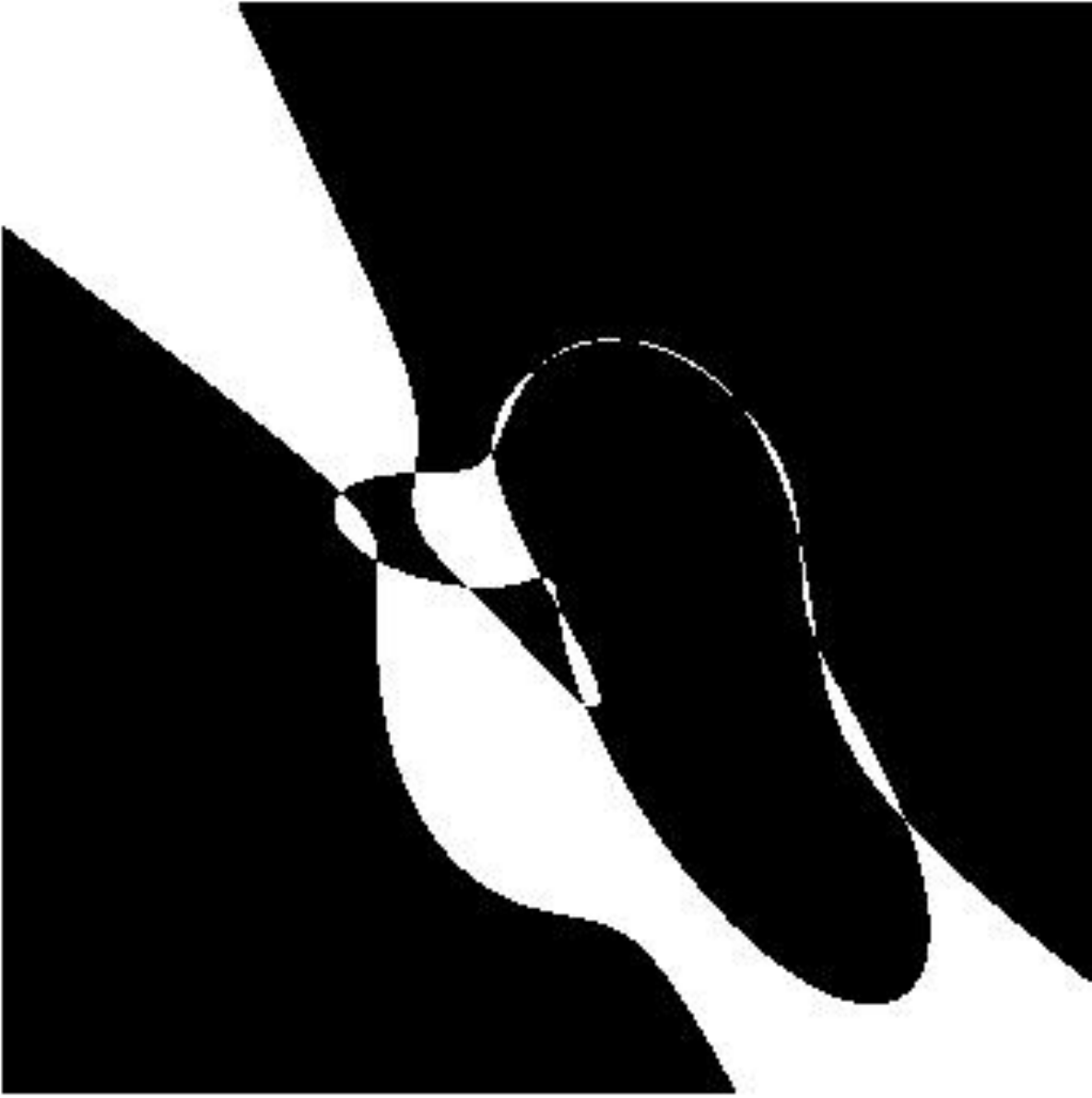}}
\hfill
\subfloat[]{\includegraphics[width=0.3\linewidth]{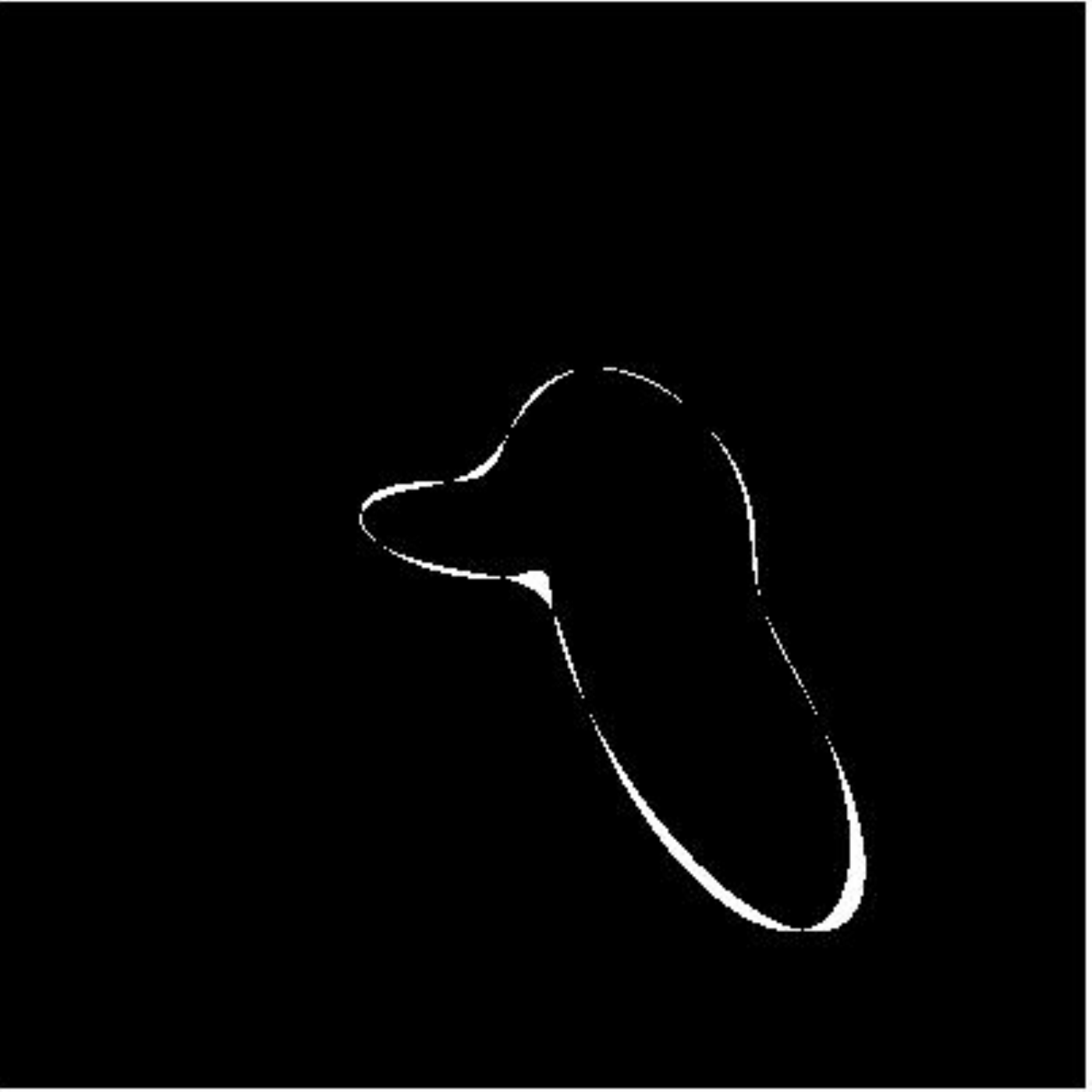}}
\hfill
\subfloat[]{\includegraphics[width=0.3\linewidth]{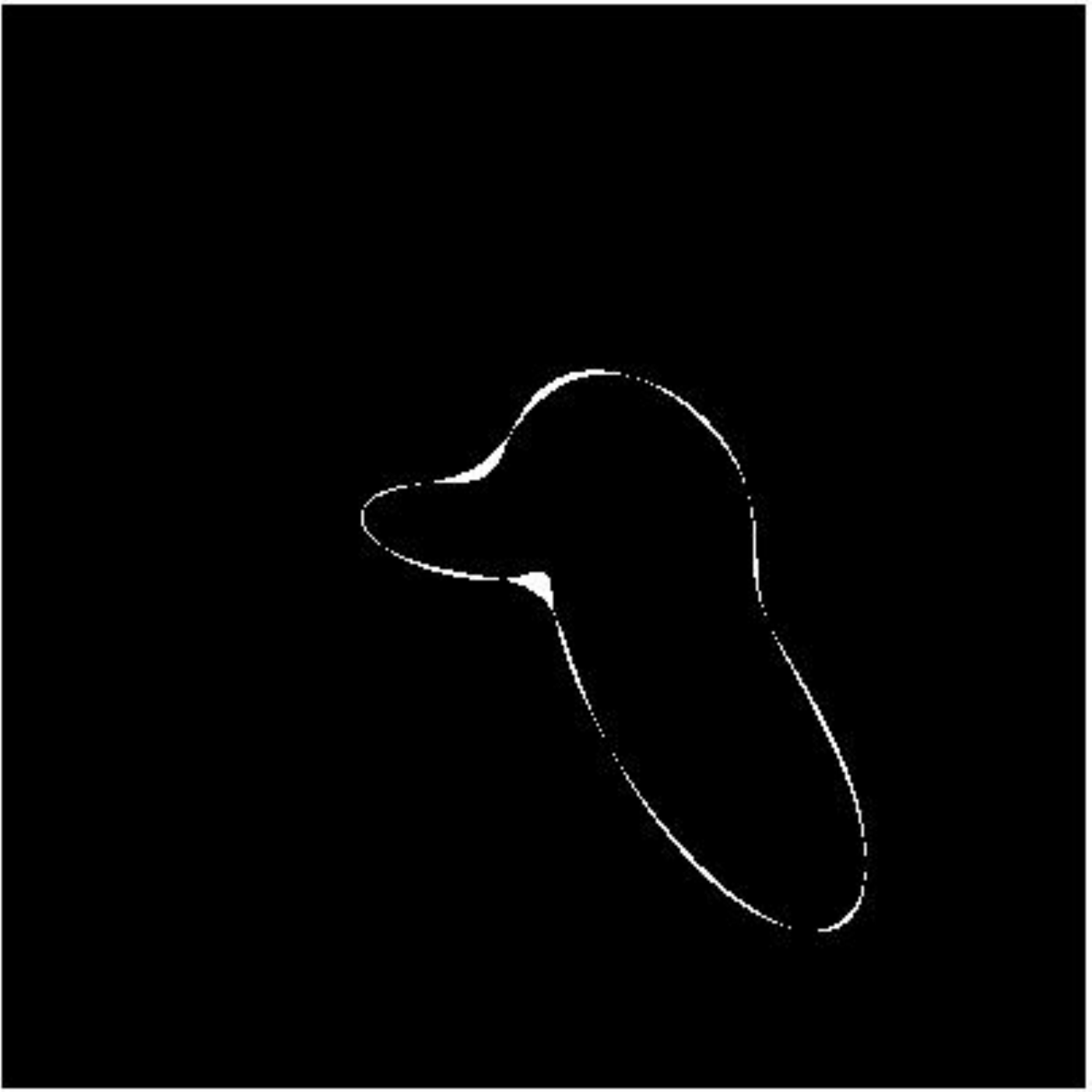}}
\caption{Reconstruction from noisy samples. (a) Original image. (b) Noisy samples of size $29\times 29$ with SNR = 22 dB. (c) Absolute error of the least squares solution. (d) Absolute error of the quadratic programming solution (PSNR = 21.0 dB). (e) Absolute error of the output of the consistency improvement algorithm (PSNR = 22.8 dB). SNR between the samples of the final reconstruction and the noisy samples (b) is 20.9 dB.}\label{fig:noisy2}
\end{figure}

In this experiment, we aim at studying the effect of each step of the algorithm on the reconstructed image from noisy samples. For this purpose, we consider two distinct algebraic shapes of degree 4 with different levels of noise in their samples and we plot each stage of the reconstruction (Figures \ref{fig:noisy1} and \ref{fig:noisy2}). The samples of both images are generated with the sampling kernel $\varphi(x,y)=\beta^{(6)}(x)\beta^{(6)}(y)$ and the annihilation equations involve generalized moments corresponding to the function $g(x)$ in Figure \ref{fig:g_functions}(a). We see that although the least squares solution might be offbeat in presence of noise, the constraints on the sign of the polynomial substantially restrain the solution and lead to satisfactory reconstructions at moderate signal to noise ratios (SNRs).

\subsection{Sampling kernel sensitivity}
\begin{figure}
\centering
\subfloat[]{\includegraphics[width=0.3\linewidth]{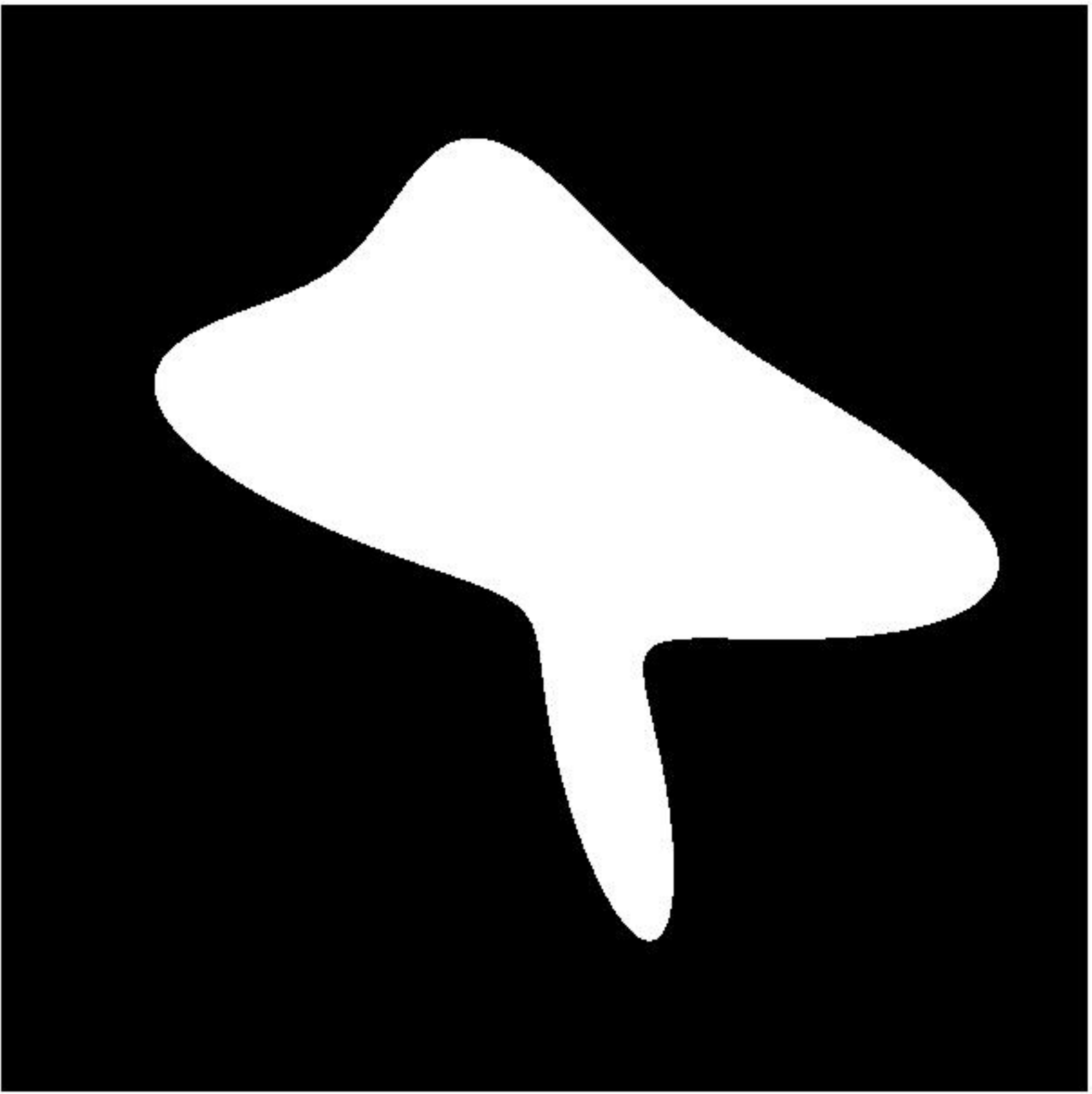}}
\\
\subfloat[]{\includegraphics[width=0.3\linewidth]{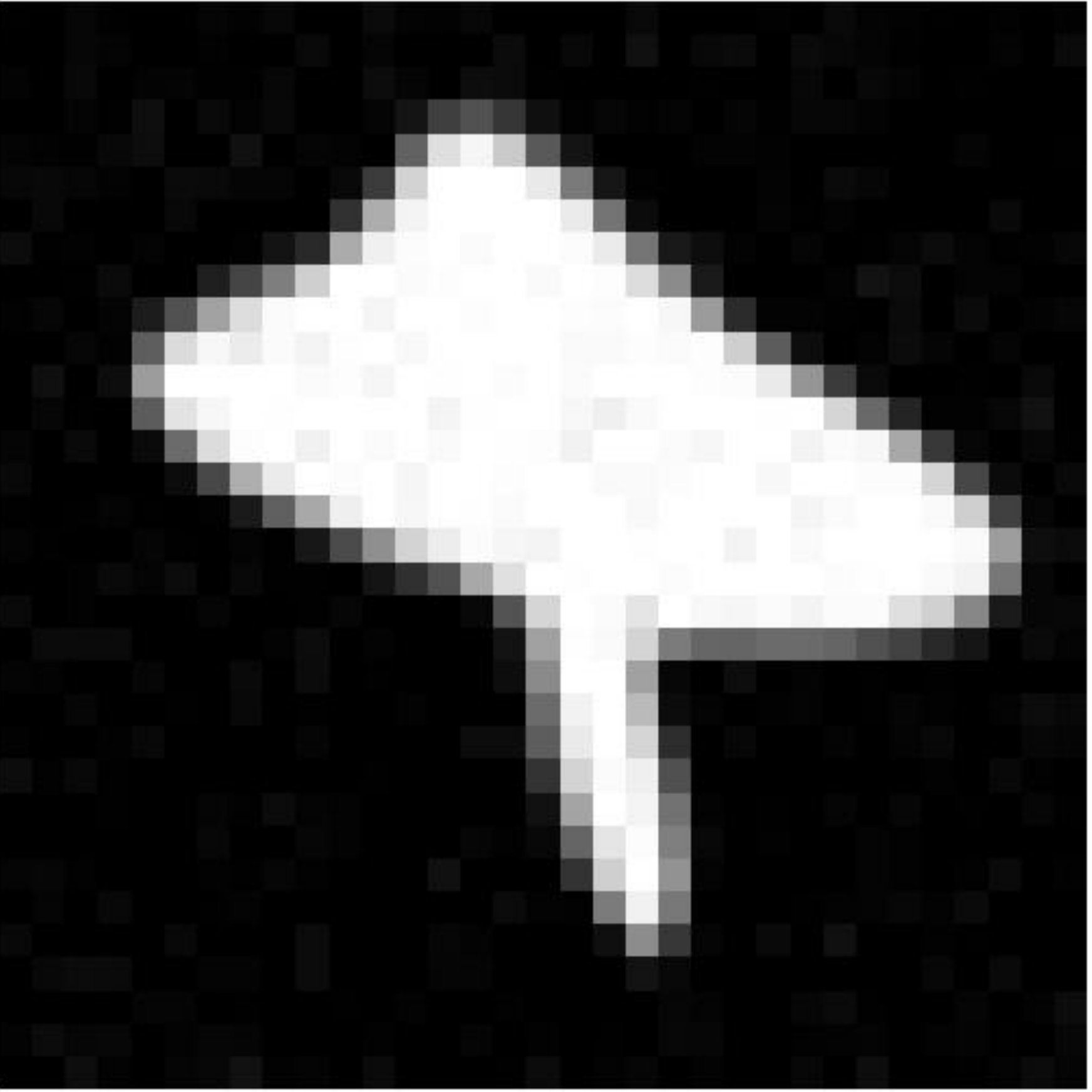}}
\hfill
\subfloat[]{\includegraphics[width=0.3\linewidth]{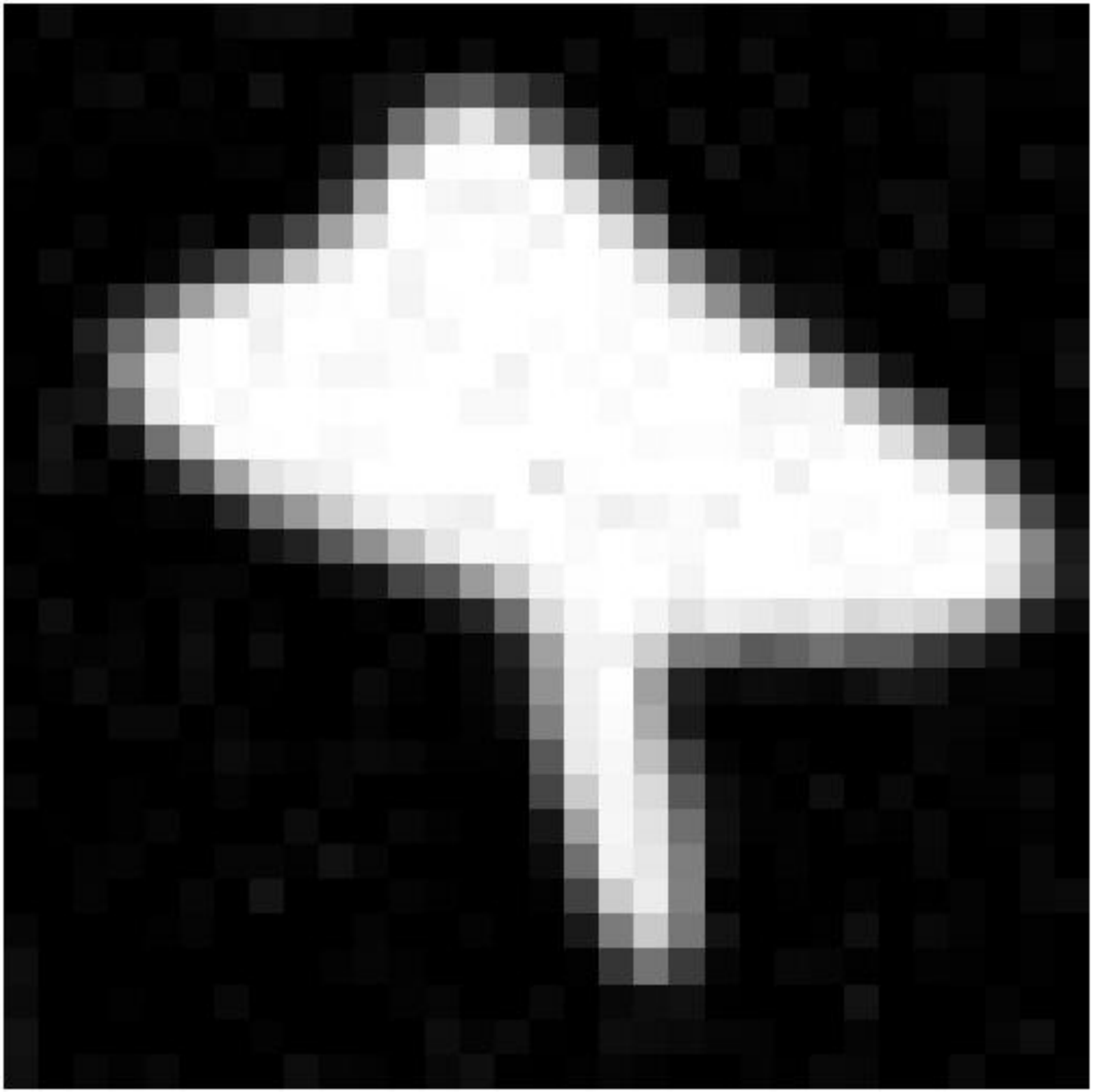}}
\hfill
\subfloat[]{\includegraphics[width=0.3\linewidth]{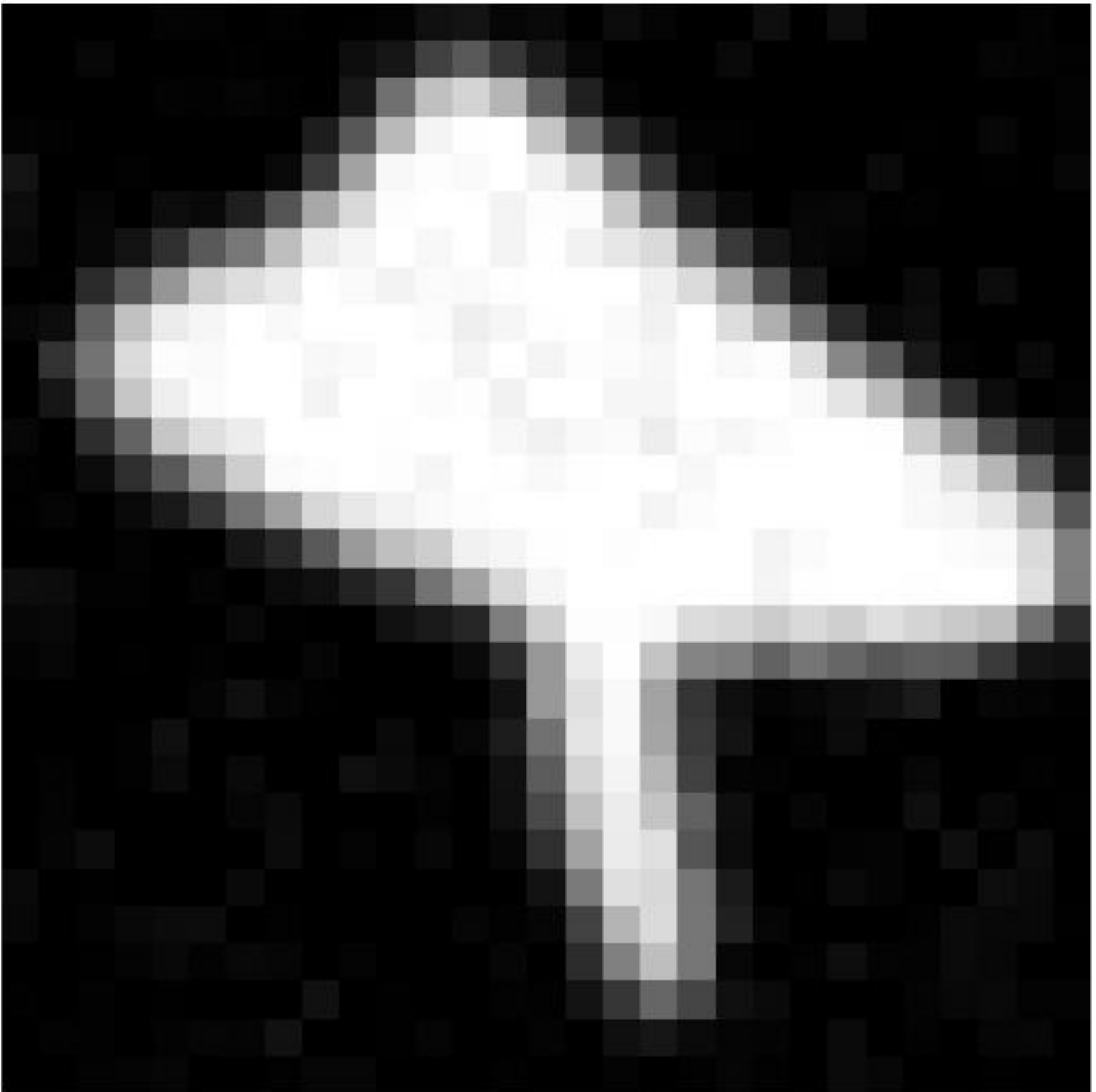}}
\\
\subfloat[]{\includegraphics[width=0.3\linewidth]{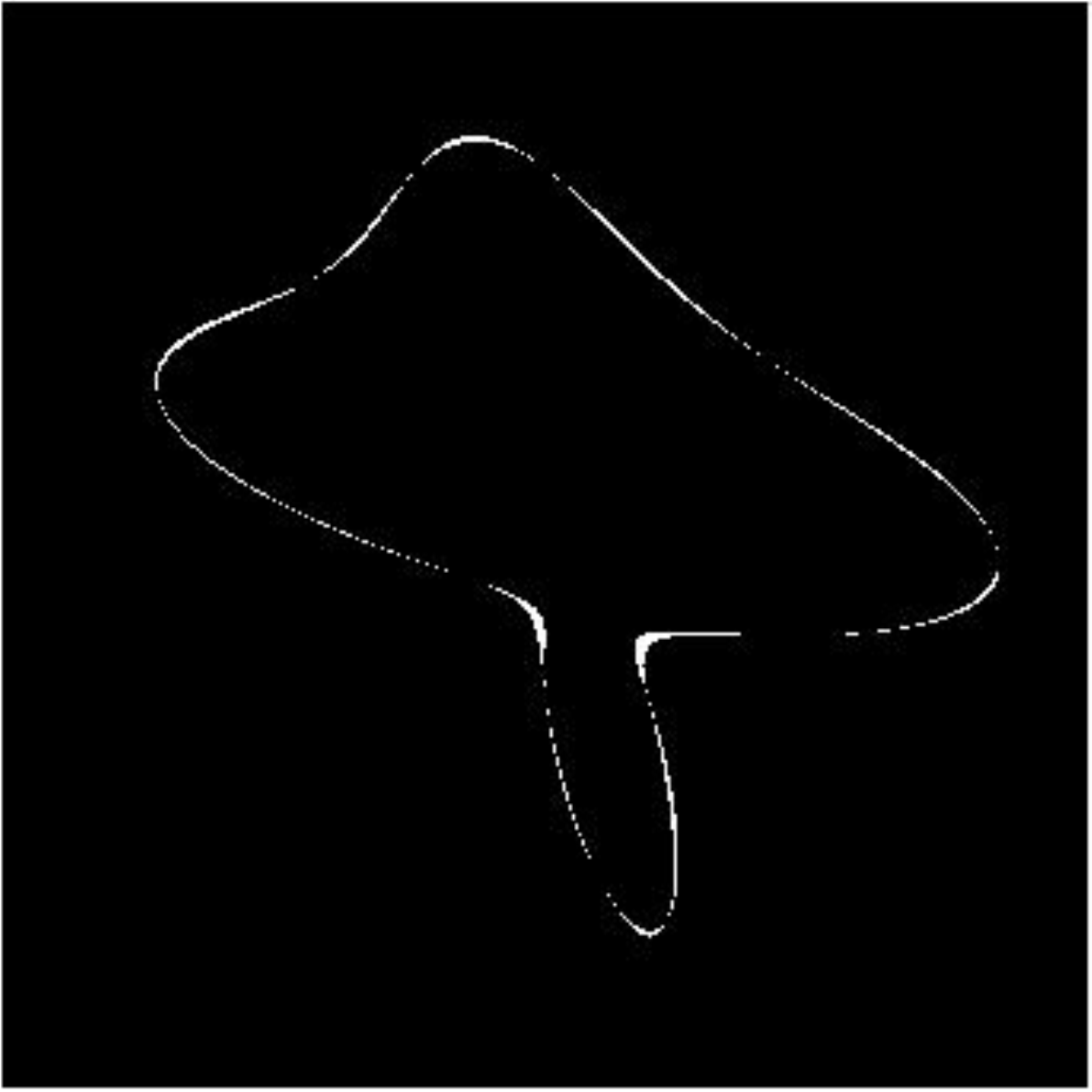}}
\hfill
\subfloat[]{\includegraphics[width=0.3\linewidth]{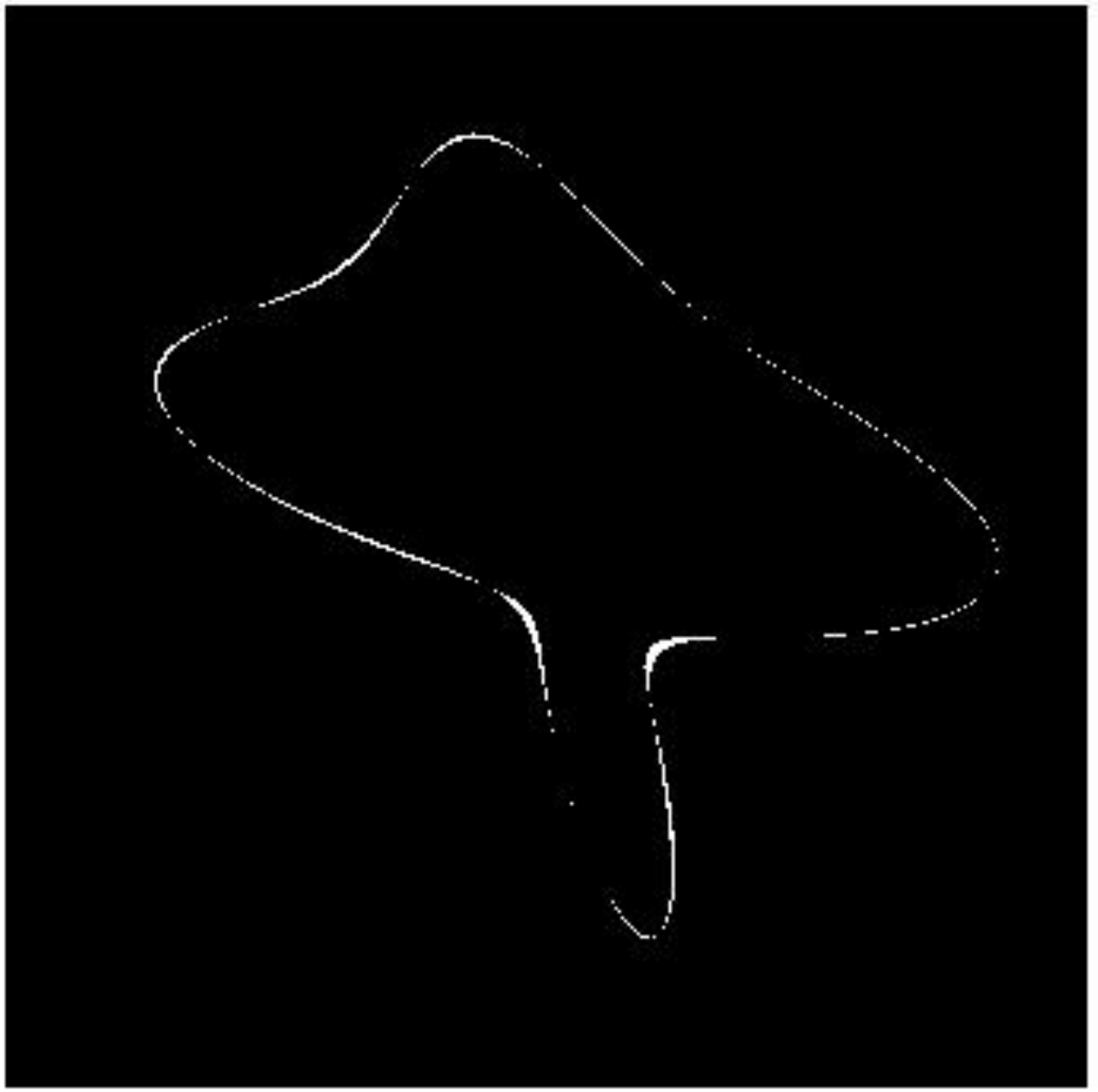}}
\hfill
\subfloat[]{\includegraphics[width=0.3\linewidth]{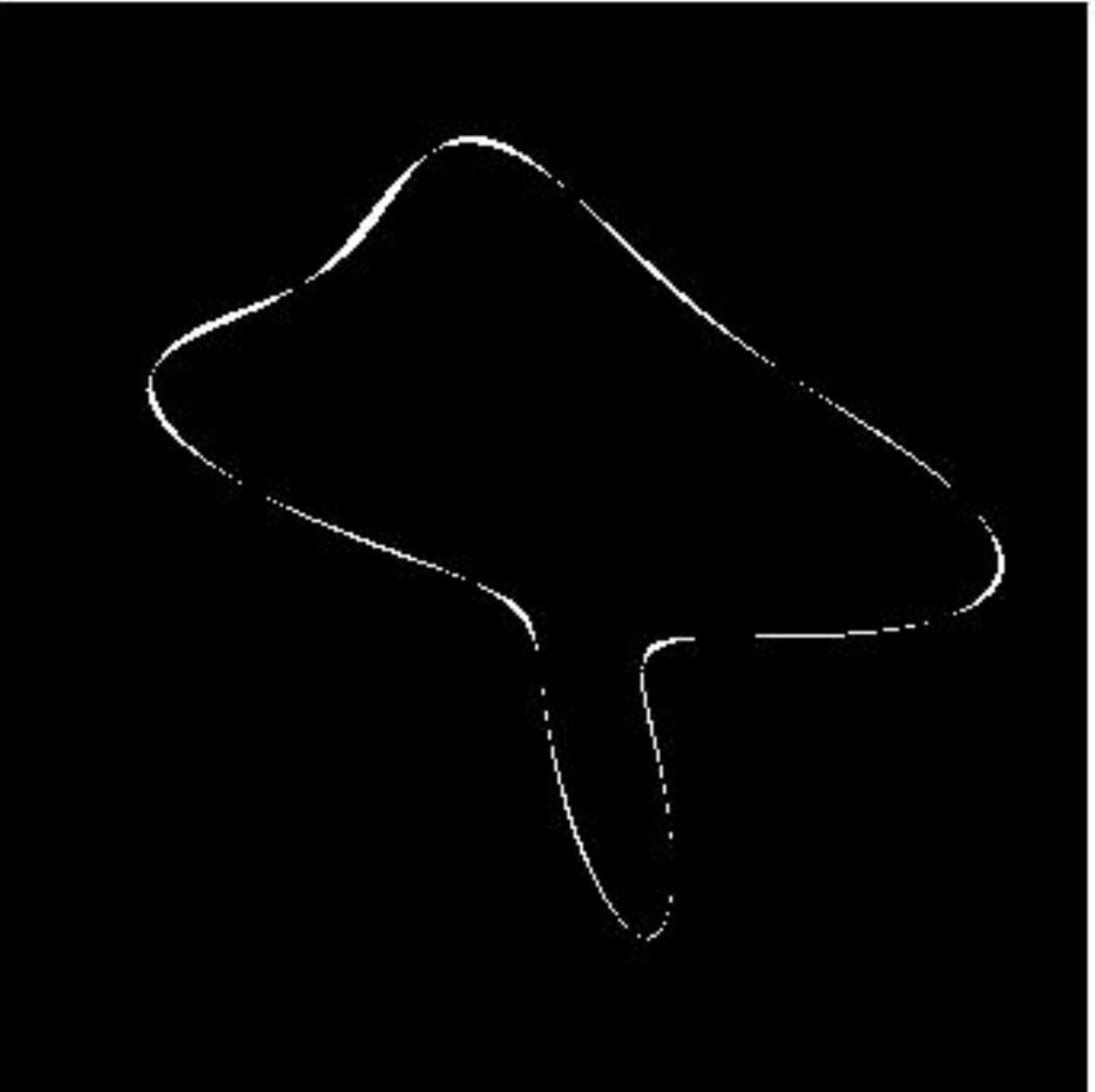}}
\caption{Sensitivity of the reconstruction to the choice of the sampling kernel. (a) Algberaic shape of degree 4. (b),(c),(d) Noisy samples (SNR = 27 dB) of size $33\times 33$, $31\times 31$ and $29\times 29$, generated with B-spline kernels of degree 2, 4, and 6, respectively. (e),(f),(g) Absolute error of the reconstructions from samples in (b),(c) and (d) with reconstruction PSNRs 22.2 dB, 23.2 dB and 21.3 dB, respectively.}\label{fig:kernel_sensitivity}
\end{figure}

Earlier, we mentioned that a consequence of replacing conventional moments with generalized moments is relaxing the restrictive polynomial-reproducing requirement on the sampling kernel. Specifically, we worked out the reproducing coefficients for the B-spline kernels of order 2, 4, and 6 that generate stable generalized moments of order less than or equal to 6 (see Figure \ref{fig:g_functions}). This, for example, allows us to recover algebraic shapes of degree 4 from samples generated with the sampling kernel $\varphi(x,y)=\beta^{(2)}(x)\beta^{(2)}(y)$. In this experiment, we study the sensitivity of the reconstruction to the choice of the kernel. Figure \ref{fig:kernel_sensitivity} displays the absolute difference between an image and its reconstructions from samples generated with different sampling kernels and similar signal-to-noise-ratios. The results are comparable irrespective of the choice of the sampling kernel (note the expected difference in the sample sizes that calls for different noise realizations for the three samples).

\subsection{Unbounded algebraic shapes}
\begin{figure}
\centering
\subfloat[]{\includegraphics[width=0.3\linewidth]{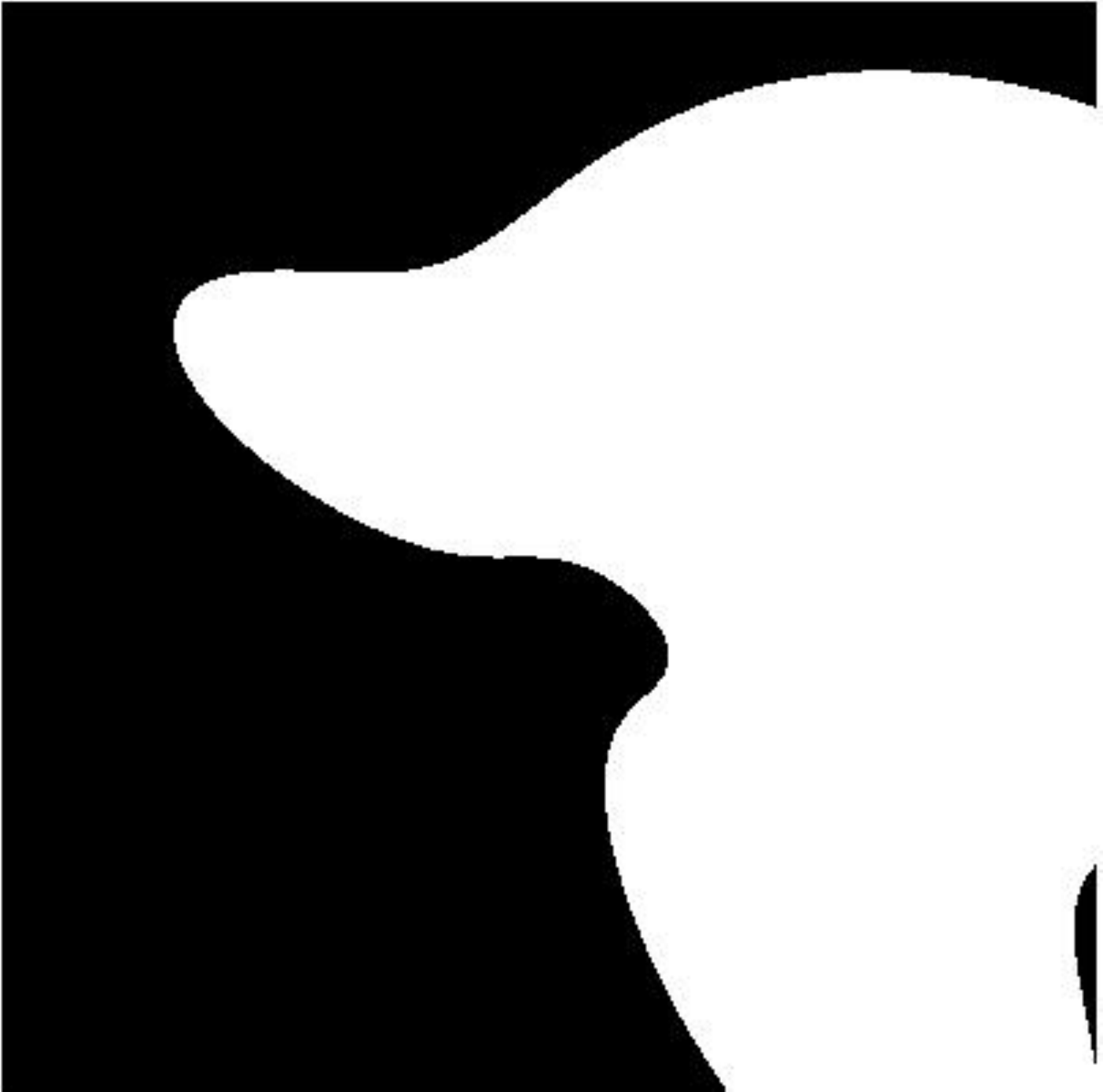}}
\hfill
\subfloat[]{\includegraphics[width=0.3\linewidth]{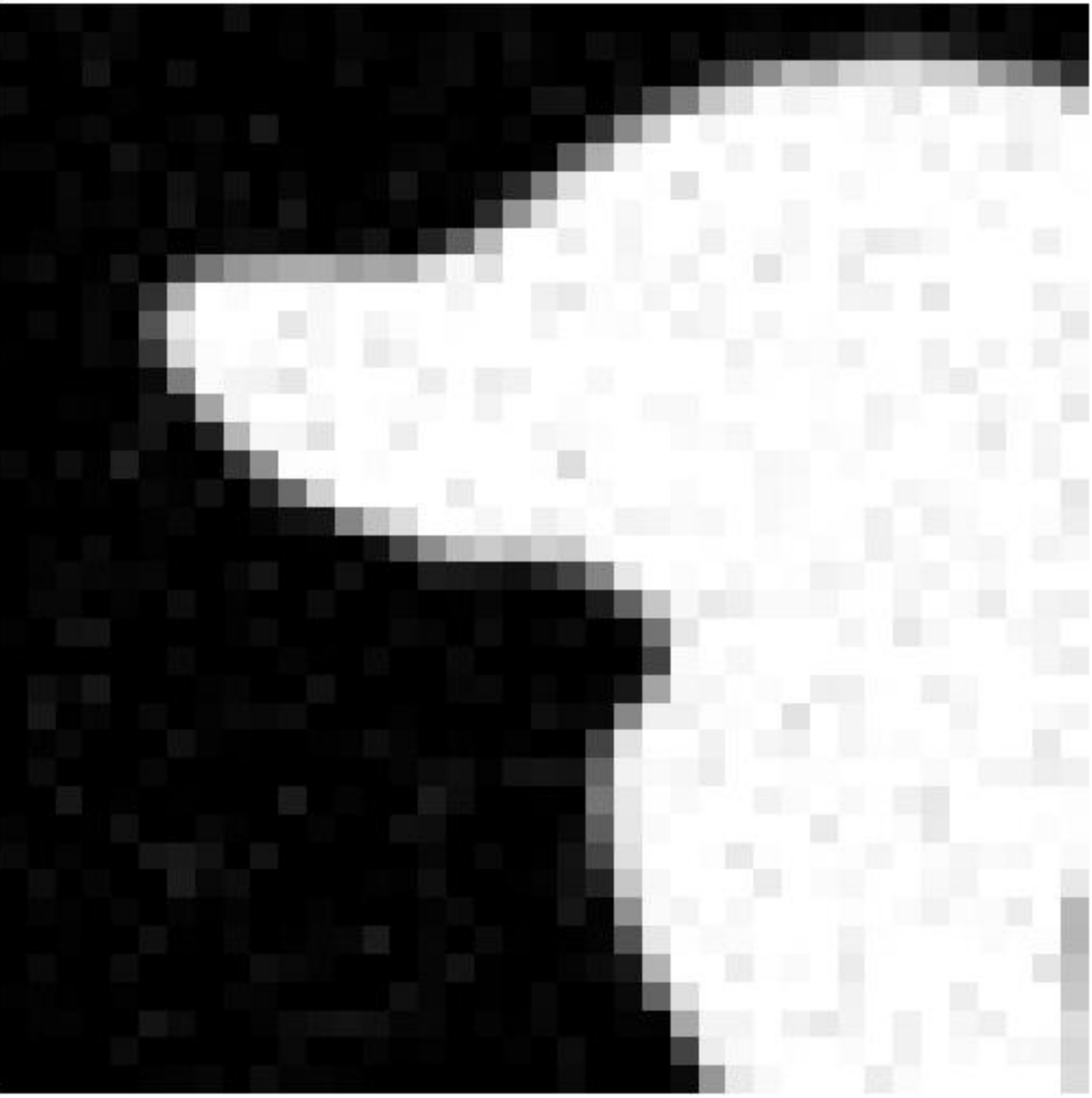}}
\hfill
\subfloat[]{\includegraphics[width=0.3\linewidth]{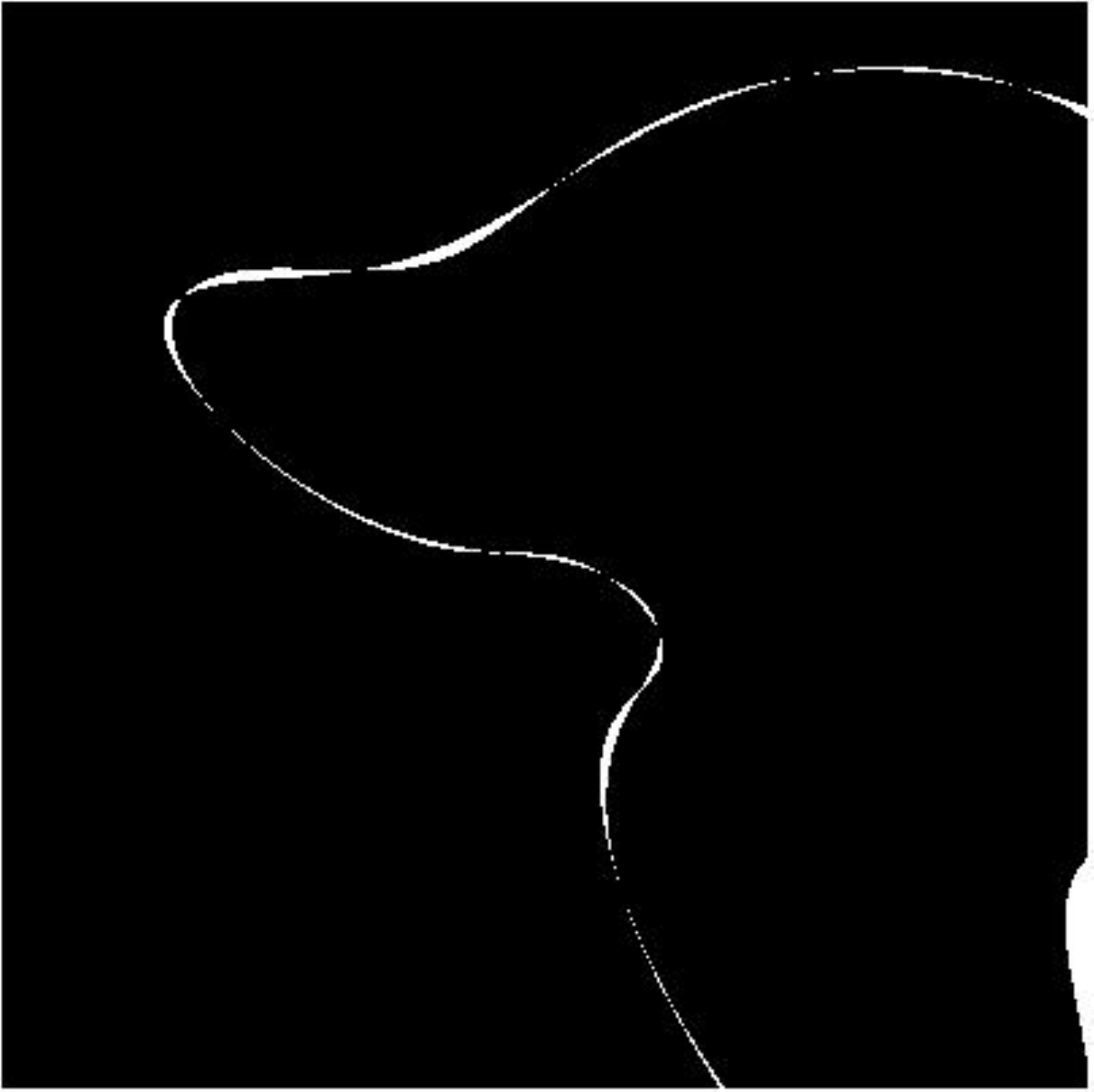}}
\caption{(a) Unbounded algebraic shape of degree 4. (b) Noisy samples of size $39\times 39$ with SNR = 25 dB. (c) Absolute reconstruction error.}\label{fig:unbounded}
\end{figure}

Introducing generalized moments to the annihilation equations facilitated sampling and reconstruction of algebraic shapes with open boundaries (also referred as unbounded algebraic shapes). This additionally allows the reconstruction to enjoy oversampling by forming annihilation equations for each sample window, without caring about the image content of the window. Figure \ref{fig:unbounded} shows the reconstruction of an unbounded image from its noisy samples, where the sampling kernel is the tensor product of 2nd order B-splines. The peak signal-to-noise ratio (PSNR) of the reconstructed image is 20.1 dB and SNR between its samples and the original noisy samples (sample consistency) is equal to 23.1 dB. These numbers clearly indicate the success of our proposed algorithm for reconstructing unbounded shapes.

\subsection{Overfitting}
\begin{figure}
\centering
\subfloat[]{\includegraphics[width=0.3\linewidth]{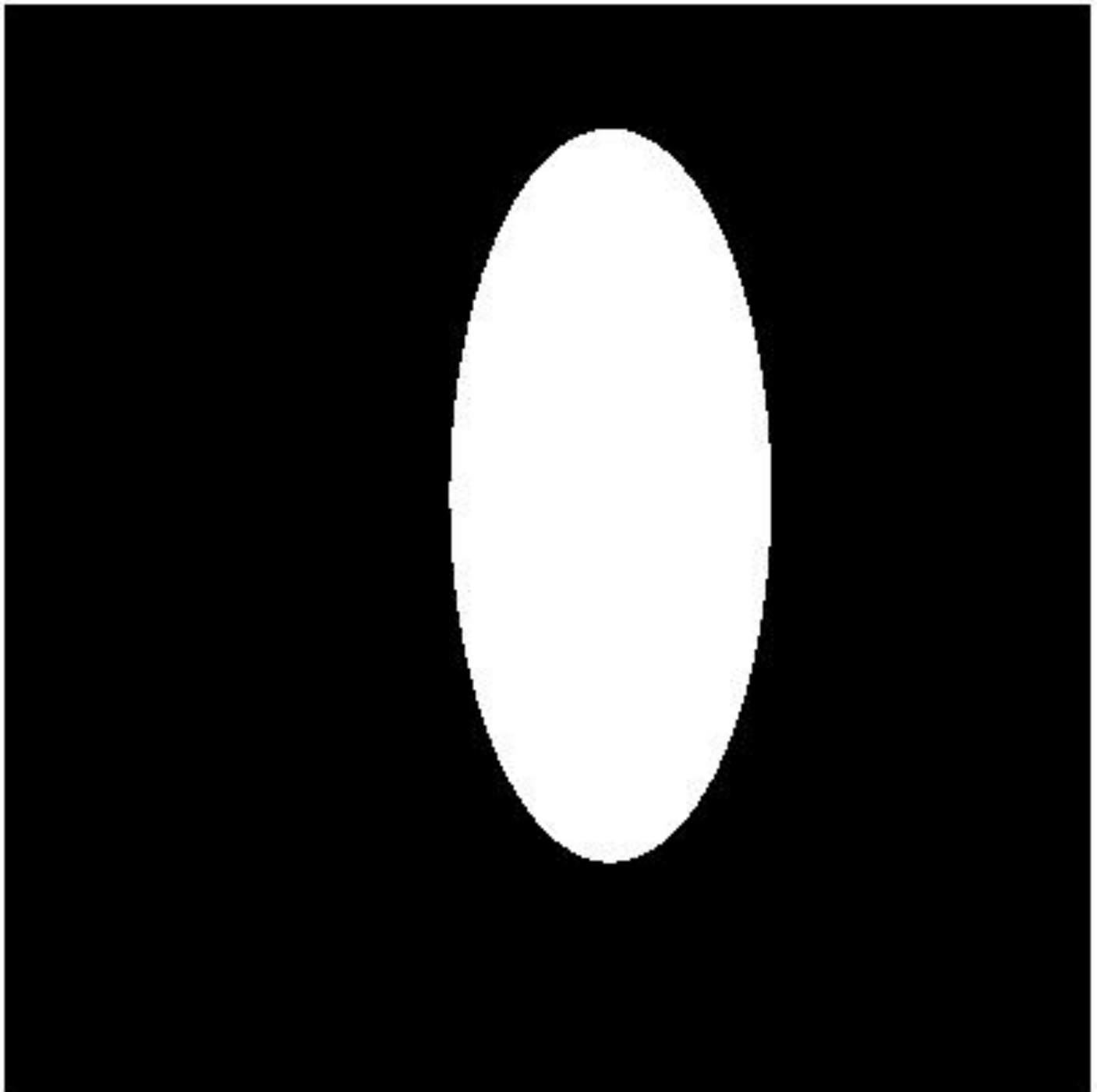}}
\\
\subfloat[]{\includegraphics[width=0.3\linewidth]{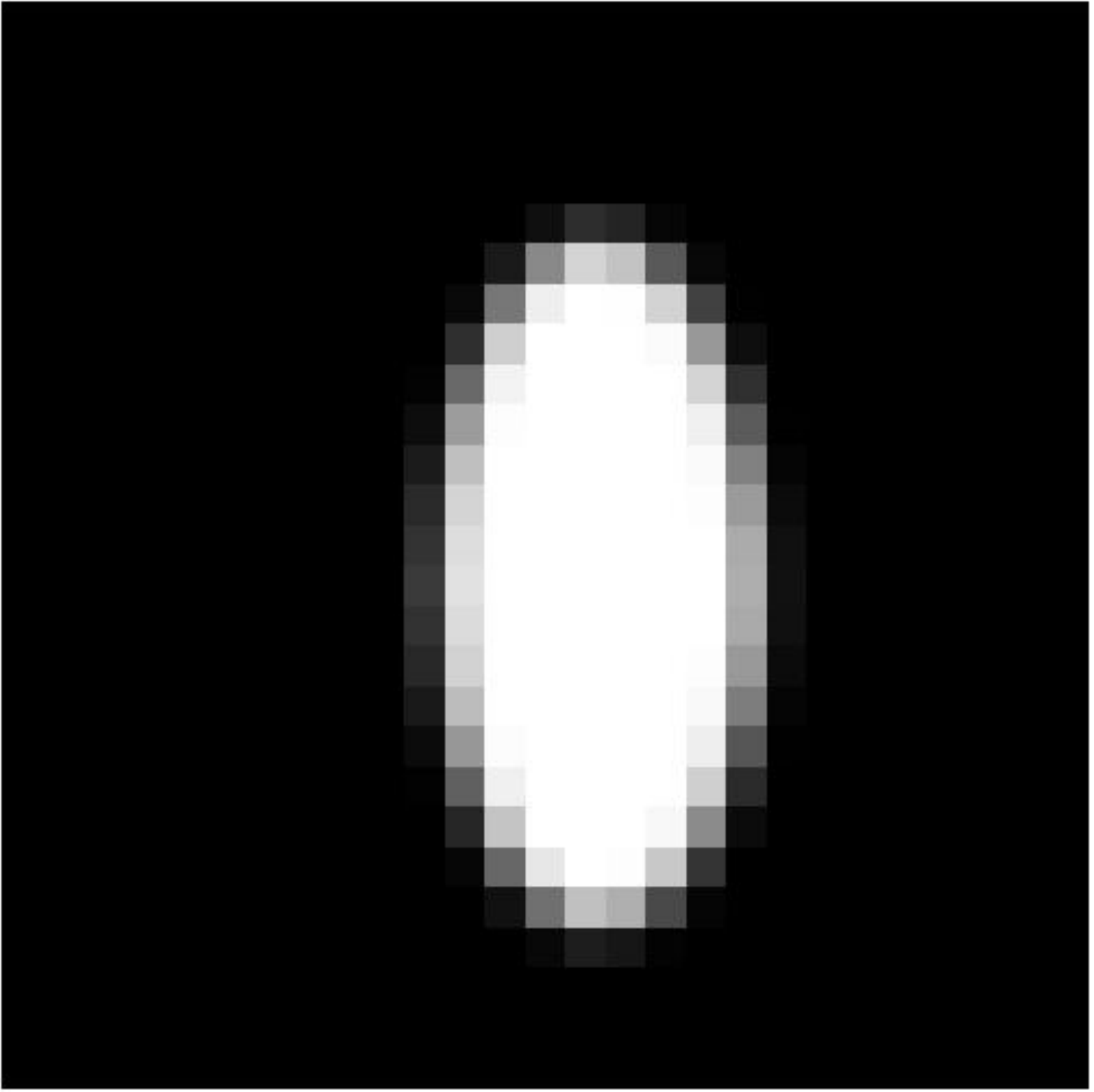}}
\hfill
\subfloat[]{\includegraphics[width=0.3\linewidth]{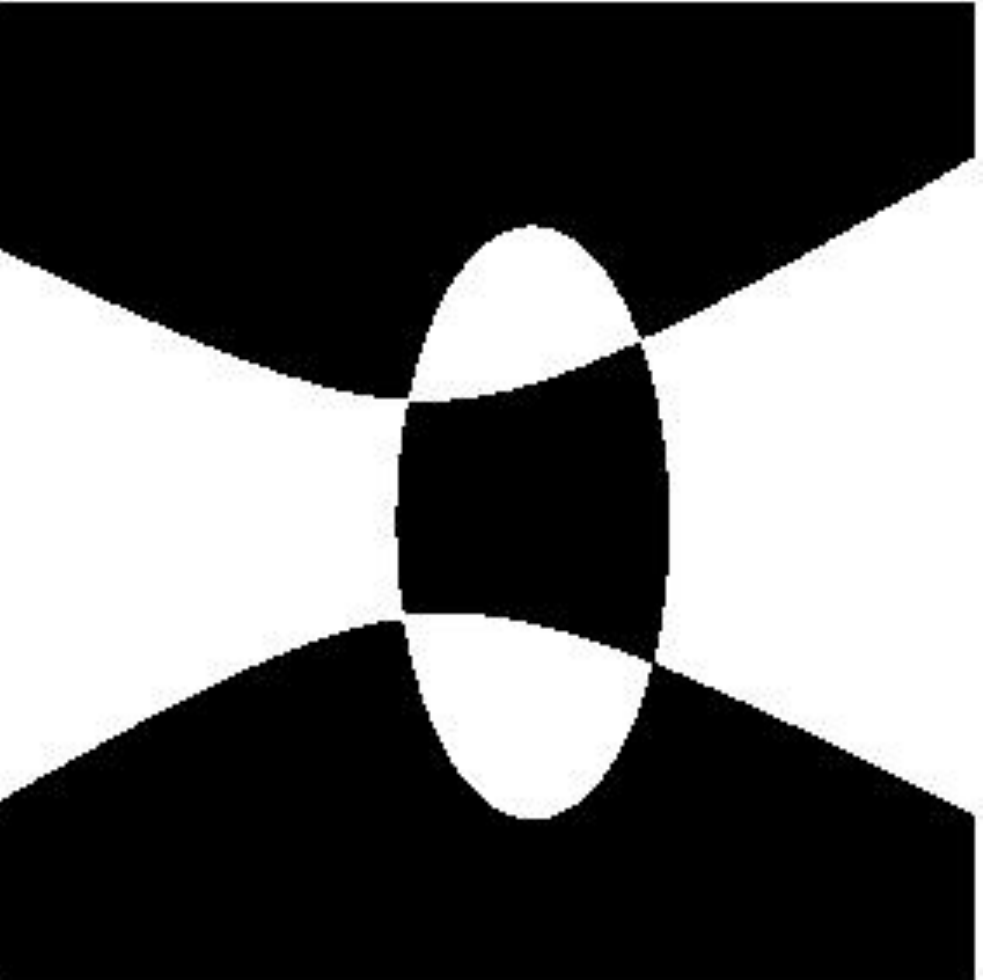}}
\hfill
\subfloat[]{\includegraphics[width=0.3\linewidth]{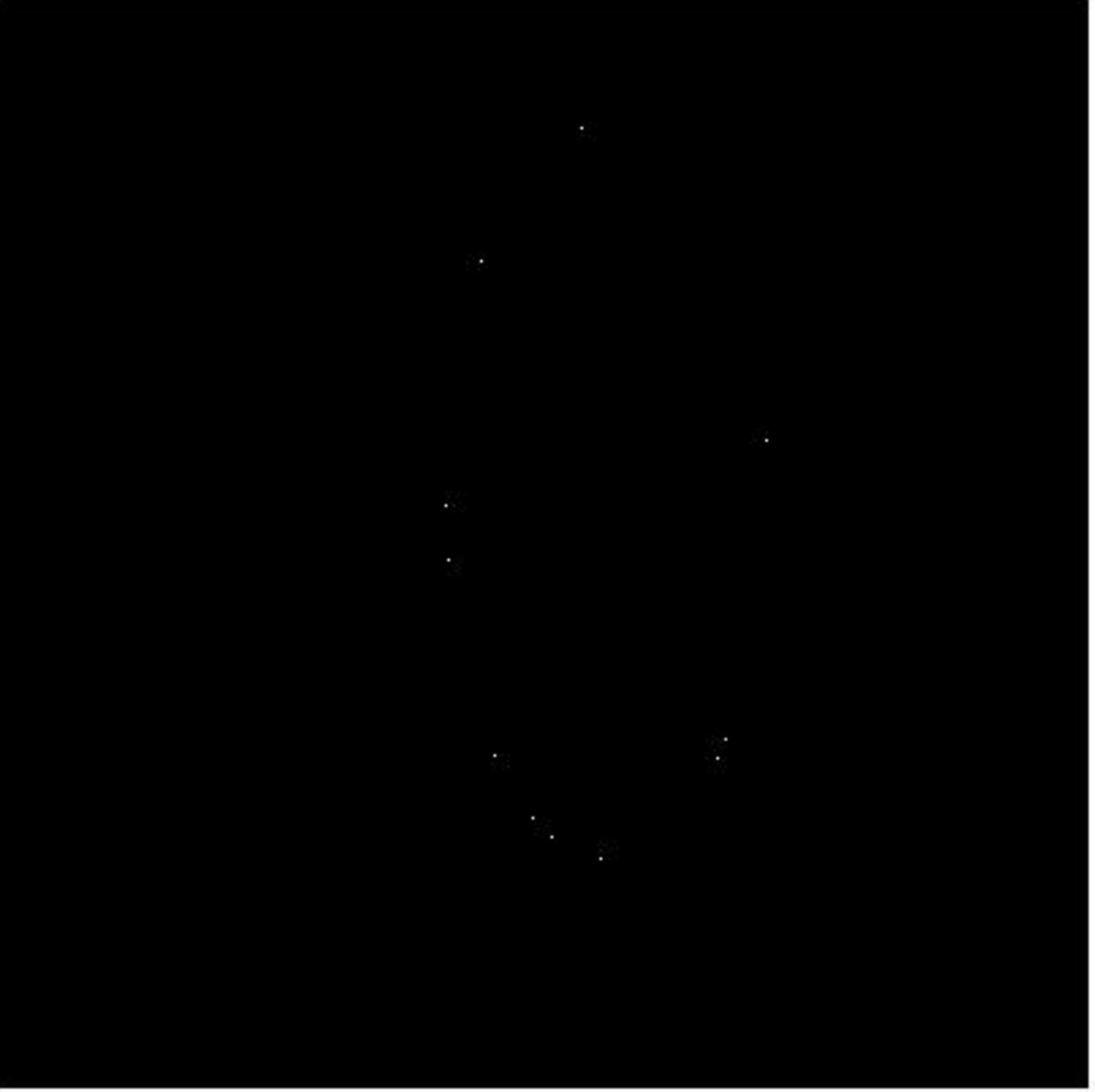}}
\\
\subfloat[]{\includegraphics[width=0.3\linewidth]{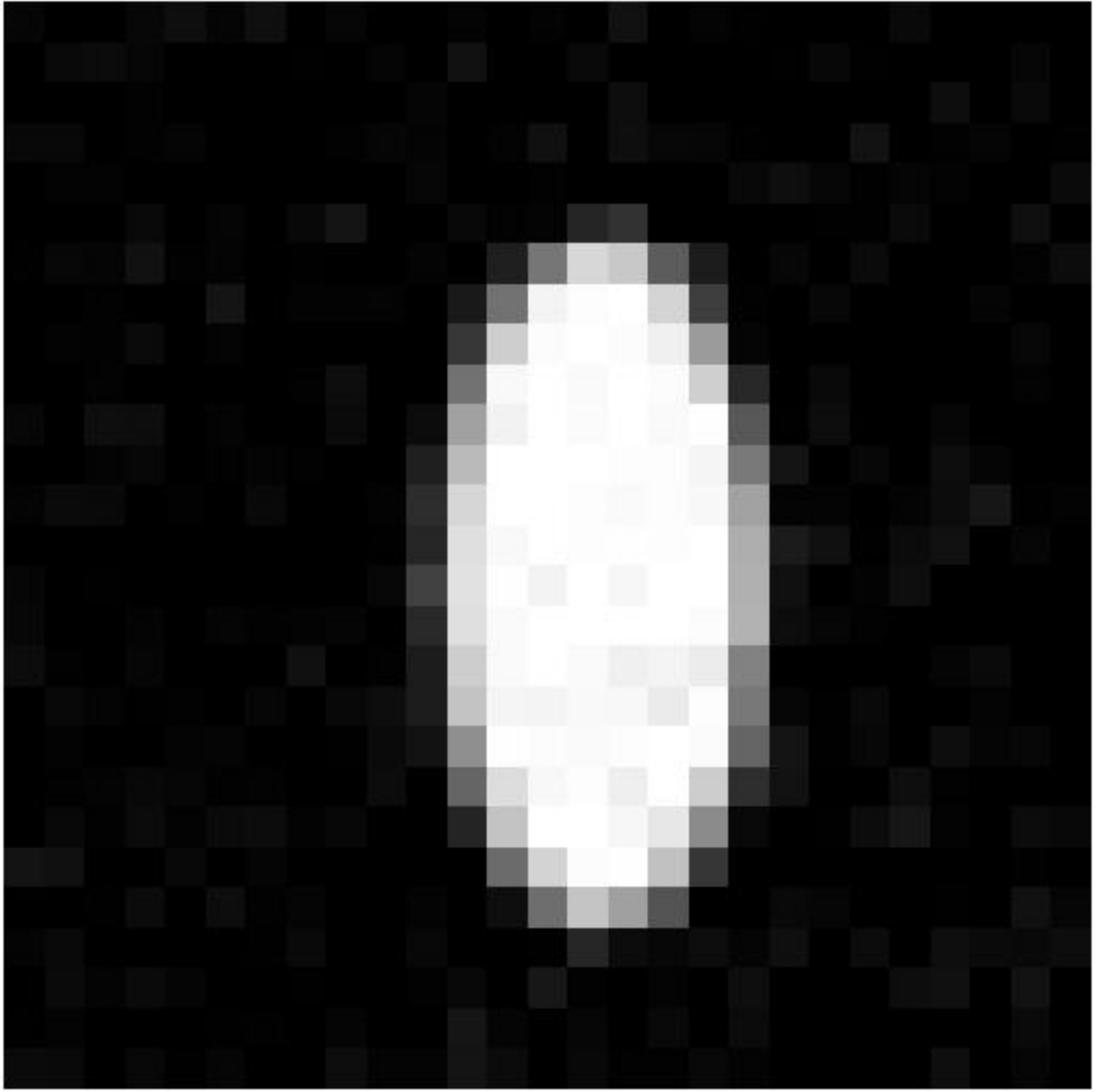}}
\hfill
\subfloat[]{\includegraphics[width=0.3\linewidth]{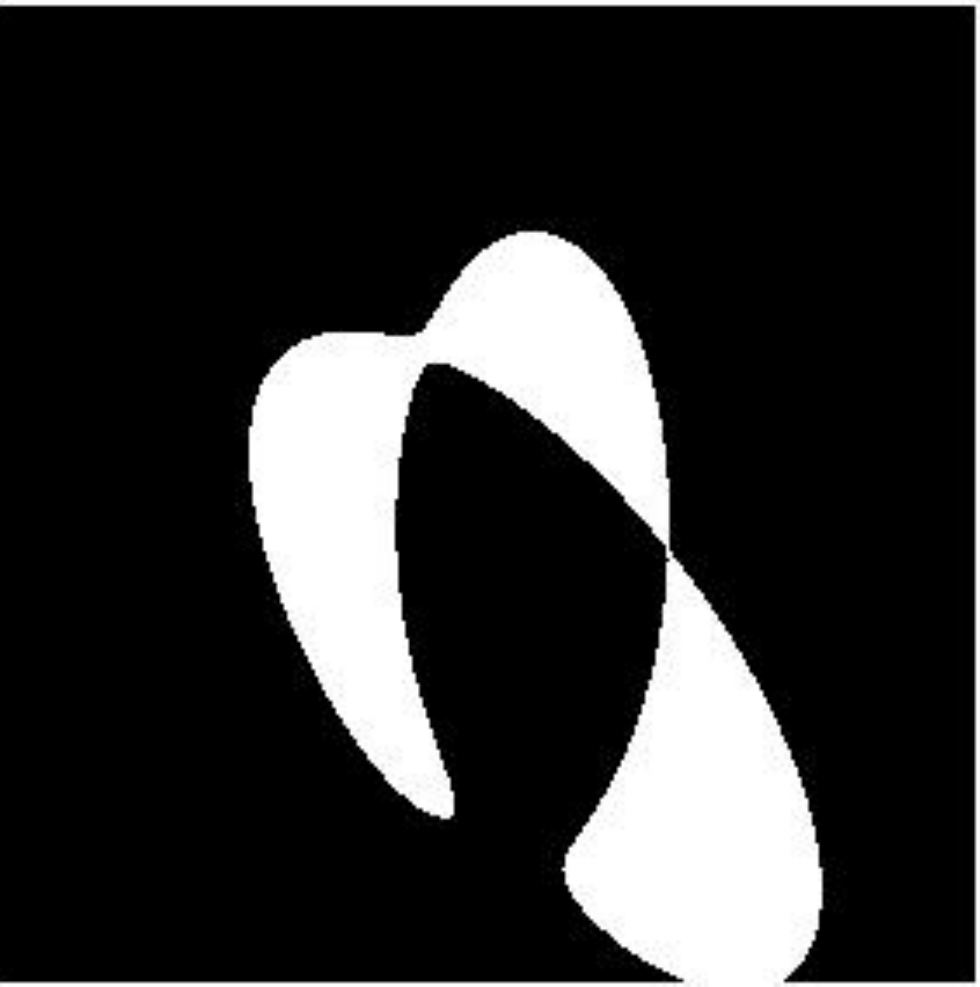}}
\hfill
\subfloat[]{\includegraphics[width=0.3\linewidth]{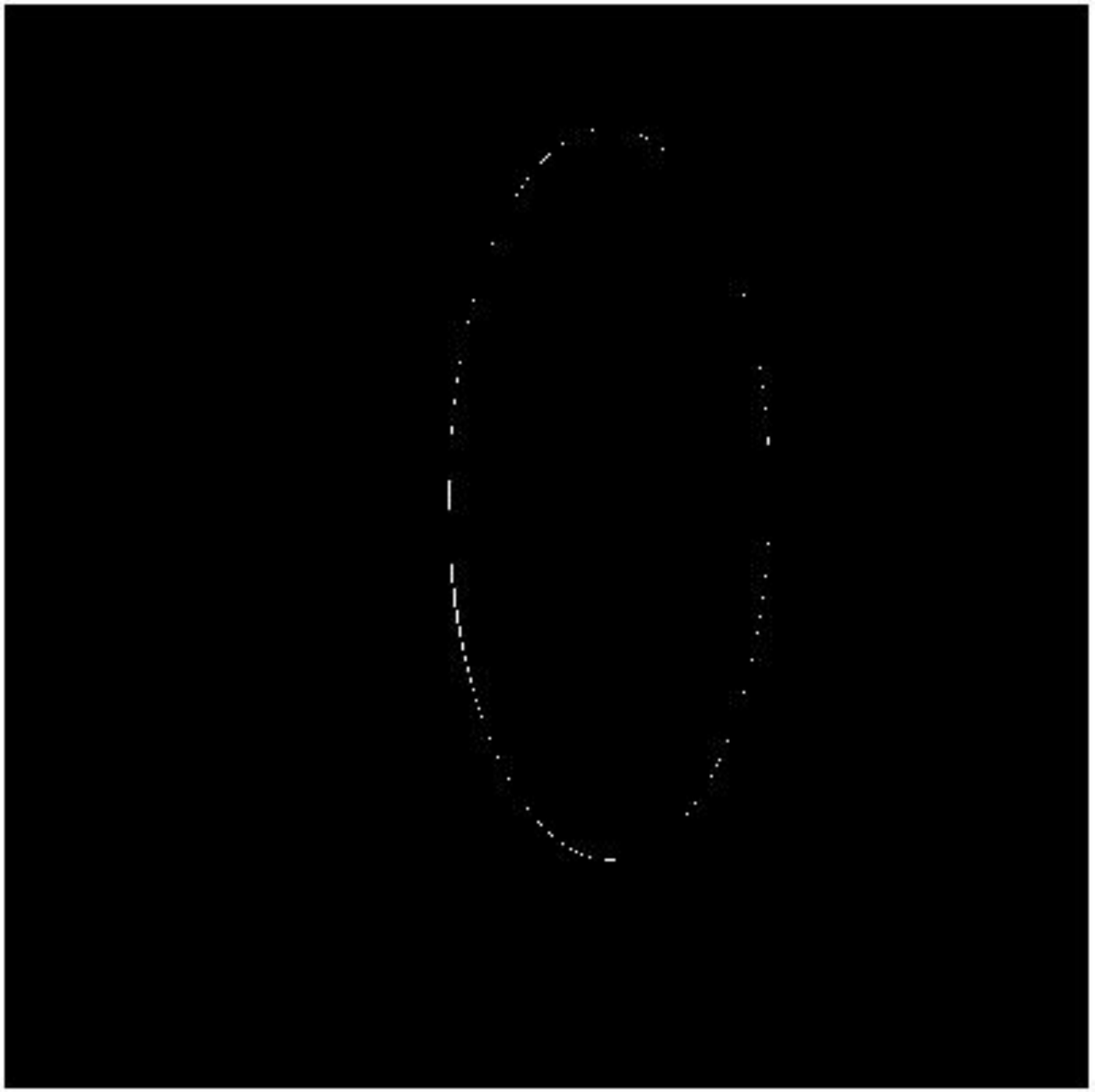}}
\caption{Approximation of an ellipse with algebraic shapes of degree 4. (a) Original ellipse. (b) Noiseless samples of size $27\times 27$, generated with B-splines of degree 2. (c) Least squares solution for noiseless samples. (d) Absolute error of the final reconstruction of the algorithm. (e) Noisy samples with SNR = 22 dB. (f) Least squares solution for noisy samples. (g) Absolute error of the final reconstruction from noisy samples.}\label{fig:overfitting}
\end{figure}

In the last two experiments, we address uncertainties in the image model. First, we study the situation where we overestimate degree of an algebraic shape. Recalling Theorem \ref{theo:mainTheorem} of the previous section, we expect the recovered polynomial from the annihilation equations to vanish on the boundaries of the original shape in the noiseless scenario. Figure \ref{fig:overfitting} displays the results when we approximate an ellipse with algebraic shapes of degree 4 from its noiseless and noisy samples. Figures \ref{fig:overfitting}(c) and \ref{fig:overfitting}(f) show the least squares solutions for noiseless and noisy samples, respectively. Both figures indicate that the boundaries of the recovered images contain the boundary of the original ellipse. Equivalently, the recovered polynomials are factors of the original polynomial of degree 2. The extra factors are resolved in the next steps of the algorithm, resulting in exact reconstructions in both scenarios.

\subsection{Algebraic shape approximation}
\begin{figure}
\centering
\subfloat[]{\includegraphics[width=0.3\linewidth]{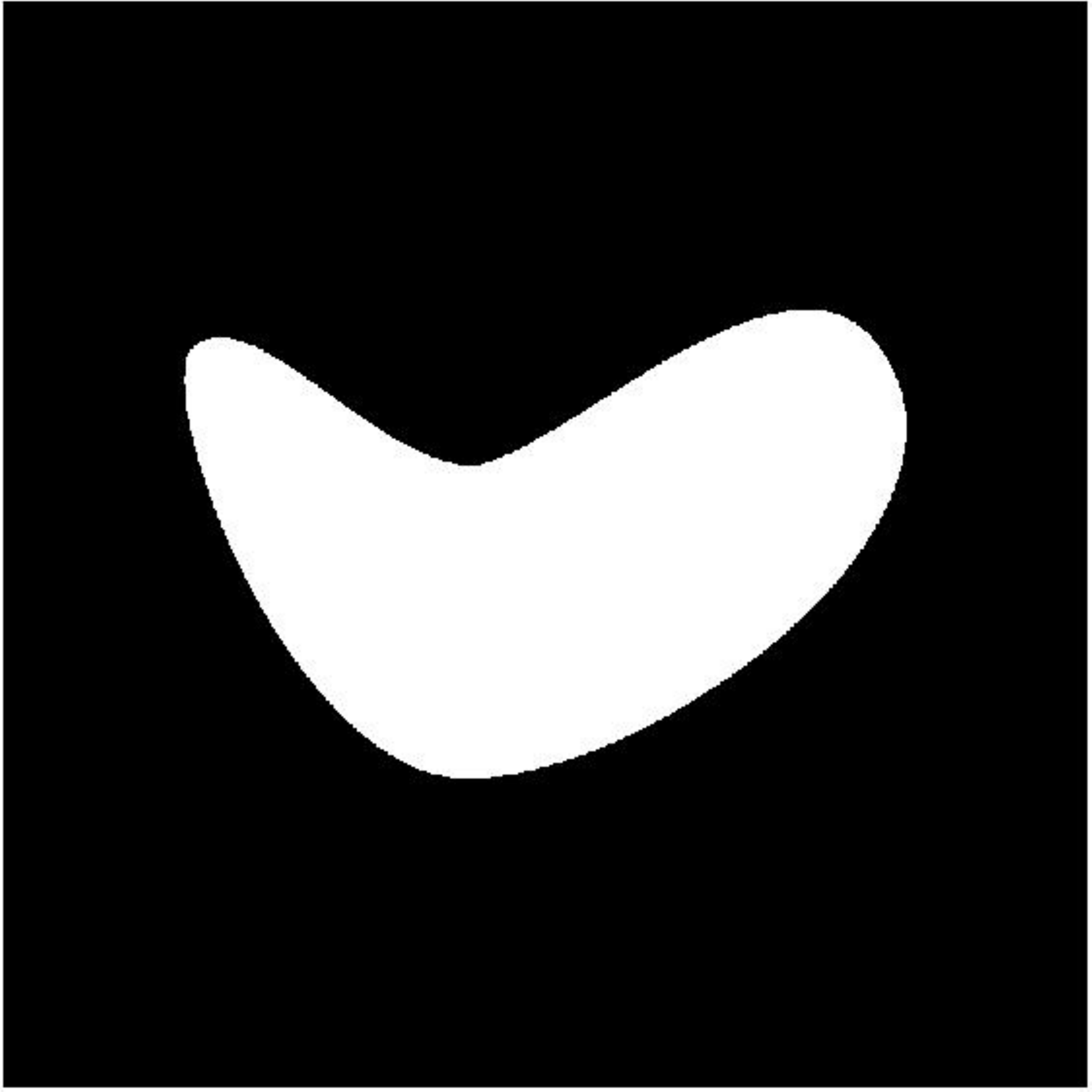}}
\hfill
\subfloat[]{\includegraphics[width=0.3\linewidth]{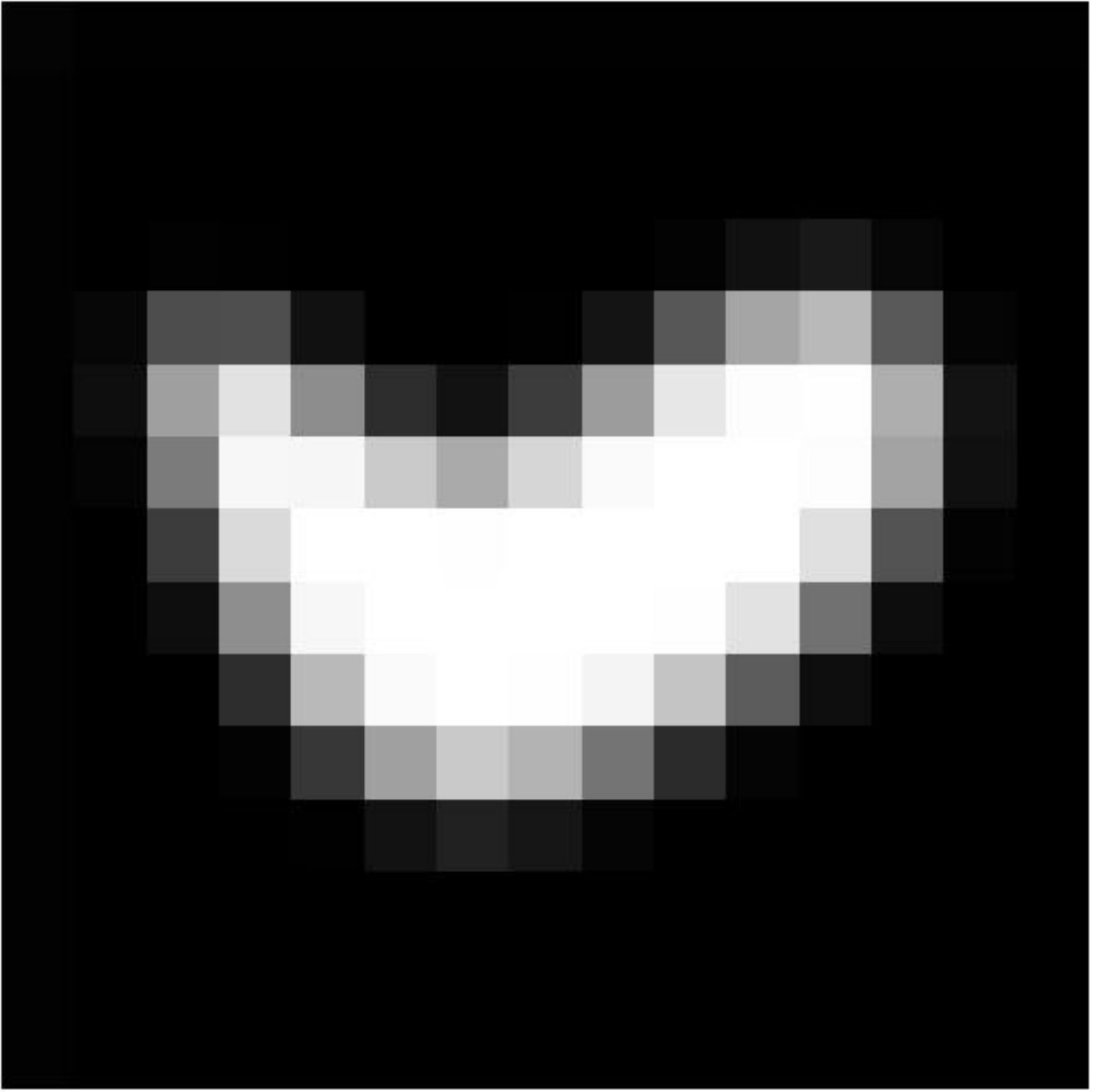}}
\hfill
\subfloat[]{\includegraphics[width=0.3\linewidth]{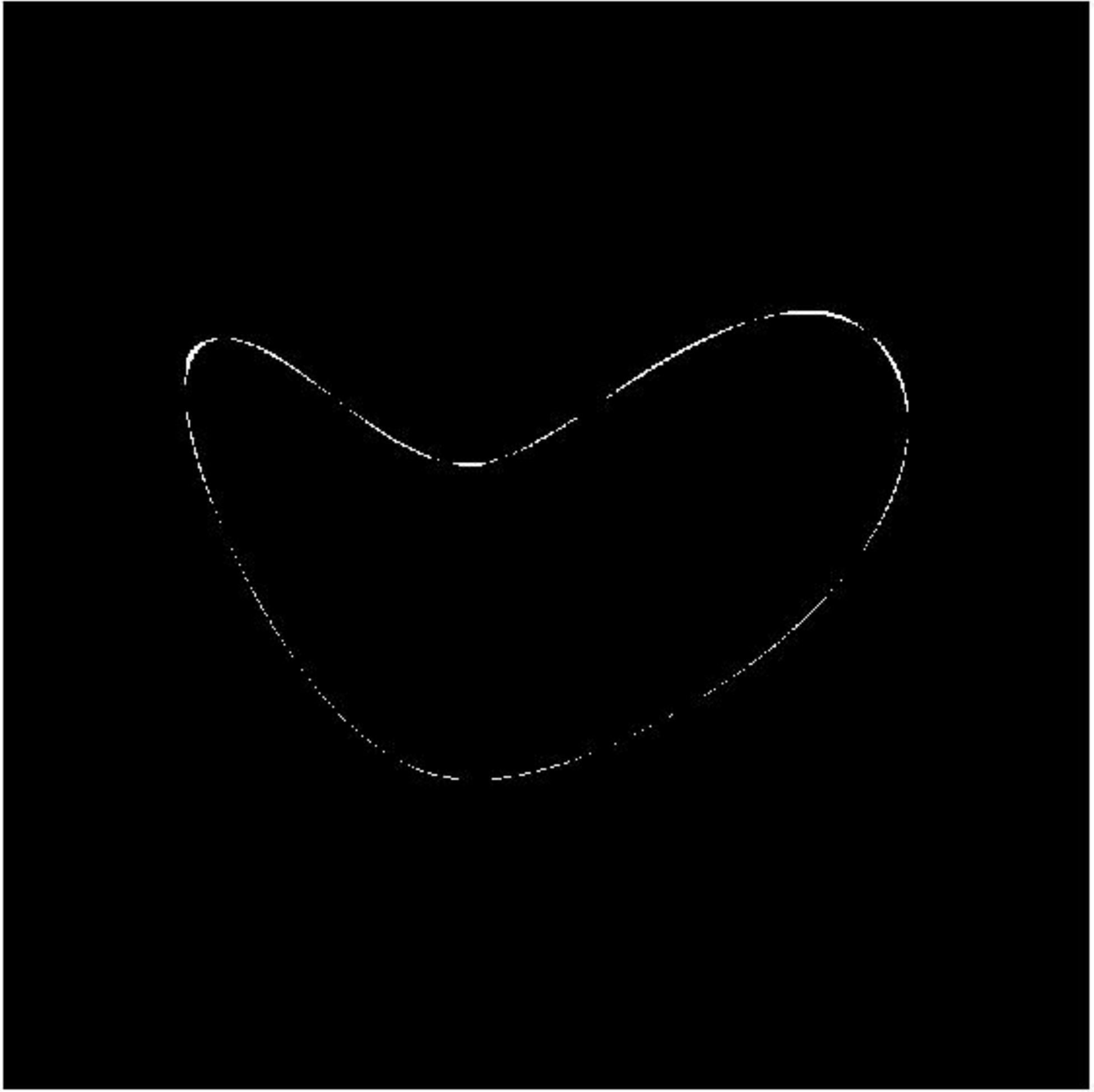}}
\caption{Approximation of non-annihilable curves with algebraic curves. (a) A shape with a B\'ezier curve boundery. (b) $15\times 15$ noiseless samples. (c) The absolute error between the original shape and its approximation with an algebraic shape of degree 4. The reconstruction PSNR is 19.8 dB.}\label{fig:approximation}
\end{figure}

Another type of uncertainty in the image model happens when the image boundary is not an algebraic curve. Regarding the descriptive power of algebraic curves, we still expect to find a good approximation of the image. To investigate this, we generated a shape with a B\'ezier curve boundary with four control points and generated its $15\times15$-samples with 2nd order B-spline sampling kernels. Then, we obtained the approximate algebraic shape from the noiseless samples.  The original image and the absolute error of its algebraic approximation are depicted in Figure \ref{fig:approximation}. We observe that the reconstructed curve is a rather accurate descriptor of the original B\'ezier curve.

\section{Conclusion}\label{sec:conclusion}
Designing sampling schemes for images with arbitrary edge geometries is still a challenging research problem. In this paper, we proposed a sampling and reconstruction algorithm for binary images with boundary curves that are zeros of an implicit bivariate polynomial. We developed a set of linear annihilation equations from the image samples and proved that every solution of the equations restores the image boundaries, in the noiseless scenario. The primary equations involve 2D moments of the image. To make the reconstruction robust against noise, we replaced conventional moments with generalized moments associated with a compact-support 2D function that is paired with the given sampling kernel. This leads to a reconstruction algorithm from more realistic samples and extends the model to images with open boundaries.

The image model we considered in this paper is very rich and may be used for the approximation of general shapes from their samples. Also, the idea of replacing conventional moments with generalized moments might find applications in other image processing tasks which use moments as the image descriptors.

\section*{Acknowledgment}
The authors would like to thank P. Th\'evenaz for his helpful hints and suggestions. Also, they are thankful to L. Baboulaz for sharing MATLAB codes and P.~L.~Dragotti for the valuable discussion in the beginning of this work.

\appendix[Proof of Theorem \ref{theo:mainTheorem}]
We prove by contradiction. Assume the zero level set of $\tilde{p}$ does not fully include $\mathcal{C}$; thus, $p(x,y)$ can be factorized as 
$$p(x,y) = \zeta(x,y)\, h(x,y),$$ 
where $h(x,y)$ is coprime with $\tilde{p}$ and $\zeta$, and has a non-trivial zero level set $\mathcal{C}_h$. Meanwhile, the zero level set $\mathcal{C}_{\zeta}$ of $\zeta(x,y)$ is included in that of $\tilde{p}$. Roughly speaking, $h$ and $\zeta$ stand for parts of $\mathcal{C}$ that are excluded and included in the zero level set of $\tilde{p}$, respectively. Further, let $r(x,y)$ be a polynomial with minimum degree such that $\ind_{h(x,y)\leq 0} = \ind_{r(x,y)\leq 0}$. If $h$ is irreducible, we shall have $r(x,y)=h(x,y)$, otherwise, $r$ might be different from $h$. In either case, we have $\textrm{deg}\, r \leq \textrm{deg}\, h$.

The validity of annihilation equations \eqref{eq:annihilation_generalized1} and \eqref{eq:annihilation_generalized2} imply 
\begin{align*}
\iint_\Omega g(x)g(y)x^r y^s \tilde{p}(x,y)\frac{\partial I(x,y)}{\partial x} \,\dd x \dd y=0
\end{align*}
for all $0\leq r,s,~ r+s\leq 2n-1$. By linearly combining these equalities, we conclude that
\begin{align}\label{eq:comb_annihilation}
\iint_\Omega g(x) g(y) q(x,y) \tilde{p}(x,y)\frac{\partial I(x,y)}{\partial x} \,\dd x \dd y=0
\end{align}
holds for any polynomial $q(x,y)$ of degree no higher than $2n-1$. From this point on, we set $q$ as
\begin{align}
q(x,y) = \tilde{p}(x,y) \frac{\partial}{\partial x} \Big(\zeta(x,y)  r(x,y) \Big).
\end{align}
Because $\textrm{deg}\,q \leq \textrm{deg}\,\tilde{p} + \textrm{deg}\,p-1=2n-1$, this choice of $q$ fulfills the degree constraint.

Let $y^{*}$ be such that the line $y=y^{*}$ intersects $\mathcal{C}$. According to Bezout's theorem, the number of intersections $m^{*}$ shall be limited to $n$. We assume the intersections are at $x\in \{x^{*}_i\}_{i=1}^{m^{*}}$ and conclude that
\begin{align}\label{eq:deriv_I}
\frac{\partial }{\partial x} I(x,y^{*}) = \sum_{i=1}^{m^{*}} s_i \, \delta(x-x^{*}_i),
\end{align}  
where $\delta(\cdot)$ is the Dirac's delta function and $\{s_i\}_i$ are sign values; $s_i=1$ ($s_i=-1$) if $p(x , y^{*}_i)$ is positive (negative) at $x=x^{*}_i-\epsilon$ and negative (positive) at $x=x^{*}_i+\epsilon$ for small enough $0<\epsilon$. Hence,
\begin{align*}
-s_i &= \lim_{\epsilon\rightarrow 0^{+}} \textrm{sign}\bigg(\frac{ p(x^{*}_i+\epsilon\,,\, y^{*})  - p(x^{*}_i-\epsilon\,,\, y^{*})  }{2\epsilon} \bigg)\nonumber\\
&= \lim_{\epsilon\rightarrow 0^{+}} \textrm{sign} \bigg(\frac{ \big(\zeta \cdot r\big)(x^{*}_i +\epsilon, y^{*})  -  \big(\zeta \cdot r\big)(x^{*}_i - \epsilon, y^{*})  }{2\epsilon} \bigg).
\end{align*}
This shows that the value of $\tfrac{\partial}{\partial x} \big(\zeta \cdot r\big)(x,y)$ at $(x^{*}_i, y^{*})$ is either $0$ or has the opposite sign as $s_i$. This implies that
\begin{align}\label{eq:Sign_ineq}
s_i \, \tfrac{\partial}{\partial x} \big(\zeta \cdot r\big)(x^{*}_i, y^{*}) \leq 0,
\end{align}
where equality happens only if $\tfrac{\partial}{\partial x} \big(\zeta \cdot r\big)(x^{*}_i, y^{*})=0$. By taking advantage of \eqref{eq:deriv_I}, we can rewrite the inner integral in \eqref{eq:comb_annihilation} as
\begin{align}
&\int g(x) g(y^{*}) q(x,y^{*}) \tilde{p}(x,y^{*})\frac{\partial I(x,y^{*})}{\partial x} \,\dd x \nonumber\\
&= \sum_{i=1}^{m^{*}} s_i g(x^{*}_i) g(y^{*}) q(x^{*}_i,y^{*}) \tilde{p}(x^{*}_i,y^{*}) \nonumber\\
&= \sum_{i=1}^{m^{*}} s_i g(x^{*}_i) g(y^{*}) \big(\tilde{p}(x^{*}_i,y^{*}) \big)^2 \tfrac{\partial}{\partial x} \big(\zeta \cdot r\big)(x^{*}_i, y^{*}) \leq 0.
\end{align}
Thus, for \eqref{eq:comb_annihilation} to hold, $q(x,y)$ needs to vanish at all points on $\mathcal{C}$, and in particular, at points on $\mathcal{C}_h$. As $h$ and $\tilde{p}$ are coprime, $\tilde{p}(x,y)$ can vanish only on a finite number of points on $\mathcal{C}_h$ (Bezout's theorem) . This forces the zero level set of $\tfrac{\partial}{\partial x} \big(\zeta \cdot r\big)$ to include $\mathcal{C}_h$ (inclusion of $\mathcal{C}_h$ except finitely many points implies inclusion of the whole $\mathcal{C}_h$). 

For any $(x^{*},y^{*})\in C_h$, because of $r(x^{*},y^{*})=h(x^{*},y^{*})=0$ we have that
\begin{align}
\tfrac{\partial}{\partial x} \big(\zeta \cdot r\big)(x^{*},y^{*}) = \zeta(x^{*},y^{*})    \tfrac{\partial}{\partial x} r(x^{*},y^{*}).
\end{align}
Again, since $h$ and $\zeta$ are coprime, $\zeta(x^{*},y^{*})=0$ can happen only for a finite number of points $(x^{*},y^{*})\in C_h$. Therefore, $\tfrac{\partial}{\partial x} r(x^{*},y^{*})=0$ should hold for all $(x^{*},y^{*})\in C_h$; \emph{i.e.}, the zero level set of  $\tfrac{\partial}{\partial x} r(x,y)$ includes the zero level set of $r(x,y)$. This, however, contradicts our initial assumption that $r$ is a polynomial with minimum degree that satisfies this property. $\hspace{\stretch{1}}\blacksquare$

\bibliographystyle{IEEEtran}
\bibliography{../../../refs}

\end{document}